\documentclass[twocolumn,twocolappendix]{aastex631}

\shorttitle{Formation of episodic jets and associated flares}
\shortauthors{Čemeljić et al.}
\graphicspath{{./}{figures/}}

\begin{document}

\title{Formation of episodic jets and associated flares from black hole accretion systems}

\correspondingauthor{Feng Yuan}
\email{fyuan@shao.ac.cn}

\author{Miljenko Čemeljić}
\affiliation{Shanghai Astronomical Observatory, Chinese Academy of Sciences; 80 Nandan Road, Shanghai 200030, China}
\affiliation{Nicolaus Copernicus Astronomical Center, Polish Academy of Sciences; Bartycka 18, 00-716 Warsaw, Poland.}
\affiliation{Academia Sinica, Institute of Astronomy and Astrophysics; P.O. Box 23-141, Taipei 106, Taiwan}

\author{Hai Yang}
\altaffiliation{Co-first author}
\affiliation{Shanghai Astronomical Observatory, Chinese Academy of Sciences; 80 Nandan Road, Shanghai 200030, China}
\affiliation{University of Chinese Academy of Sciences; 19A Yuquan Road, Beijing 100049, China}


\author{Feng Yuan}
\affiliation{Shanghai Astronomical Observatory, Chinese Academy of Sciences; 80 Nandan Road, Shanghai 200030, China}
\affiliation{University of Chinese Academy of Sciences; 19A Yuquan Road, Beijing 100049, China}

\author{Hsien Shang}
\affiliation{Academia Sinica, Institute of Astronomy and Astrophysics; P.O. Box 23-141, Taipei 106, Taiwan}

\begin{abstract}
Episodic ejections of blobs (episodic jets) are widely observed in black hole sources and usually associated with flares. In this paper, by performing and analyzing three dimensional general relativity magnetohydrodynamical numerical simulations of accretion flows, we investigate their physical mechanisms. We find that magnetic reconnection occurs in the accretion flow, likely due to the turbulent motion and differential rotation of the accretion flow, resulting in flares and formation of flux ropes. 
Flux ropes formed inside of 10-15 {gravitational radii} 
are found to mainly stay within the accretion flow, while flux ropes formed beyond this radius are ejected outward by magnetic forces and form the episodic jets. These results confirm the basic scenario proposed in \citet{Yuan09}. Moreover, our simulations find that the predicted velocity of the ejected blobs is in good consistency with observations of Sgr A*, M81, and M87.  The whole processes are found to occur quasi-periodically, with the period being the orbital time at the radius where the flux rope is formed. The predicted period of flares and ejections is consistent with those found from the light curves or image of Sgr A*, M87, and PKS 1510-089.  The possible applications to protostellar accretion systems are  discussed.  
\end{abstract}
\keywords{jet, flares,episodic jets --- black hole physical : jet --- hydrodynamics}

\section{Introduction} \label{sec:intro}

Jets are ubiquitous in astrophysical accreting systems. Two types of jets have been observed, namely continuous jets and episodic ones. In addition to the temporal difference,  the radiation spectrum, polarization, and velocities of these two types of jet are also different, as summarized in \citet{Fender04}. Episodic jets and radiation flares are often found to be physically associated with each other. One such example is Sgr A*, the supermassive black hole in the Galactic center. Flares are observed in multi-wavebands, ranging from radio, submm, infrared, and X-ray \citep{Baganoff01,Genzel2003,Eckart06,Trippe2007,Yusef-Zadeh2006,Marrone2008}. The flares are strongest in IR and X-ray \citep{Baganoff01,Genzel2003} and they occur simultaneously \citep{Genzel2003}, followed by flares at submm and radio wavebands \citep{Eckart06,Marrone2008}. The peak of the light curve at 43 GHz is found to lead that at 22 GHz by 20-40 minutes \citep{Yusef-Zadeh2006}. This time lag is regarded as the evidence for the ejection and adiabatic expansion of plasmoid from the accretion flow \citep{Yusef-Zadeh2006,Marrone2008}. This interpretation is confirmed by the 7 mm VLBA observations which found possible image of the ejected blob 4.5 hour later after the IR flare, and the speed of the blob is estimated to be 0.4 speed of light \citep{Rauch2016}. 

The phenomena of multi-waveband flares with frequency-dependent time lag and associated ejection are common in black hole sources. For example, large radio flares were detected in three AGNs, namely M81  \citep{King2016}, blazar PKS 1510–089 \citep{Park2019}, and M87 \citep{Hada2014}, preceded by an X-ray flare, $\gamma$-ray flare, and $\gamma$-ray flare in these three sources, respectively. Discrete blobs associated with flares were clearly detected by VLBA or VLBI, with the speed of the blob measured to be 0.5 and 0.2 speed of light in the case of M81 \citep{King2016} and M87 \citep{Hada2014}. Similar phenomenon occurs in a black hole X-ray binary, Cygnus X-1 \citep{Wilms2007}.  

Flares and episodic ejection (as jet knots) have also been observed in protostellar accretion systems \citep{Lee2020,Wolk2005,Feigelson1999}. Although consensus has not been reached on the mechanisms of the ejection and flares, both of them are believed to be powered by magnetic activities in the accretion flow \citep{Feigelson1999}, so it is natural to conjecture that they are physically associated with each other. More interestingly, even in the case of the Sun, large solar flares and coronal mass ejections (CMEs) also occur together \citep{2000JGR...105.2375L,Zhang2005}. 

Many works have been published to explain the origin of black hole flares, usually by invoking synchrotron radiation or Compton scattering by relativistic electrons. These electrons are either accelerated by shocks in jets \citep{Markoff2001} or by magnetic reconnection in accretion flows \citep{Yuan2004,Yuan09,Dodds2009,Dodds2010,Nathanail20,Nathanail2021,Ripperda20,Ripperda2021,Dexter20,Dexter2021,2020MNRAS.499.1561Z}. 
These works usually focus only on the flares, with little discussions on episodic ejections and their associations with flares.

An exception is \citet{Yuan09}. In this work, by analogy with the coronal mass ejections in the Sun, they propose that the turbulent motion and differential rotation of the black hole accretion flow will deform the magnetic field lines; consequently, magnetic reconnection occurs and flux ropes are formed in the coronal region of the accretion flow, which will then be ejected by the magnetic compression force. 

These processes must be highly dynamic but assumptions and simplifications have to be adopted in the analytical treatment of \citet{Yuan09}. 
In the present work, we perform three dimensional  magnetohydrodynamical numerical simulations to examine the  scenario proposed in \citet{Yuan09}, investigating the formation of episodic jets and their association with flares. The paper is structured as follows. In {\S\ref{sec:model}}
, we briefly describe our model of general relativity magnetohydrodynamical simulations of hot accretion flows. In {\S\ref{sec:floats}}
, we present the main results obtained from our simulations, including the formation of flux rope, its ejection, and the periodicity of the whole processes. In {\S\ref{sec:observation}}
, we discuss the applications of our model to observations, including the periodic flares detected in Sgr A* and other sources, the hot spot in Sgr A* detected by the GRAVITY observations, the velocity and the possible periodicity of the ejected blobs in Sgr A*, M81, and M87, and finally the periodic ejection and flares in protostellar accretion systems. 
The comparison between our present work and previous ones are presented in {\S\ref{sec:comp}}. Finally, we summarize our work in {\S\ref{summary}}.

\begin{figure*}[htbp!]
 
   \plottwo{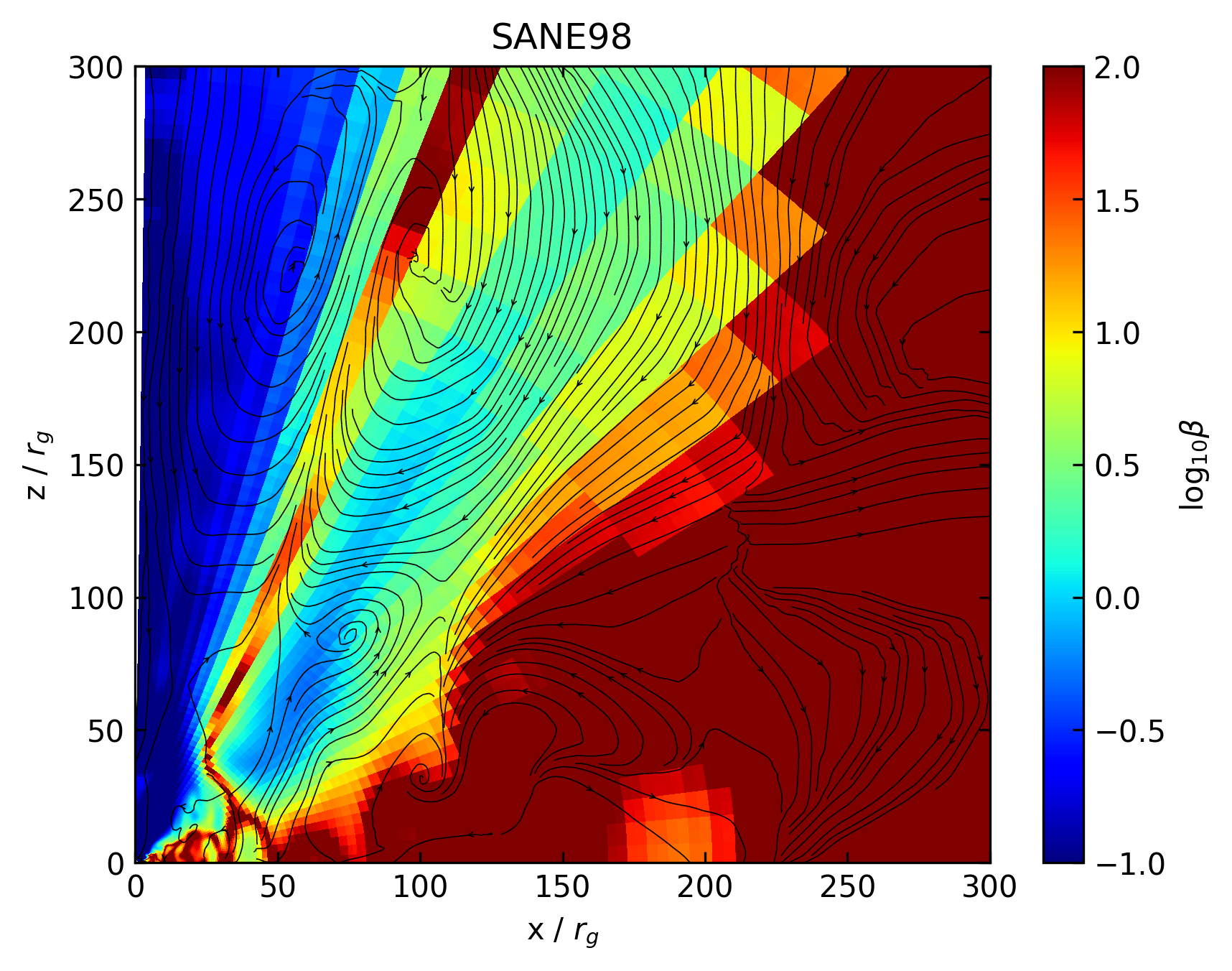}
   {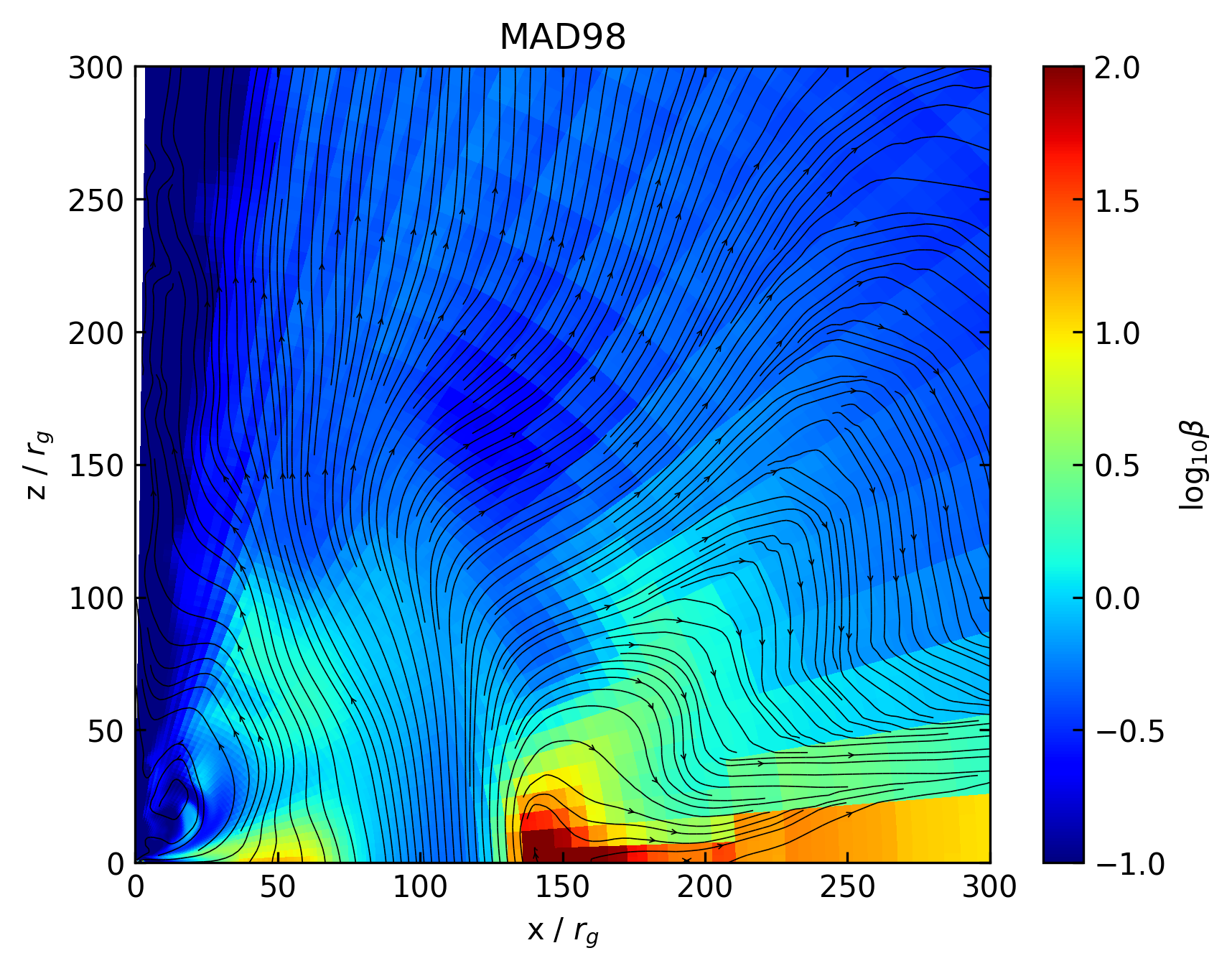}
    \caption{Snapshots of properties of SANE and MAD in the $x-z$ plane. The color indicates {plasma} $\beta$, black lines denote magnetic field lines .}
    \label{fig:figs1}
\end{figure*}

\section{Model} \label{sec:model}
All black hole sources mentioned in the introduction section are powered by hot accretion flows \citep{Ho2008,YuanNar14}. We therefore perform a set of three-dimensional general relativity magnetohydrodynamical numerical simulations of a hot accretion flow around a black hole using the ATHENA++ code. Spherical coordinates are adopted. We have simulated two kinds of accretion flows, namely SANE (standard and normal evolution) and MAD (magnetically arrested disk), with the magnetization of the accretion flow in MAD much stronger than SANE \citep{Igum03,Nar03,Tchek11}. Two values of black hole spin are considered, $a=0$ and $a=0.98$. The details of the simulations are presented in Appendix A. We find that the value of $a$ is not important for our problem, so in the following we simply refer the models as SANE and MAD. Snapshots of the distributions of magnetic field lines and plasma $\beta$ (defined as the gas-to-magnetic pressure ratio) of the models are shown in Figure~\ref{fig:figs1}.


\section{Results} \label{sec:floats}
\subsection{Formation  of flux ropes}
 Solar flares and CMEs have been intensively studied, and the physical mechanism is known to be magnetic reconnection \citep{Zhang2005,Gou2019}. After reconnection, some plasma will be enclosed within helical magnetic field lines and form a “flux rope’’. The flux rope will then be ejected by strong magnetic pressure force and form the CMEs. It was proposed in \citet{Yuan09} that similar processes should also occur in the black hole accretion flows and result in the formation of episodic jets. 
 
\begin{figure*}
\includegraphics[width=.305\linewidth]{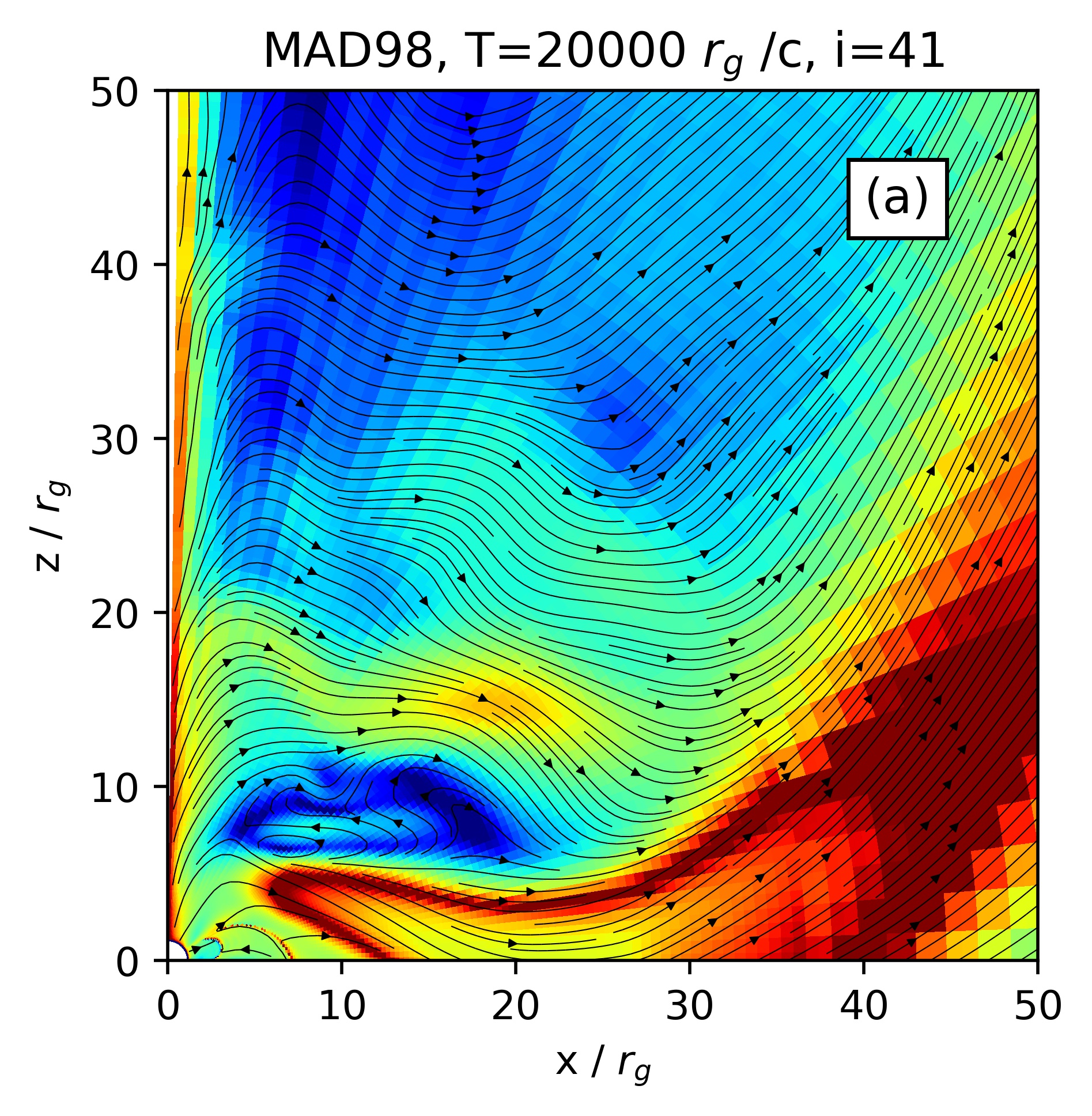}\quad\includegraphics[width=.305\linewidth]{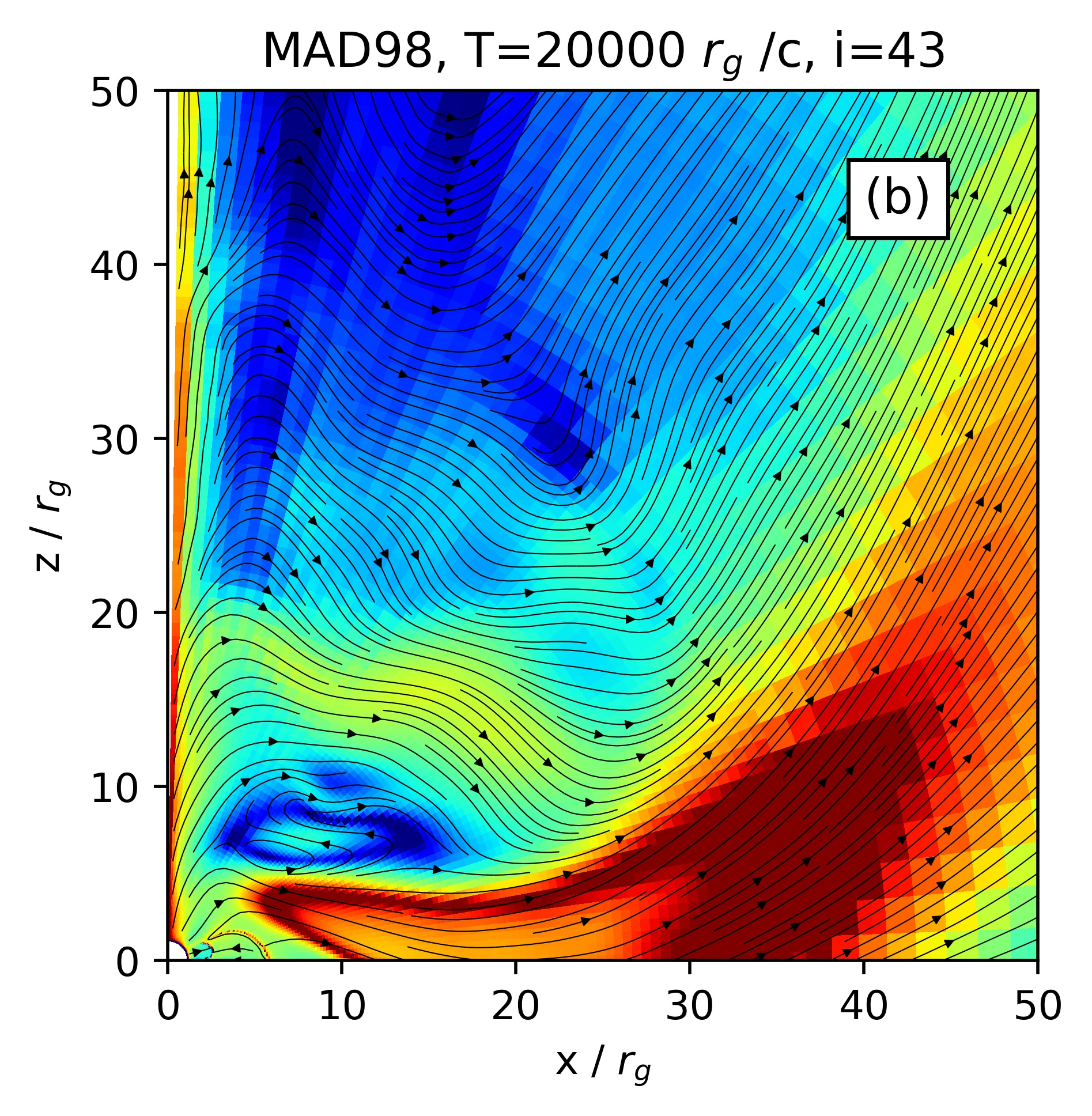}\quad\includegraphics[width=.39\linewidth]{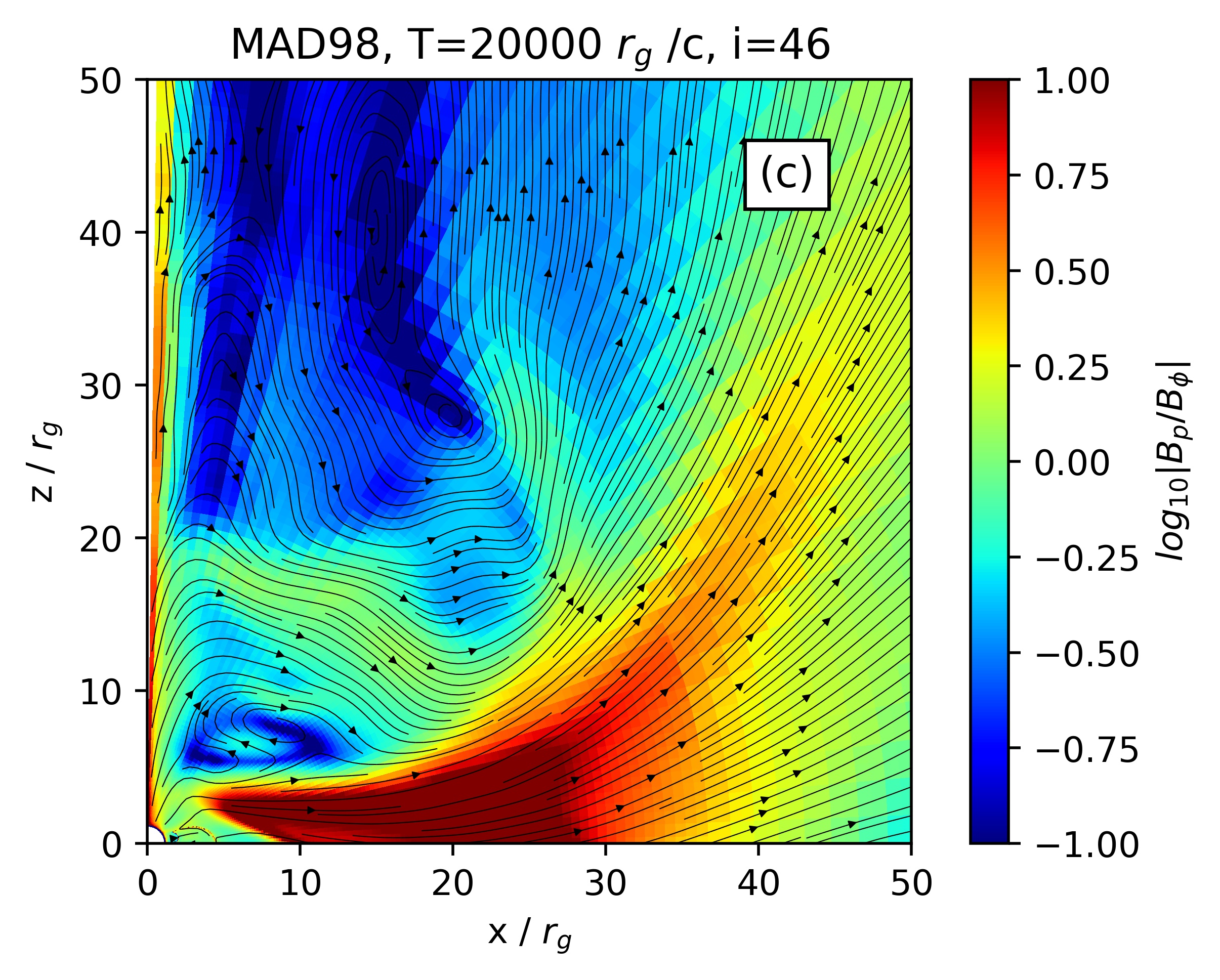}
\includegraphics[width=.305\linewidth]{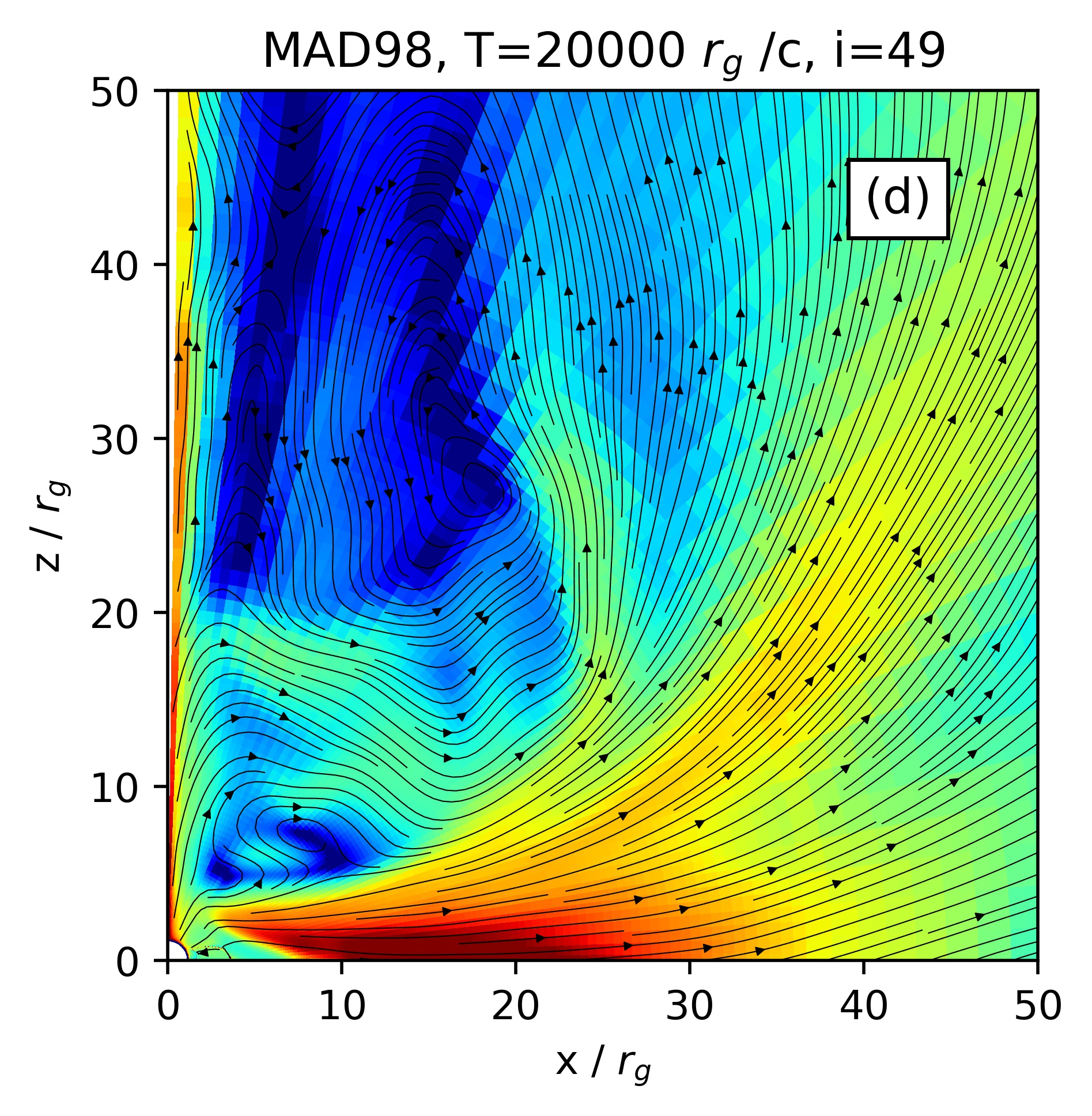}\quad\includegraphics[width=.305\linewidth]{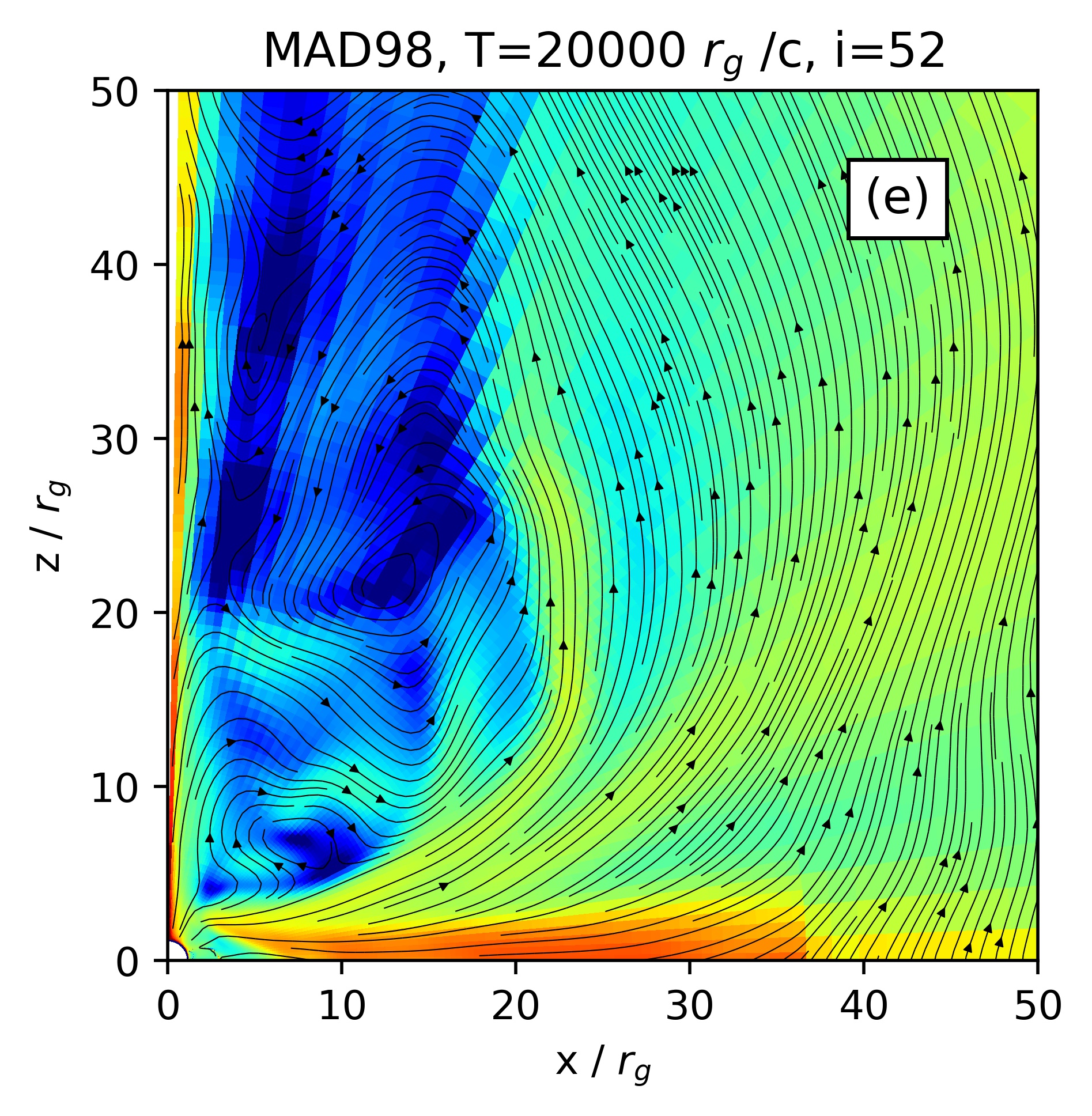}\quad\includegraphics[width=.39\linewidth]{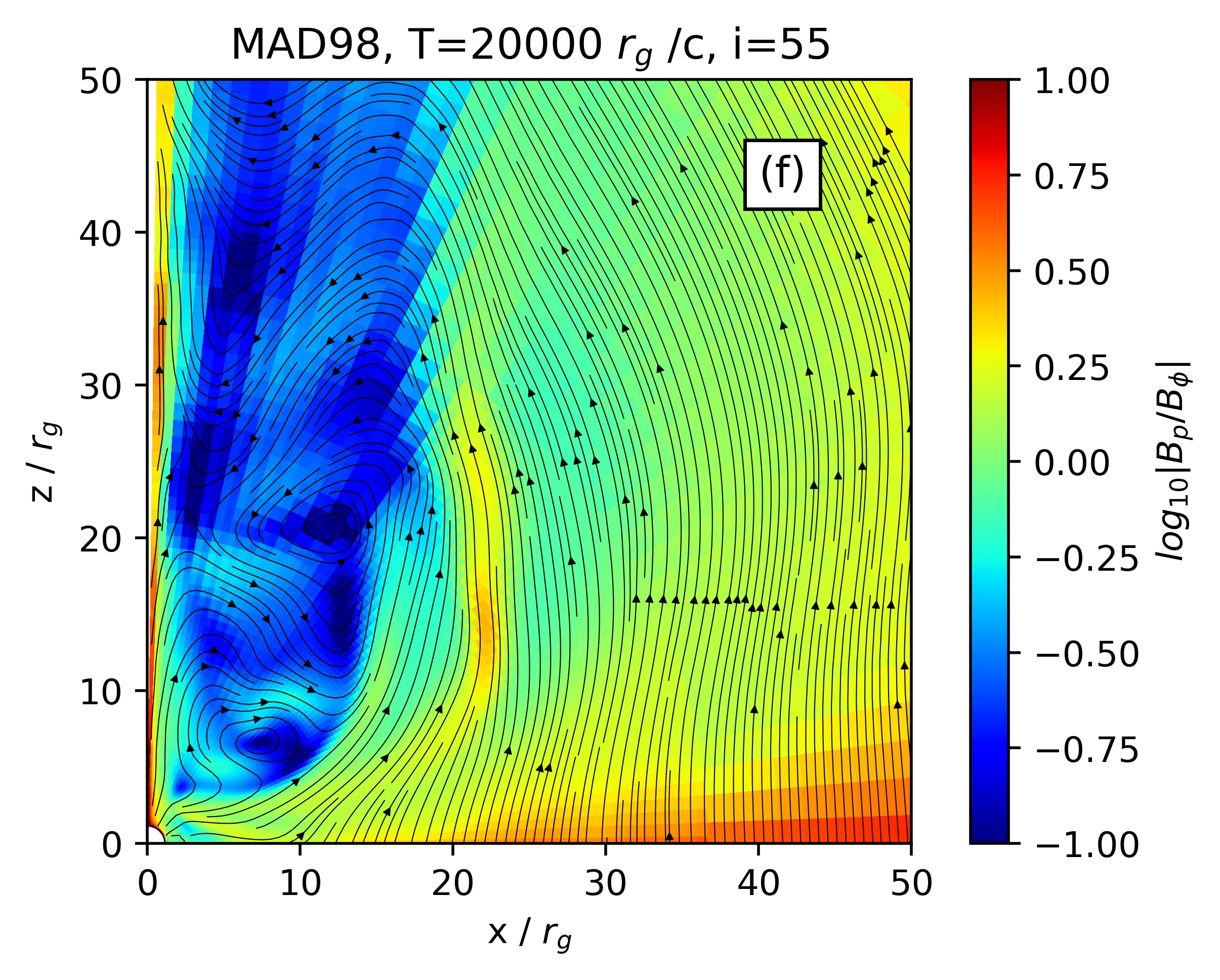}

\includegraphics[width=.305\linewidth]{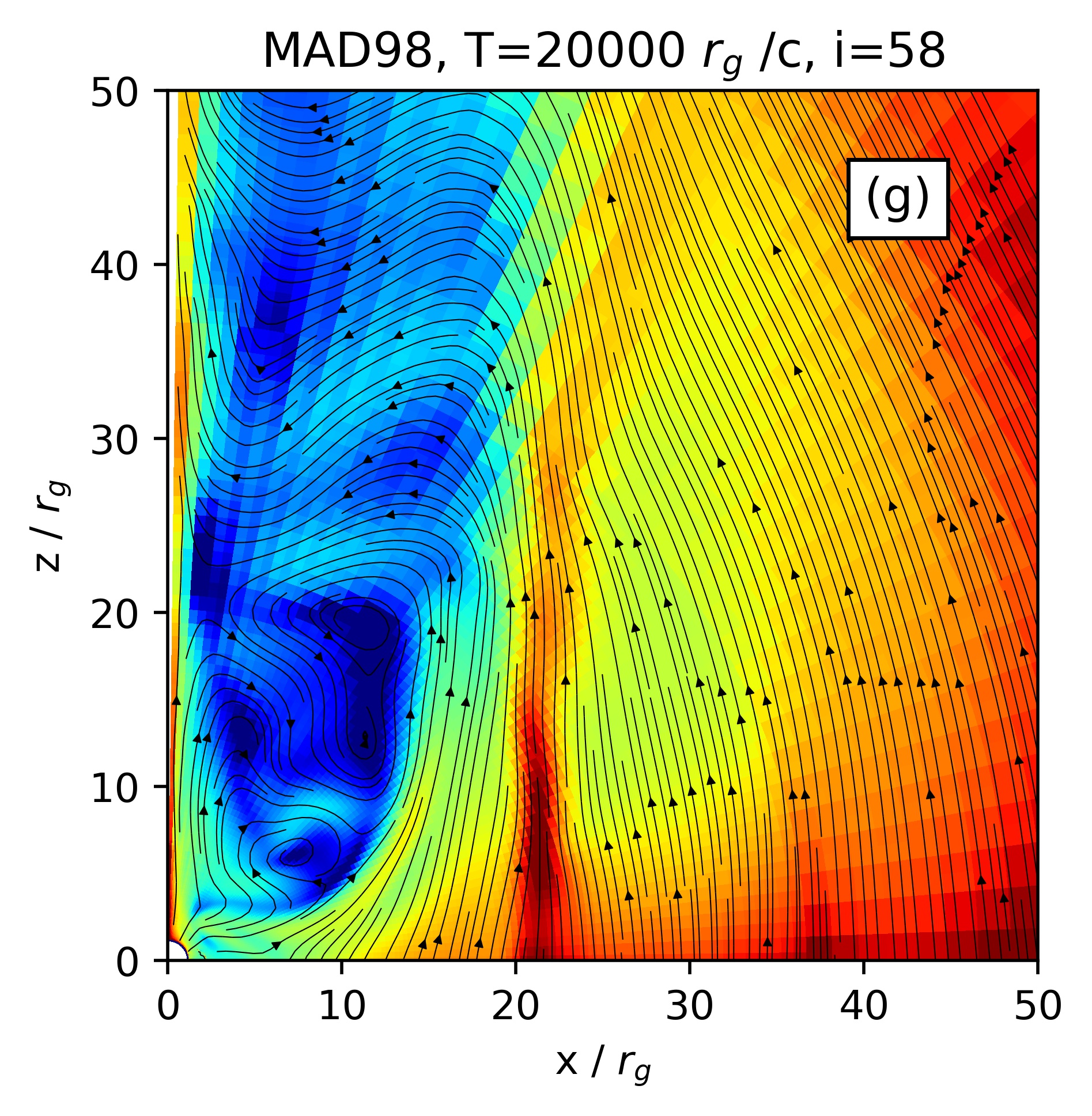}\quad\includegraphics[width=.305\linewidth]{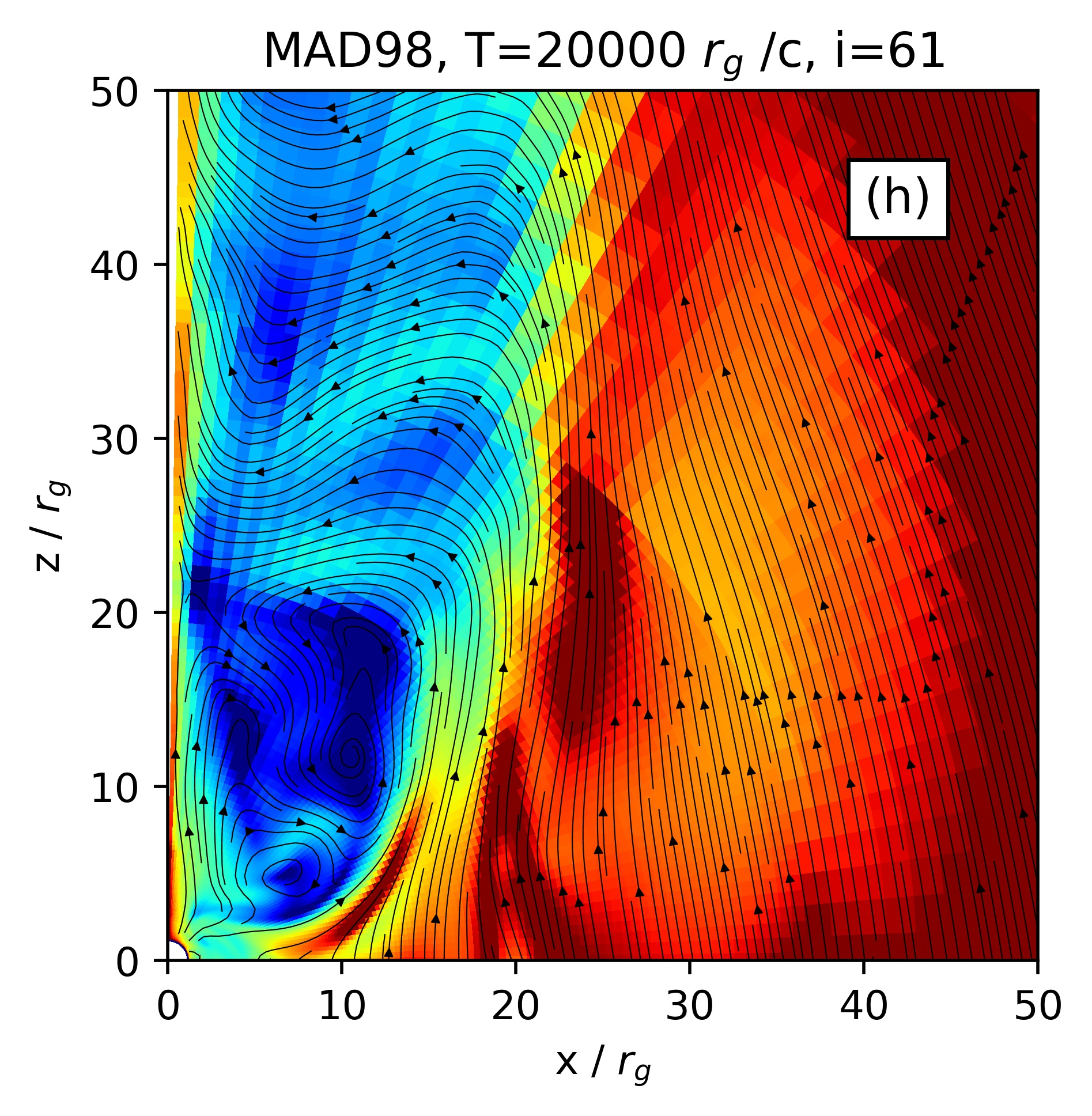}\quad\includegraphics[width=.39\linewidth]{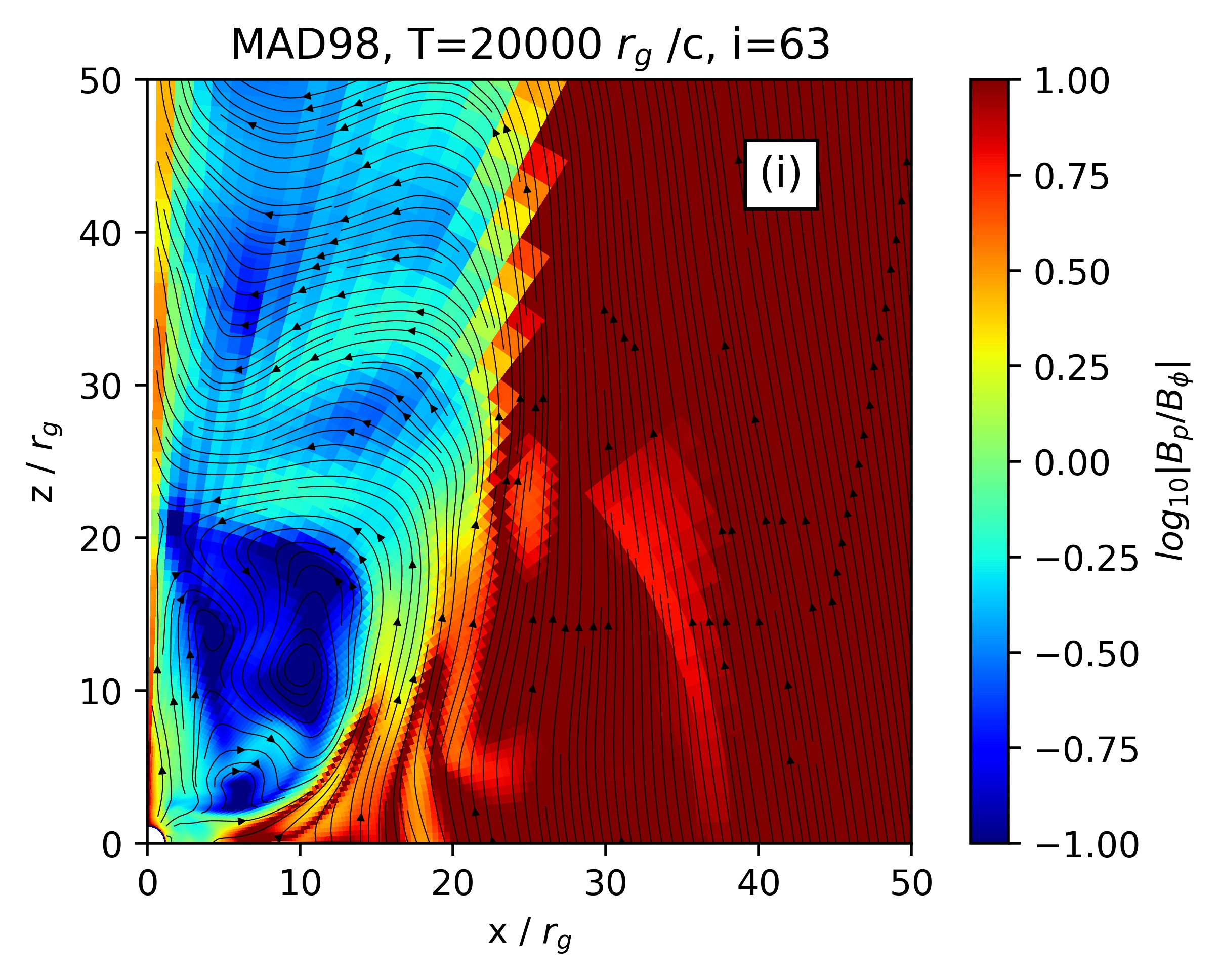}

\caption{The poloidal magnetic field structure of a flux rope identified in MAD and at simulation time of T=20000{$r_g/c$} at various azimuthal angles $\varphi$ (denoted by the grid cell number $i$ at the top of each plot) in the $x-z$ plane. The magnetic field is denoted by black solid lines with arrows indicating the direction of the field while the color indicates the logarithmic value of the ratio of poloidal and toroidal magnetic field strength. The flux rope is located at $(x,z)=(5-10,5-10)$, depending on the value of $\varphi$.  In this example, the flux rope extends in the $\varphi$-direction for roughly $(63-41)\times5.6=123^{\circ}$.}
\label{fig:fluxrope}
\end{figure*}

\begin{figure*}
\includegraphics[width=0.65\columnwidth]{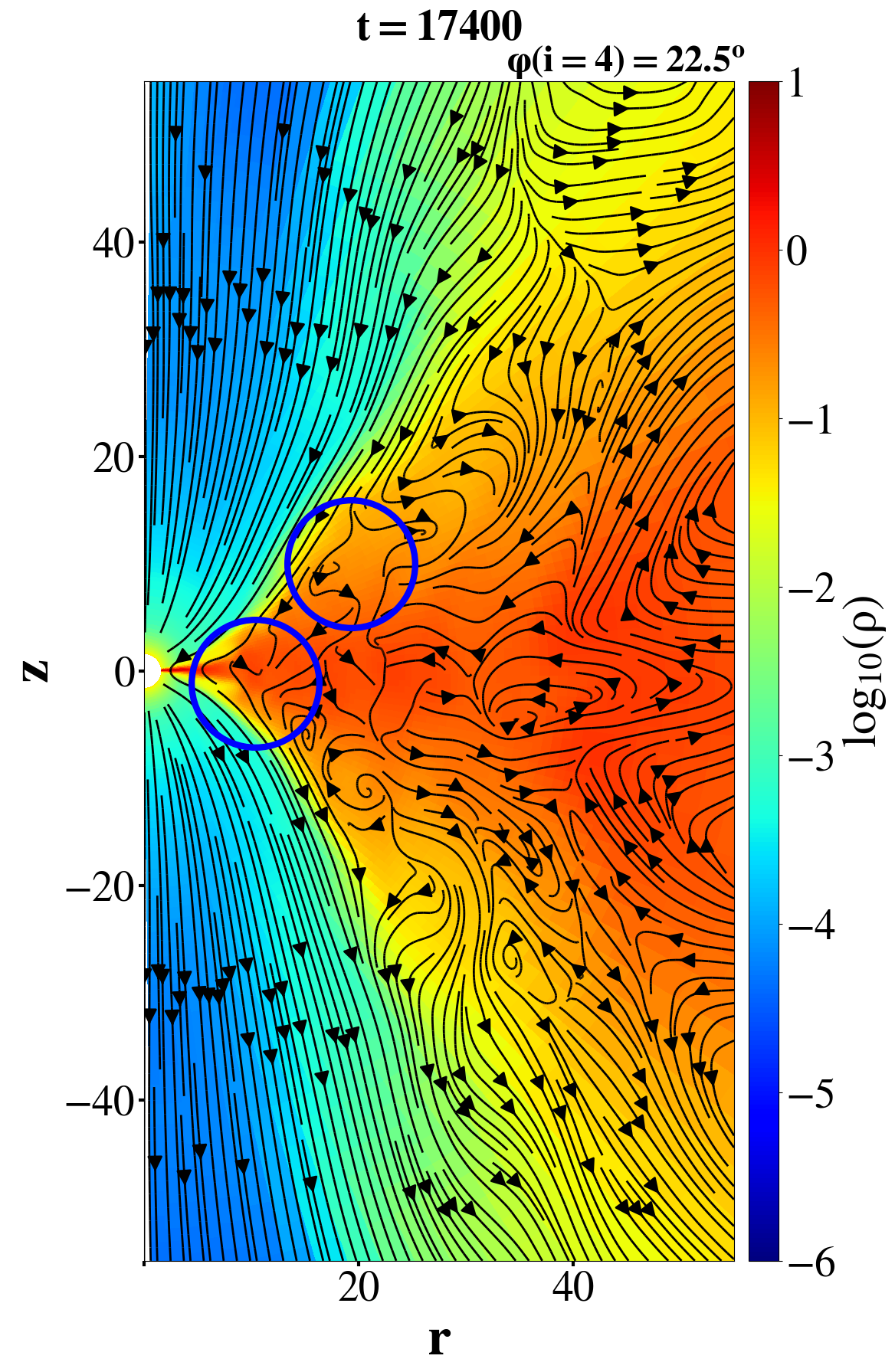}
\includegraphics[width=0.65\columnwidth]{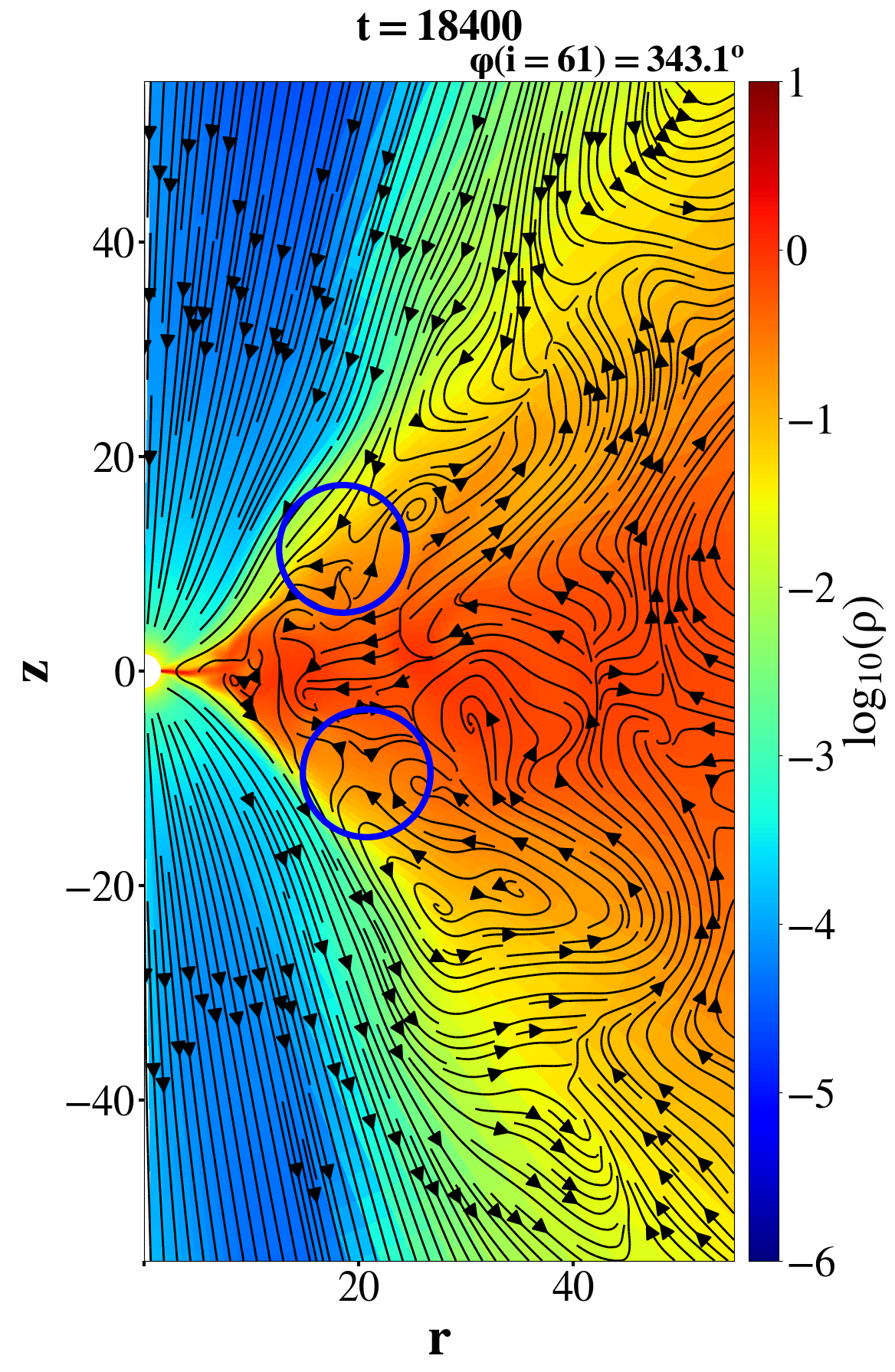}
\includegraphics[width=0.65\columnwidth]{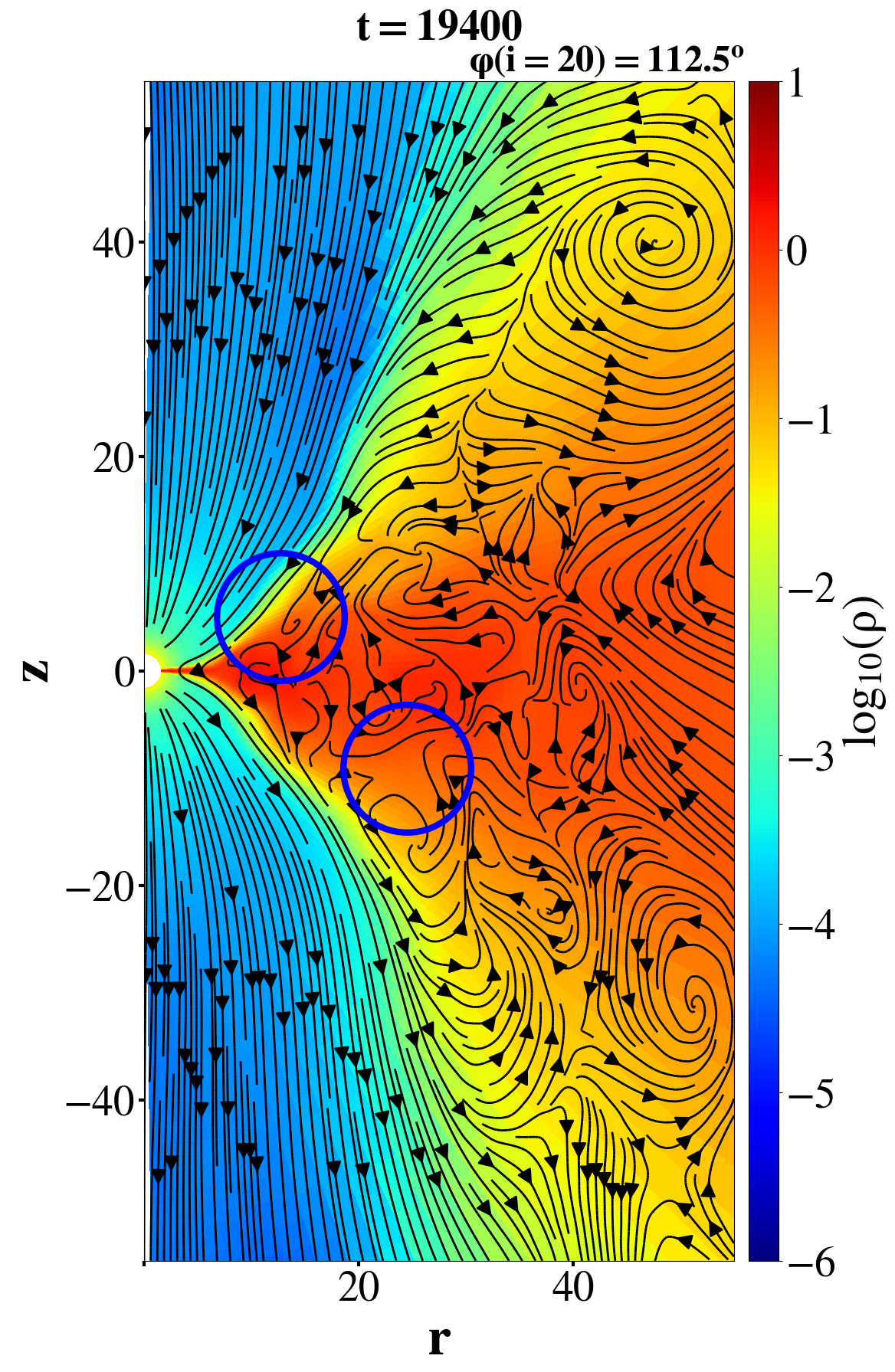}
\caption{Six reconnection layers from our SANE simulation (marked with blue circles). They will evolve and result in the occurrence of magnetic reconnection and formation of six magnetic islands in $100r_g/c$, shown by the first column of Figure \ref{fig:isla}.  The density is shown by colour and magnetic field by solid lines, with arrows indicating the direction of the field.}
\label{islarec}
\end{figure*}

\begin{figure*}
\includegraphics[width=0.65\columnwidth]{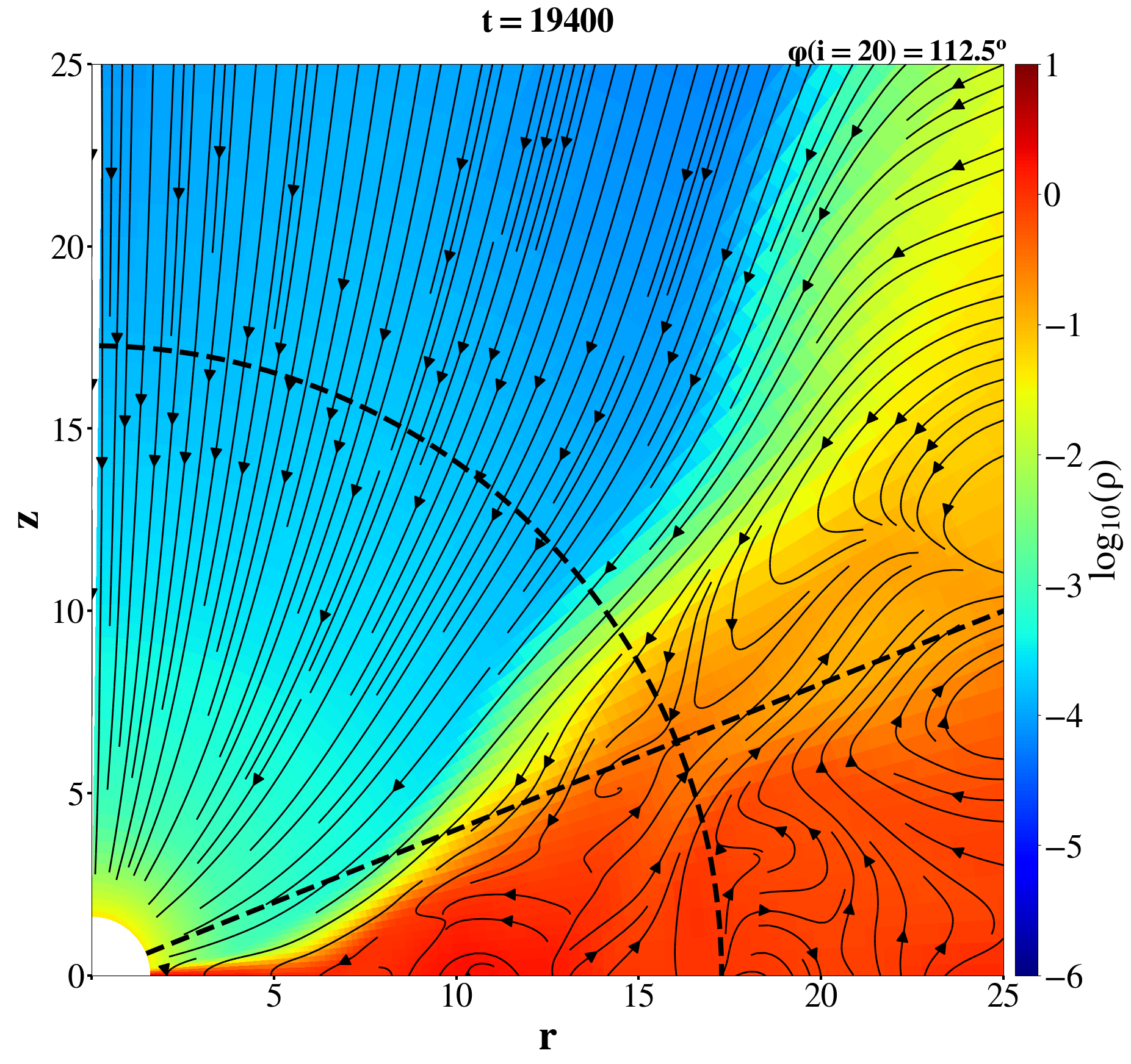}
\includegraphics[width=0.65\columnwidth]{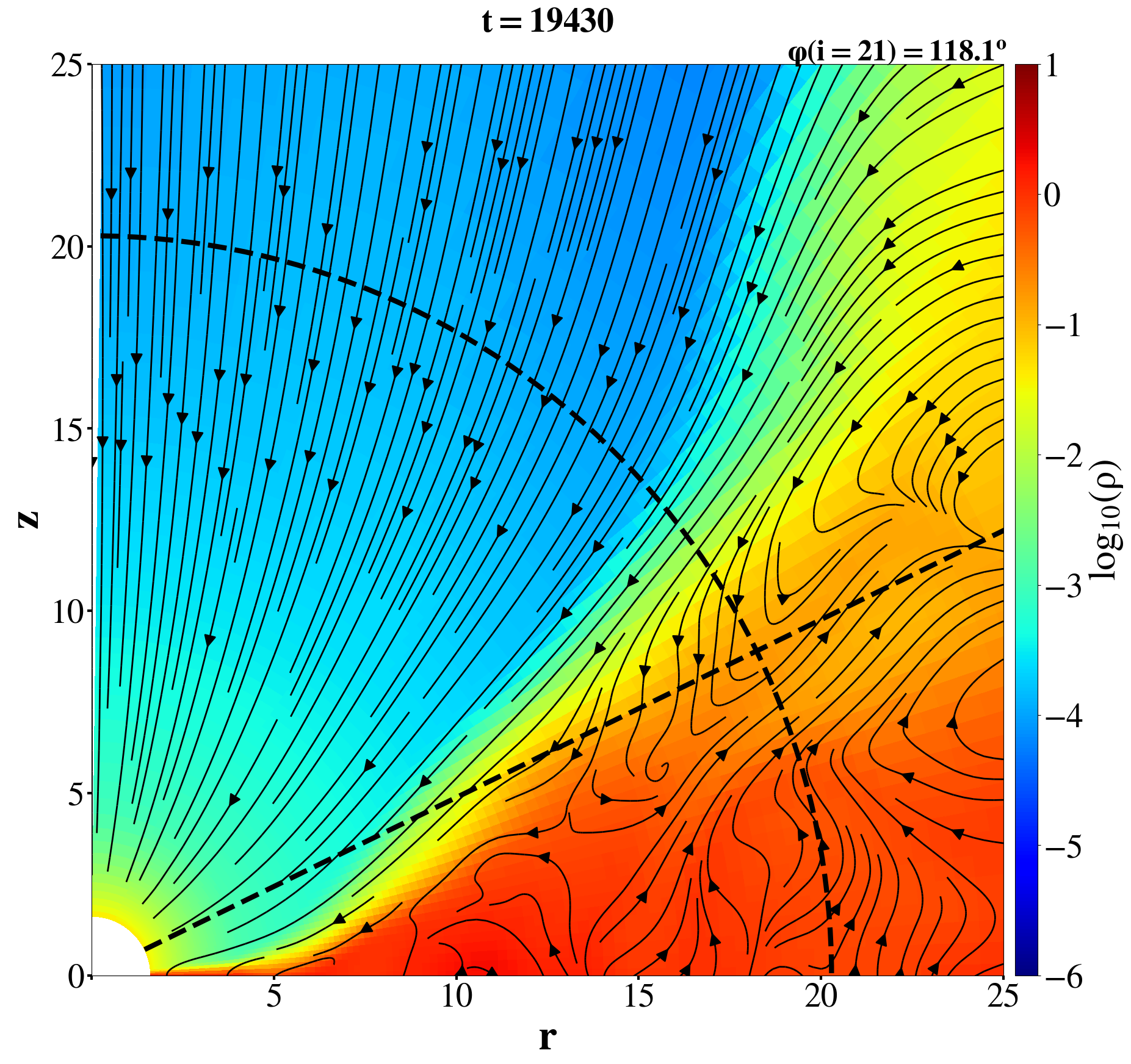}
\includegraphics[width=0.65\columnwidth]{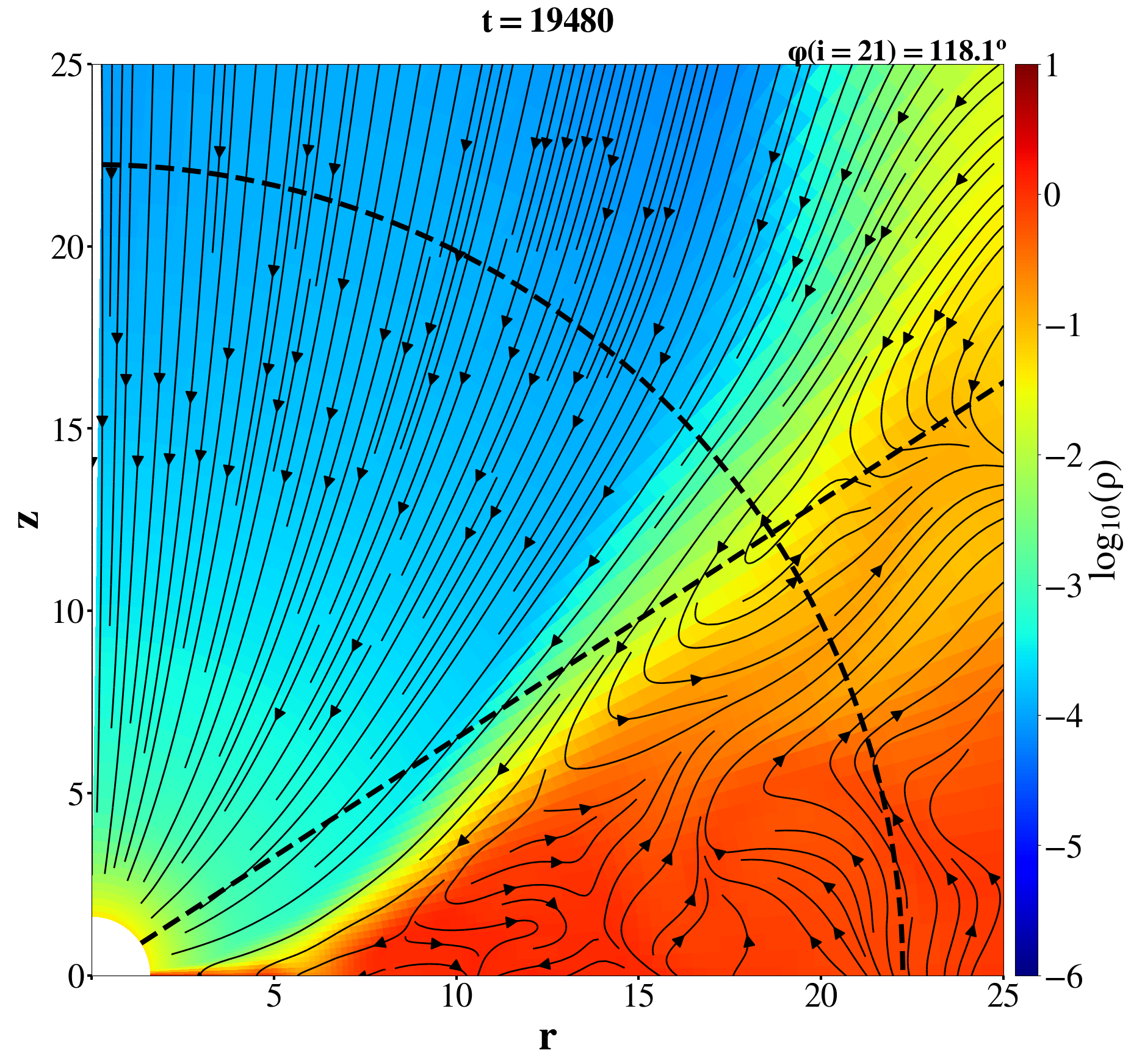}
\caption{Evolution of the magnetic field structure around a reconnection layer shown in the rightmost panel of Figure \ref{islarec}, i.e, at $t=19400r_g/c$. Formation of magnetic island and its further motion can be found in the bottom panels of Fig. \ref{fig:isla}.  }
\label{islarec2}
\end{figure*}

 Stimulated by the above scenario, we have analyzed the simulation data of accretion flows, looking for flux ropes.  We do find that many such helical structure magnetic field lines (i.e., the flux ropes) exist both within and at the coronal region of the accretion flow, from radius as small as 4$r_g$ up to $\sim 30~r_g$, where $r_g$ stands for gravitational radius. The flux ropes are found to keep forming throughout the evolution of the accretion flows in all four models once they reach the steady state. 
 In the present paper, we have tried to find as many as possible flux ropes. We then measure the properties of individual flux ropes and present the typical results. 
 
 One example of flux rope identified in the case of MAD at time T=20000~$r_g/c$ is shown in Figure~\ref{fig:fluxrope}. The poloidal magnetic field structure  is presented in the $x-z$ plane at nine various azimuthal angles\footnote{ The time-averaged magnetic field configuration at a larger scale will show an ``hourglass'' shape, as shown by Figure 5 in \citet{Yang21}.}. As we can see from the figure, in the two-dimensional plane, these flux ropes will be represented by circular magnetic field line patterns --- “magnetic islands”. Note that if a constant-$\varphi$ slice through the flux rope were not perpendicular to the axis of the rope, we would see a whirl in poloidal field lines instead of a circular magnetic island.  

\begin{figure*}
\includegraphics[width=2\columnwidth]{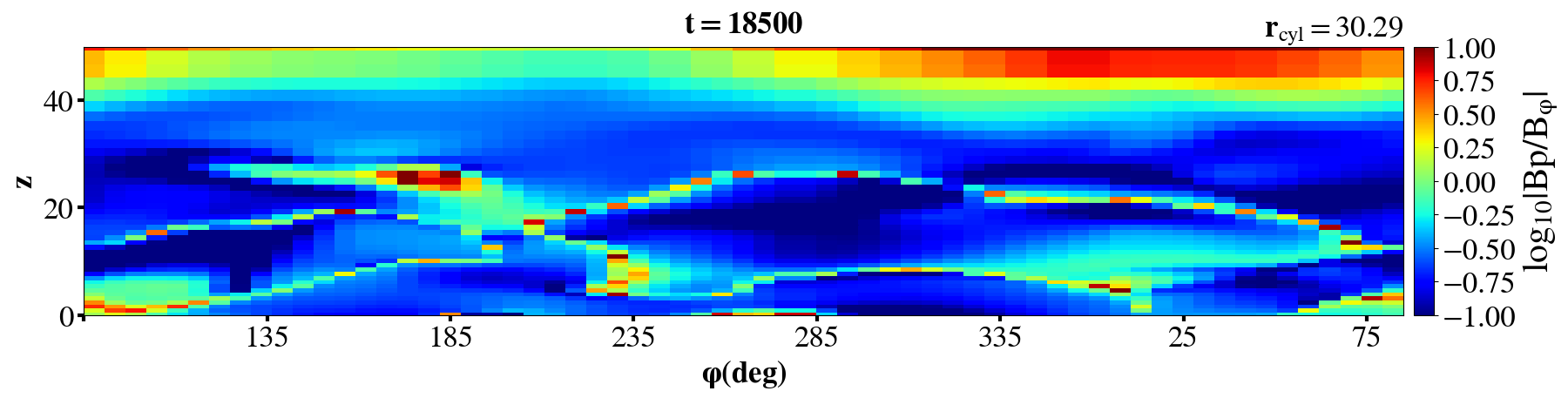}
  \includegraphics[width=2\columnwidth]{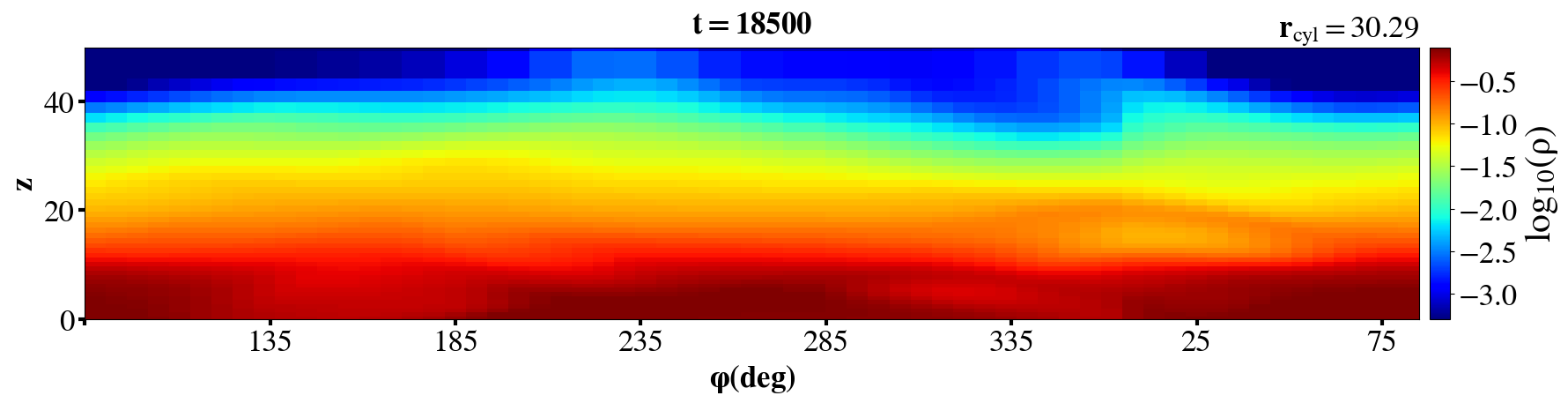}
    \caption{The spatial distribution of the magnetic field strength and density when a flux rope is present in the SANE model. Top panel: Distribution of the ratio of poloidal and toroidal magnetic field strength in the $\varphi-z$ plane with a constant cylindrical radius $r= 30.29$ above the equatorial plane at $T = 18500~r_g/c$. The poloidal field reaches a local minimum at the center of the flux rope, which is shown as a slice through the azimuthal extensions of the magnetic islands. Bottom panel: Distribution of density in the same $\varphi-z$ plane. It is clear that the density of gas in the flux rope ({\bf middle of which is located around $(\varphi,z)=(25^\circ ,20)$ in this case}) is higher than the surrounding medium. Note that only parts of the arcs of the flux ropes lie in this $\varphi-z$ plane, because the cylindrical radius of the flux rope is not constant.  }
    \label{fig:banddensity}
\end{figure*}

To understand the physical mechanism of the formation of flux ropes, after we find a flux rope, we have traced back to the time and location of the flux rope formation. The motion of the flux ropes is traced by examining their movement in different $r-\theta$ planes with various angles $\varphi$. The center of the flux rope is identified by the minimum of the poloidal magnetic field and the circular pattern of the magnetic field lines, as we will explain at the end of section 3.1. Figures \ref{islarec} \& \ref{islarec2} show the tracing results. Three panels in Figure \ref{islarec} show the magnetic field structure in the $r-\theta$ plane at three snapshots. In each panel we can see two reconnection layers, located above and below the equatorial plane of the accretion flow respectively\footnote{We note that the appearance of reconnection layer and subsequent occurence of magnetic reconnection are found to occur randomly, not necessarily simultaneously above and below the equatorial plane.}. These reconnection layers will continue to evolve, result in the occurrence of magnetic reconnection and formation of magnetic islands after a time of $\sim 100r_g/c$. As an example, the evolution process of the reconnection layer located above the equatorial plane in the rightmost (i.e, $t=19400$) snapshot of Figure \ref{islarec} is illustrated in Figure \ref{islarec2}. The six magnetic islands finally formed are shown by the first column of Figure \ref{fig:isla}. We emphasize that similar steps of evolution from the reconnection layer to the magnetic island can be traced in all the cases of island formation captured in our simulations. It is interesting to note that similar to our case, magnetic reconnection is also invoked to explain the formation of solar prominence \citep{Chenpf2020}. 

We speculate that the physical mechanism underlying the above processes is likely as follows.  In the inner region of the accretion flow, the field lines are tangled, due to the MHD turbulence driven by magnetorotational instability \citep{Balbus1991}. Thus the field lines with opposite polarity can come close enough, resulting in the occurrence of reconnection. In addition, the Parker instability and wind launched from the accretion flow will make the magnetic field lines emerge out of the accretion flow into the corona. The differential rotation and turbulent motion of the accretion flow where the field lines are rooted twist the field lines in the coronal region and make the reconnection occur there as well. This scenario of the formation of flux ropes is fully consistent with that proposed by \citet{Yuan09}. It has also been confirmed by other recent MHD numerical simulations \citep{Nathanail20,Nathanail2021,Ripperda20,Ripperda2021}, and it is confirmed again by our present analysis. The detailed comparison with these works will be presented in Section \ref{sec:comp}. 

We find that, the formation of flux rope occurs not only in the coronal region, but also within the accretion flow, as also pointed out in \citet{Nathanail20}.  The reconnection rate is proportional to the Alfven speed. Since $\beta \gg 1$ in the accretion flow, the reconnection rate is much smaller there compared to the coronal region. Combined with the fact that the energy density of the magnetic field in the corona is also stronger than that in the accretion flow \citep[e.g.,][]{Yang21}, we expect that the energy release
during reconnection occurs slowly and weakly thus may not correspond
to observed strong flares. Because of these reasons, in the present paper, we only focus on the
flux ropes formed in the coronal region of the accretion flow.

Because of the reconnection, the poloidal magnetic field at the center of the flux rope should become weaker compared to its surrounding medium. This is confirmed by the top panel of Figure~\ref{fig:banddensity}. In fact, the movement of flux ropes is followed by tracing the trajectories of those minima and the circular magnetic field line pattern. The released magnetic energy should be converted into the thermal and kinetic energy of plasma, and be used to accelerate electrons whose radiation will be responsible for flares. In addition, the density within the flux rope is larger than its surrounding medium, as shown in the bottom panel of Figure~\ref{fig:banddensity}. We speculate the reason as follows. The weakening of the magnetic field results in the decrease of magnetic pressure within the flux rope; consequently the pressure from the surrounding medium compresses the flux rope and increases its density and pressure until a new pressure balance within and outside of the flux rope is established.  Combining
the top and bottom panels of the figure, we can find that the plasma $\beta$ within the flux rope should be higher while the magnetization parameter $\sigma\equiv B^2/(\rho c^2)$ should be lower than its surrounding medium. These results are consistent with \citep{Ripperda20}.

\begin{figure}
  \includegraphics[width=1\columnwidth]{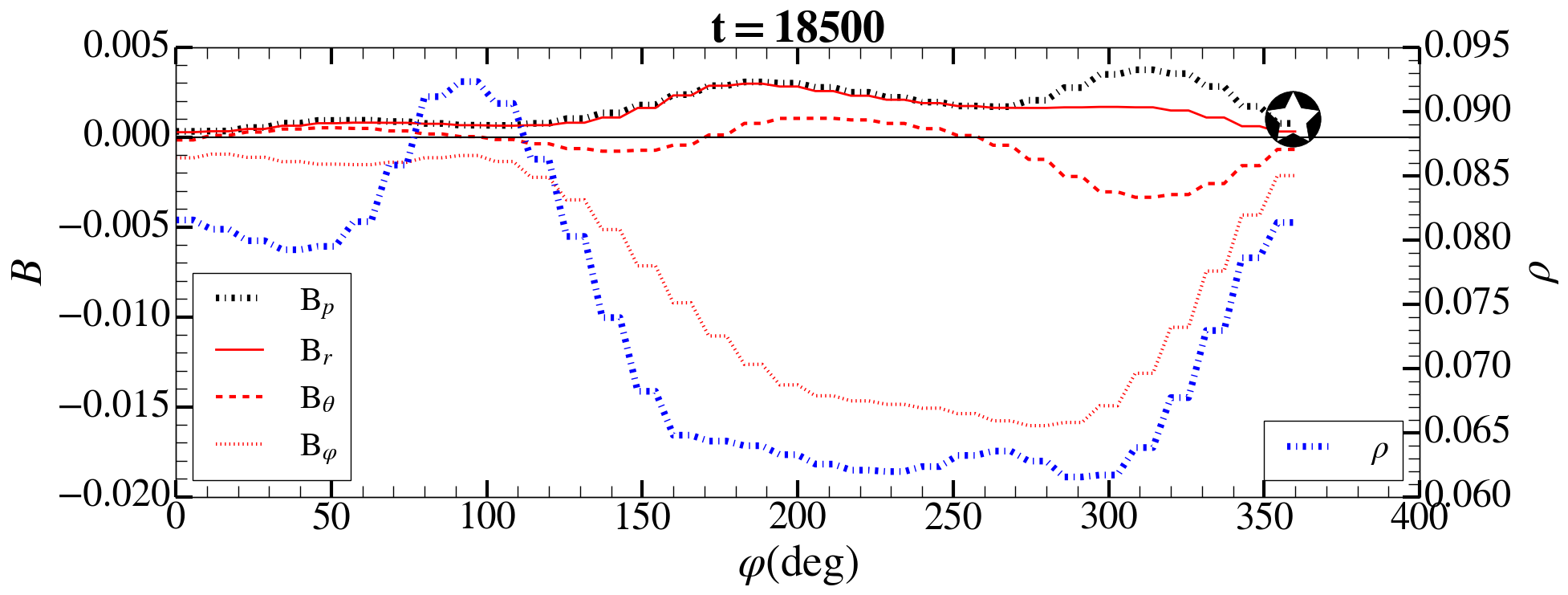}
    \caption{The $\varphi$-profiles of various components of magnetic field and density of the flux rope shown in Fig. \ref{fig:banddensity} at about $z =27r_g$
in SANE, with the star mark denoting the same
magnetic island shown in Fig. \ref{islarec}. Note that since the cylindrical radius of the flux rope at different $\varphi$ is not a constant, in this figure the flux rope is present only at two places.}
    \label{fig:densityB}
\end{figure}

To present quantitative results, the change of various components of the magnetic field and density as a function of $\varphi$ at a given cylindrical radius are given in Figure \ref{fig:densityB}. Note
that since the cylindrical radius of the flux rope is different for different $\varphi$, in this figure, the flux rope is present only at two places, i.e., at $\varphi\sim 100^{\circ}$
and $\varphi\sim 350^{\circ}$. At these two
places, the poloidal field reaches its minima while the density reaches its maxima.

\begin{figure}
  \includegraphics[width=1\columnwidth]{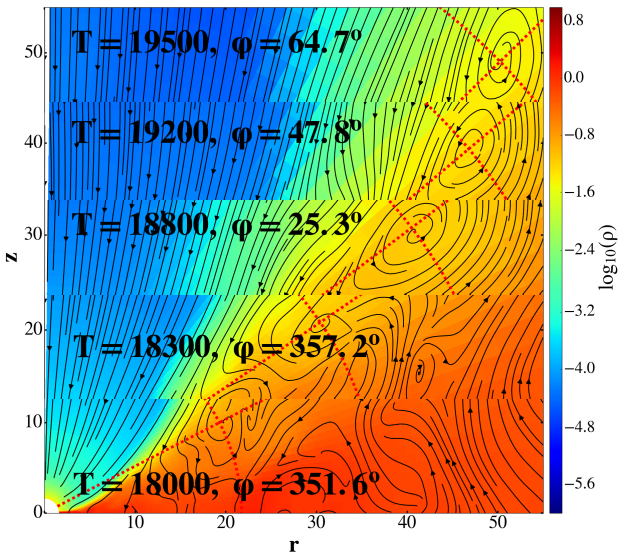}
    \caption{The movement of a flux rope (i.e., a magnetic island) in the SANE simulation, denoted by slices crossing the center of the flux rope at different times and azimuthal angle $\varphi$.  We can see that with the increase of the radial distance of the island from the black hole, its corresponding $\theta$-angle is decreasing, indicating a  ``collimation'' effect.}
    \label{fig:spiral}
\end{figure}

\begin{figure}

  \includegraphics[width=1\columnwidth]{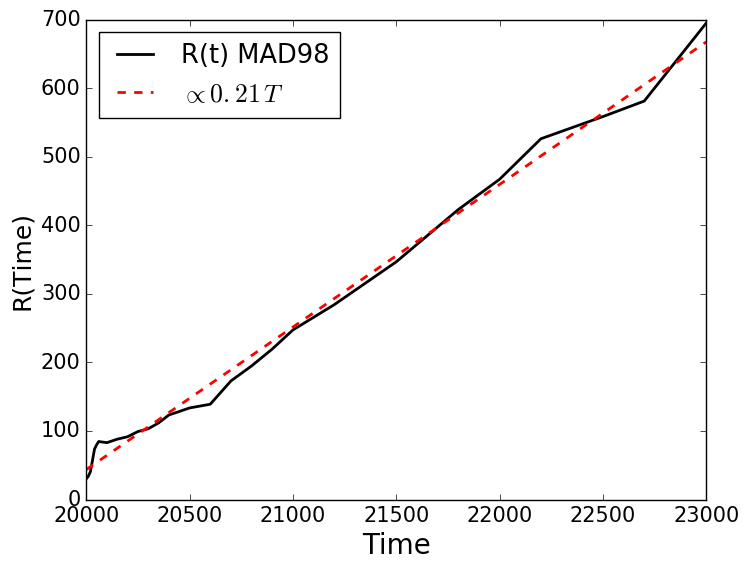}
    \caption{The time-dependent location (represented by spherical radius) of a plasmoid in the case of MAD.}
    \label{fig:movement}
\end{figure}

\begin{figure*}
 
   \plottwo
      {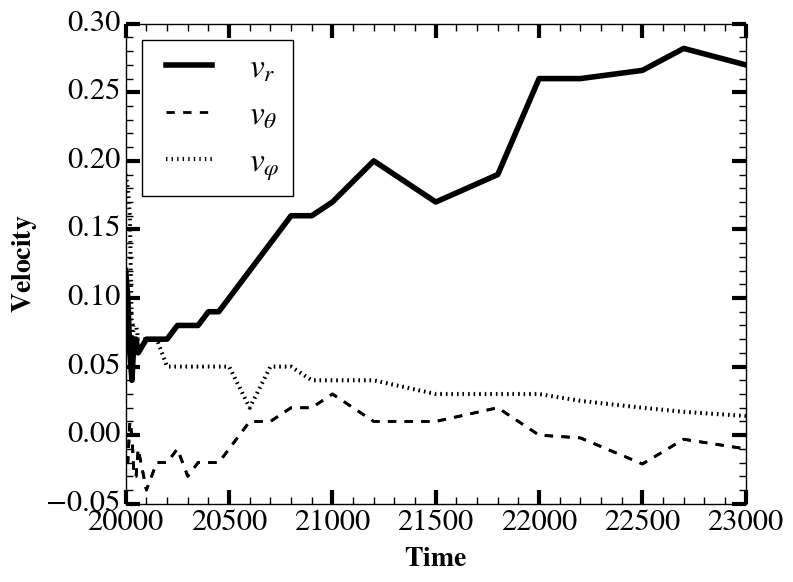}
      {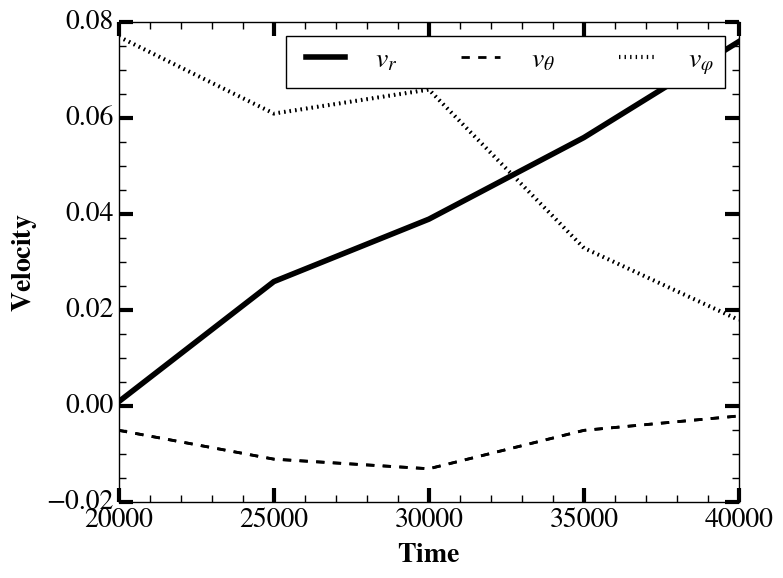}
    \caption{ (Left:) Various velocity components in units of speed of light of the plasmoid shown in Fig. \ref{fig:movement} as a function of time. (Right): Various velocity components in units of speed of light of a plasmoid in the case of SANE as a function of time.}
    \label{fig:blobvel}
\end{figure*}

\subsection{Ejection of flux ropes}

Although flux ropes can be formed at both small and large radii, we find that there is an important difference between them.  At small radii, within roughly $r\sim 10-15\, r_g$, although the flux ropes can be formed, few of them are ejected out. Many flux ropes just stay within the accretion flow and finally fall onto the black hole. This is likely because of the strong gravitational force of the black hole, similar to the existence of the ``stagnation radius’’ within which the motion of the matter is always toward the black hole. This is indicated by the direction of the velocity vectors close to the black hole shown in \citet{Yang21} (refer to Fig. 3 therein). Such a result is also consistent with the absence of wind in the innermost region of the accretion flow \citep{Yang21}. For other flux ropes, although they are found to be able to propagate outward for some distance, they quickly disappear because of the strong differential rotation of the accretion flow at small radii which tears the flux ropes. 

Beyond 10-15 $r_g$, however, we find that the flux ropes  can be ejected out. This is because both the gravity and the differential rotation become weaker. The flux rope is found to become entangled soon after its ejection, because its two end-points are anchored in two different radii of the accretion disk which have different angular velocities. Such an entanglement would lead to the ejection of the flux rope due to the kink instability within an orbital timescale, similar to the solar and space physics cases \citep{Wang2016,2021JPlPh..87f9016S}. The short lifetime is consistent with previous analytical argument \citep{Broderick2006} and perhaps also related to the short lifetime of three flares detected by high resolution GRAVITY observations \citep{Gravity2018}. 

We trace the motion of the broken flux rope (i.e., plasmoid) by examining its movement. The result is shown in Fig. \ref{fig:spiral}. We can see from the figure that the plasmoid is spiraling outward, reminding us of the helical structure of magnetic field in a continuous jet. It is interesting to note that the corresponding  $\theta$-angle of the plasmoid decreases with the increasing distance to the black hole, indicating a collimation effect, which is again similar to the behavior of continuous jets. 

To give a quantitative result, we show by Figure \ref{fig:movement}  the radial location of a plasmoid in MAD as a function of time. The left panel of Fig. \ref{fig:blobvel} shows various components of the velocity as a function of time for this plasmoid.  It reaches as high as 0.3 speed of light at T=23000~$r_g/c$ from 0.05 speed of light at T=20000~$r_g/c$. As a comparison, we show by the right panel of Figure \ref{fig:blobvel} the evolution of various component of the velocity in the case of SANE. Compared to MAD, the typical values of both the magnitude of velocity and acceleration are much smaller. 
\begin{figure}

  \includegraphics[width=1\columnwidth]{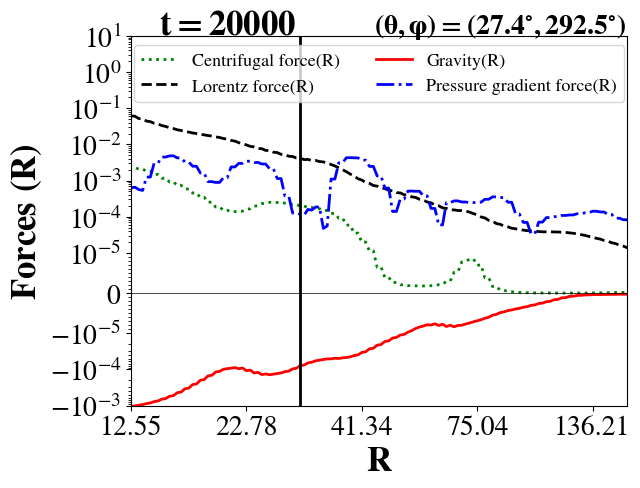}
    \caption{Forces along the radial direction acting on the magnetic island shown in the left plot of Figure~\ref{fig:blobvel}. The values of $\theta$ and $\varphi$ are indicated atop the upper right corner of the figure. The vertical solid line denotes the location of the magnetic island. We can see from the figure that the Lorentz force is the dominant force that accelerates the motion of the flux rope. }
    \label{fig:force}
\end{figure}

To understand the physical reason for the acceleration of the plasmoid, we have analyzed the forces acting on the flux rope shown in Figure~\ref{fig:movement} (and the left panel of Figure~\ref{fig:blobvel}). They include gravitational force, centrifugal force, Lorentz force, and the gradient of gas pressure. The result is shown in Figure~\ref{fig:force}. We can see from the figure that the Lorentz force is the dominant force that accelerates the plasmoid. The dominant component of the Lorentz force is the gradient of the magnetic pressure, which is enhanced by the occurrence of reconnection below the plasmoid. From this result, we can understand why the acceleration in the SANE case is much weaker than MAD; it is because the plasma $\beta$ in SANE is more than ten times larger than in MAD (refer to Figure~\ref{fig:figs1} and Figure 4 in \citet{Yang21}). Such a scenario of the magnetic acceleration of the ejected plasmoid confirms the prediction in \citet{Yuan09}, and is very similar to the case of solar CMEs \citep{Zhang2005}.

\begin{figure*}[htbp]
\includegraphics[width=0.55\columnwidth]{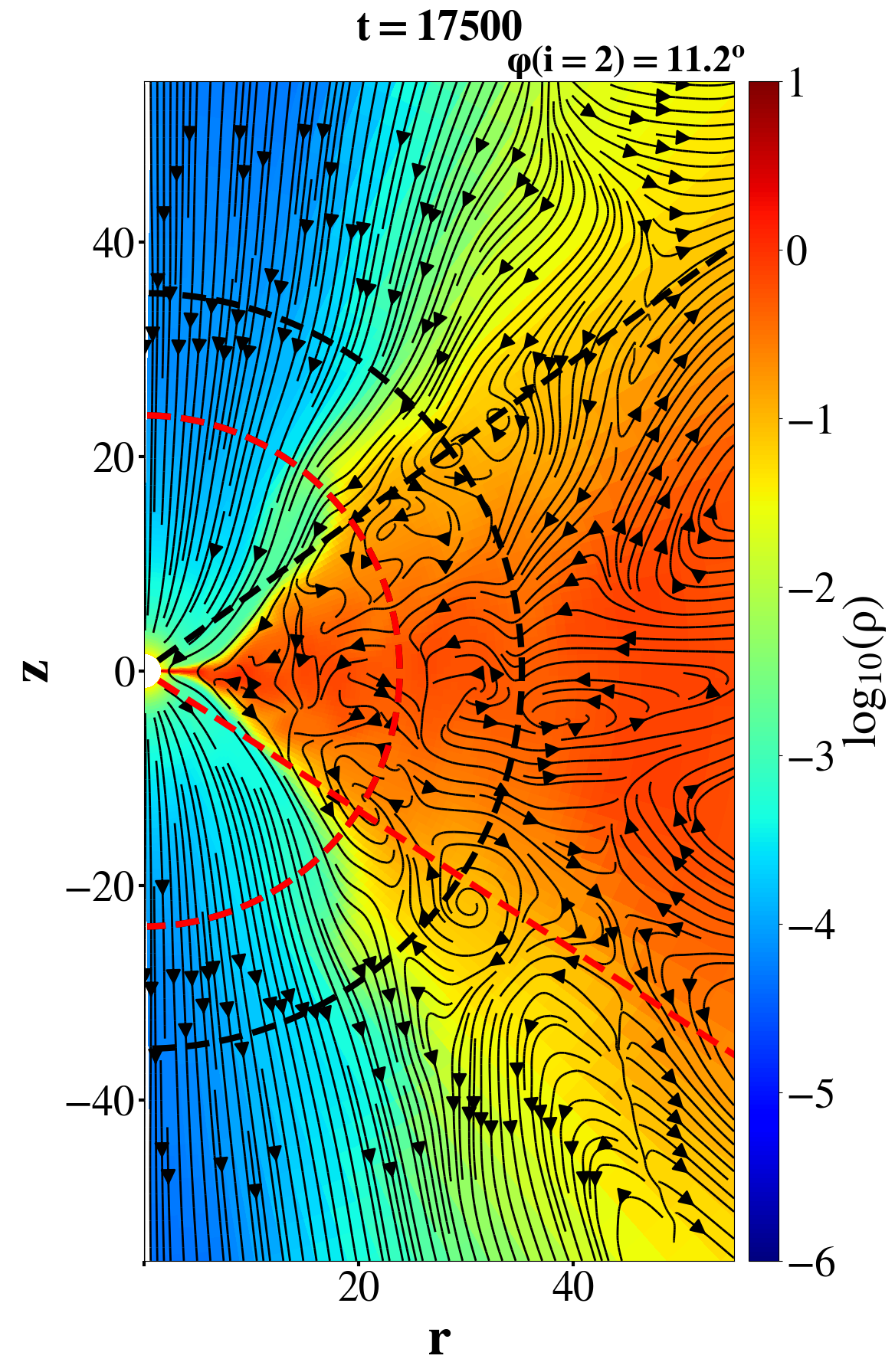}
\includegraphics[width=0.55\columnwidth]{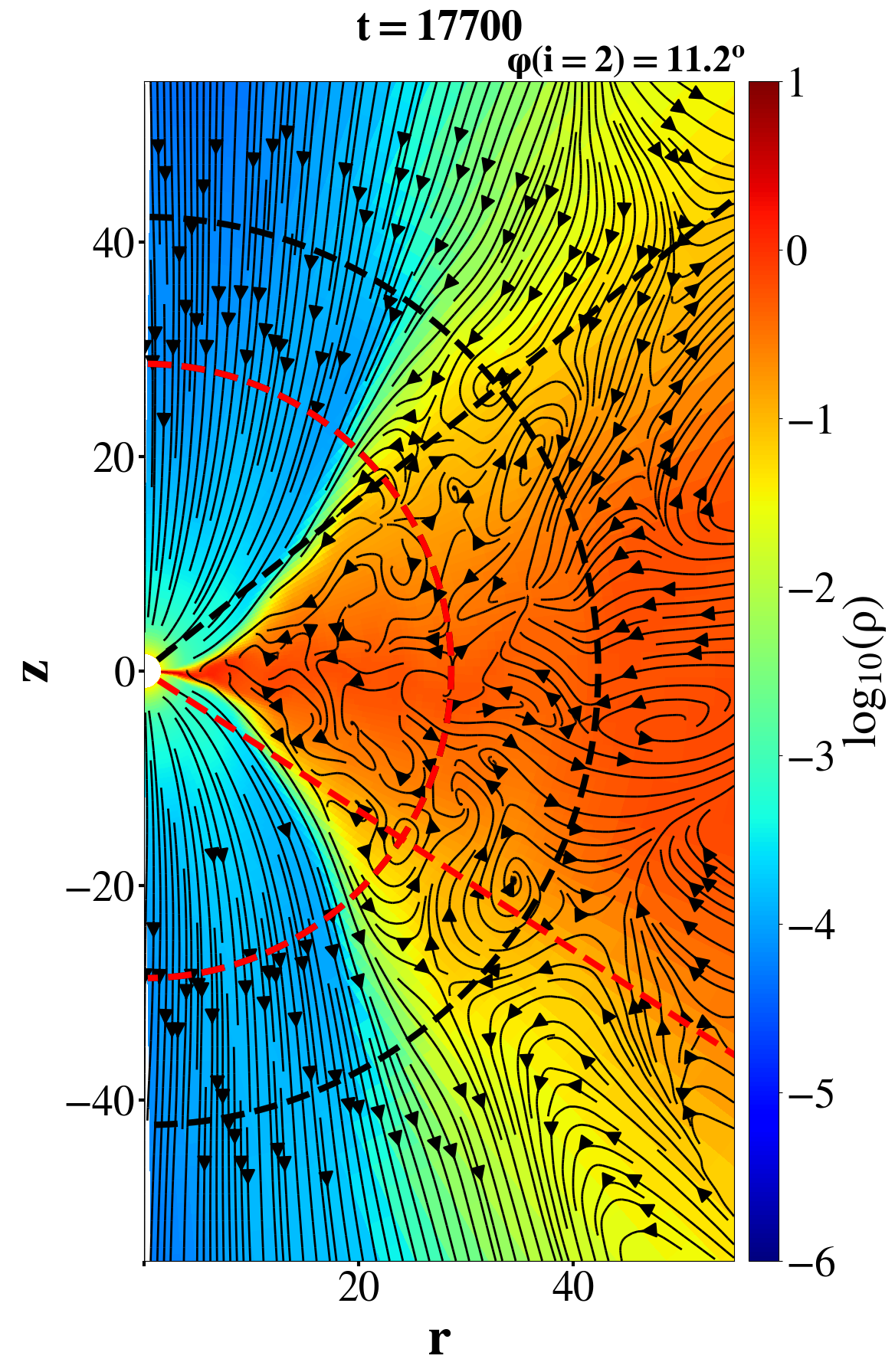}
\includegraphics[width=0.55\columnwidth]{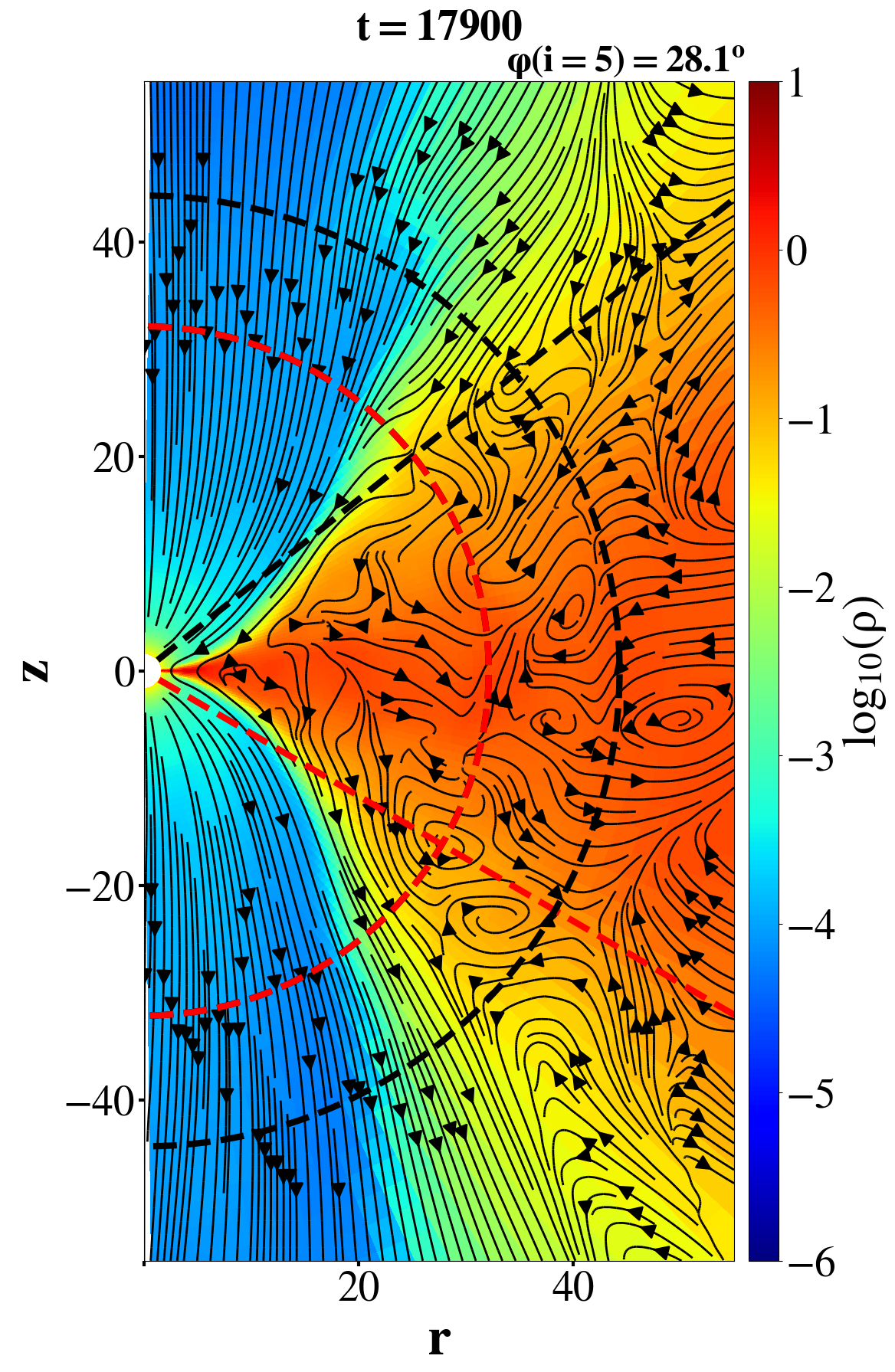}\\
\includegraphics[width=0.55\columnwidth]{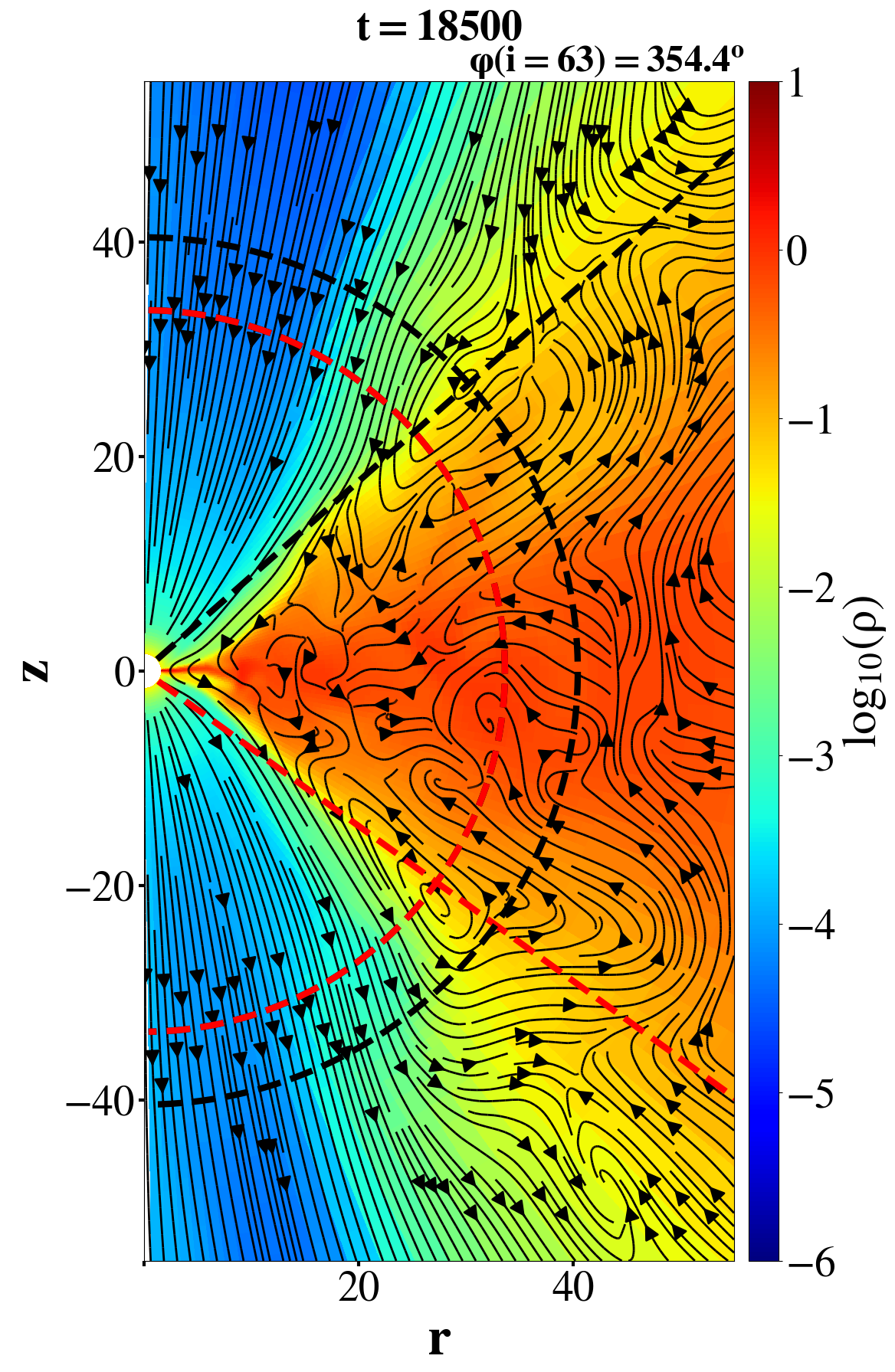}
\includegraphics[width=0.55\columnwidth]{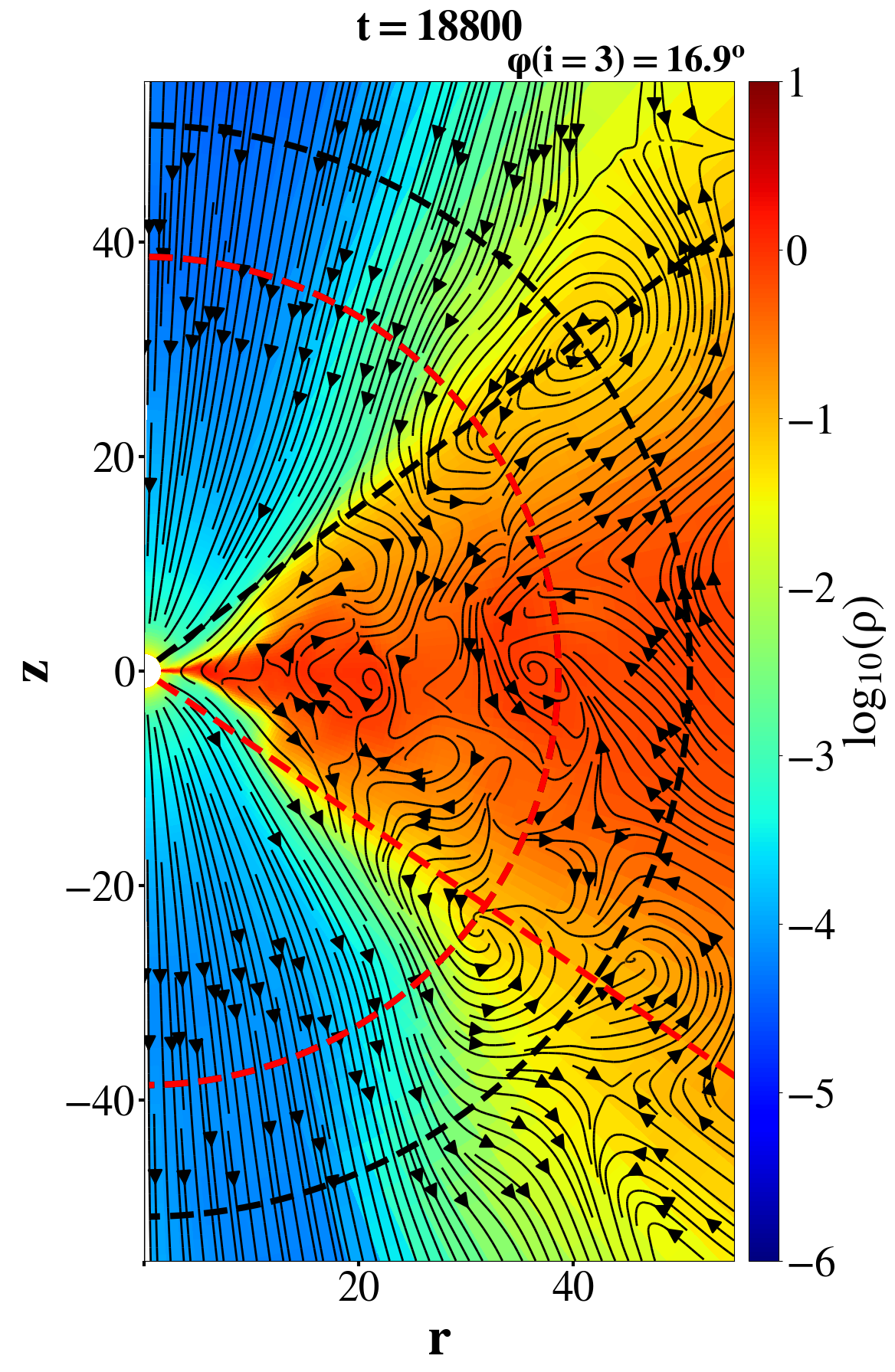}
\includegraphics[width=0.55\columnwidth]{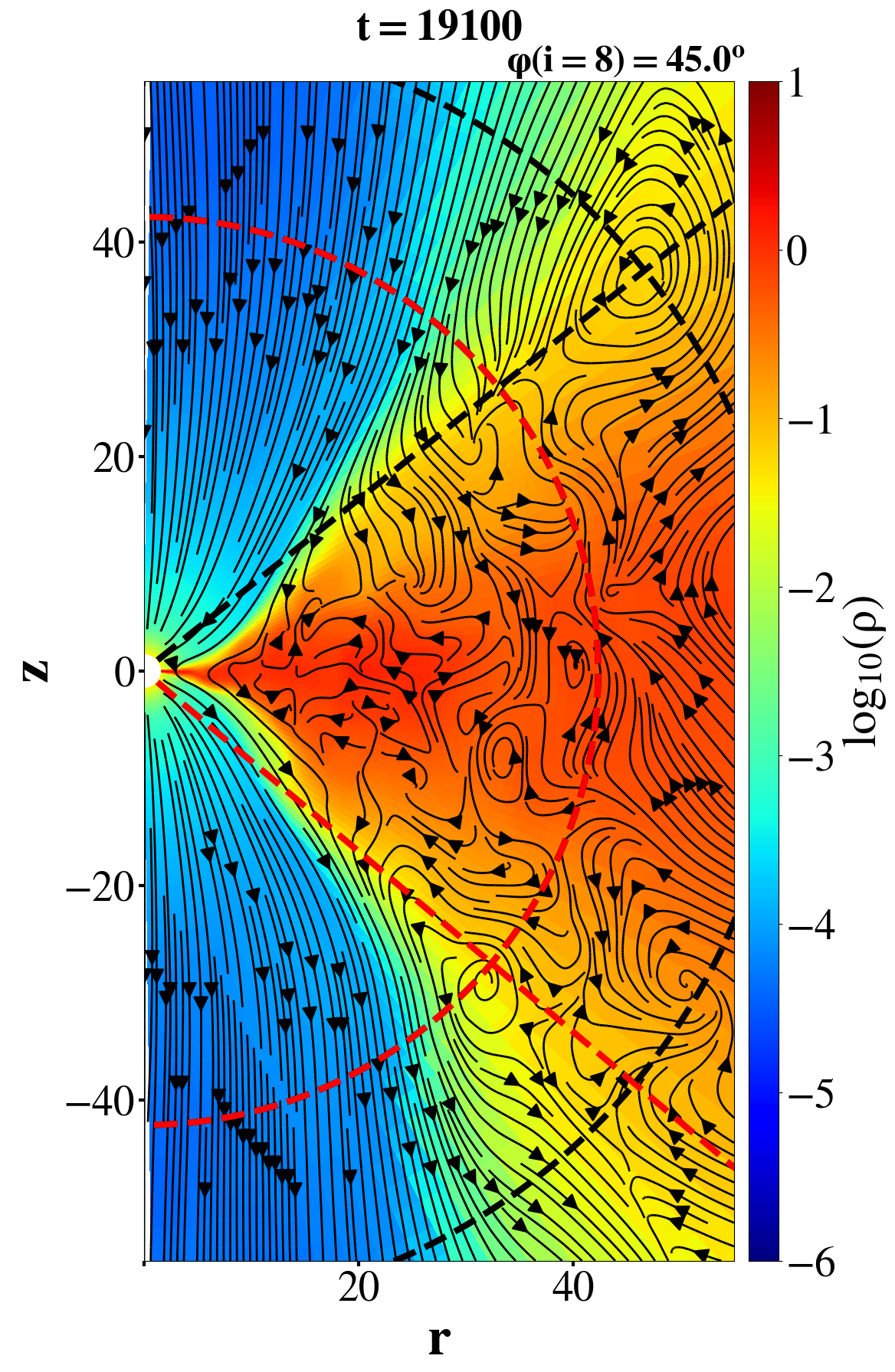}\\
\includegraphics[width=0.55\columnwidth]{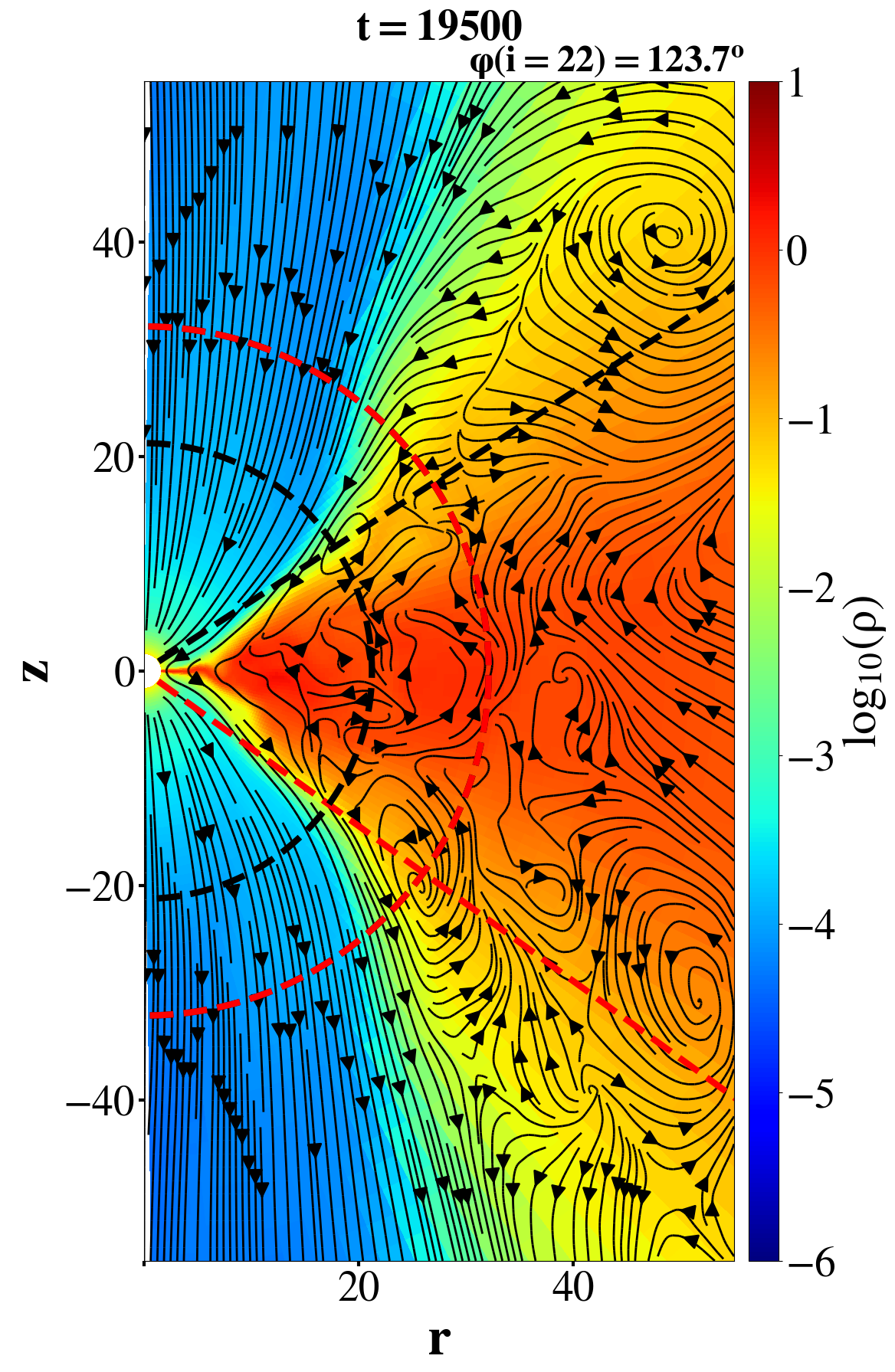}
\includegraphics[width=0.55\columnwidth]{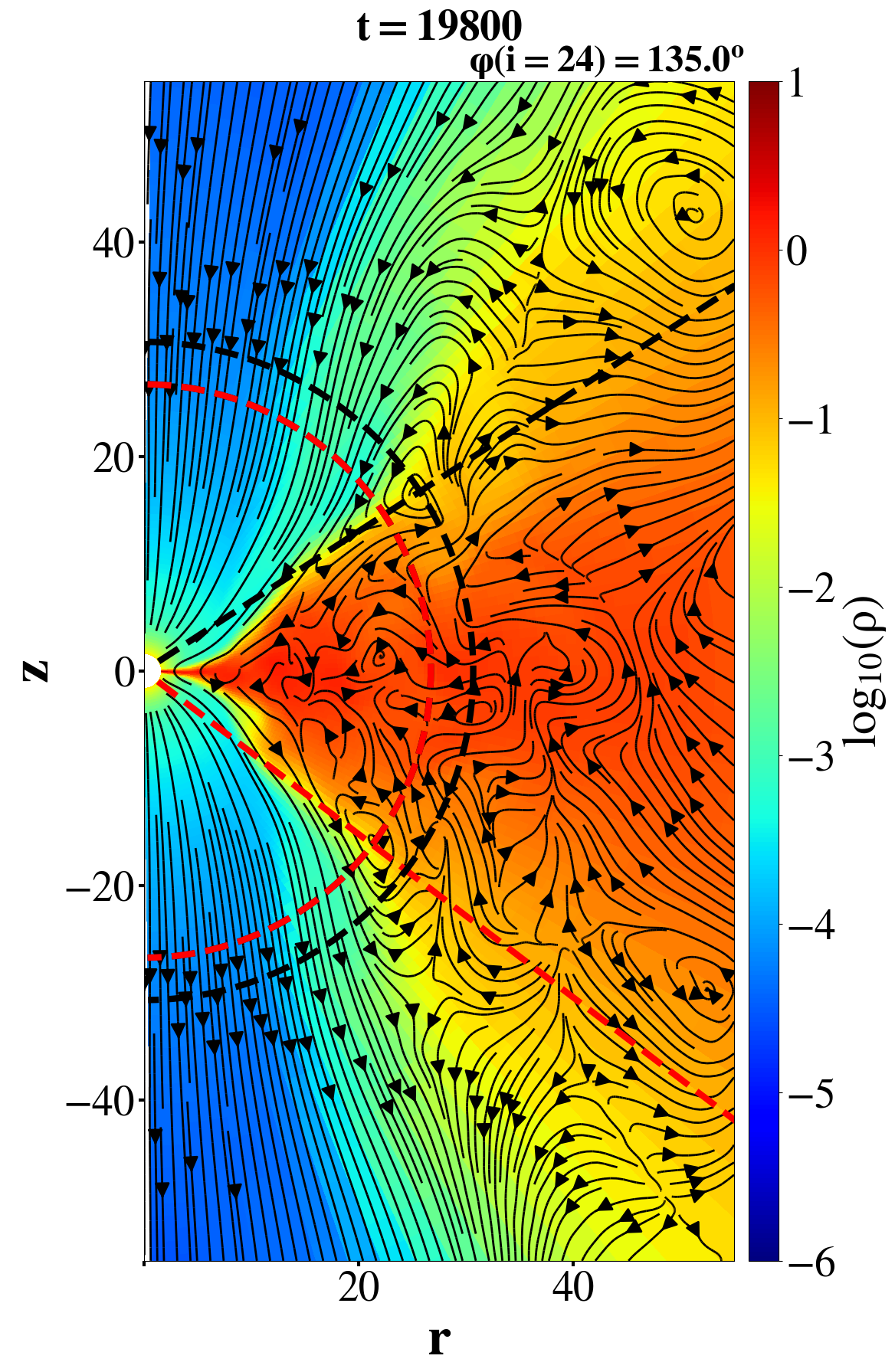}
\includegraphics[width=0.55\columnwidth]{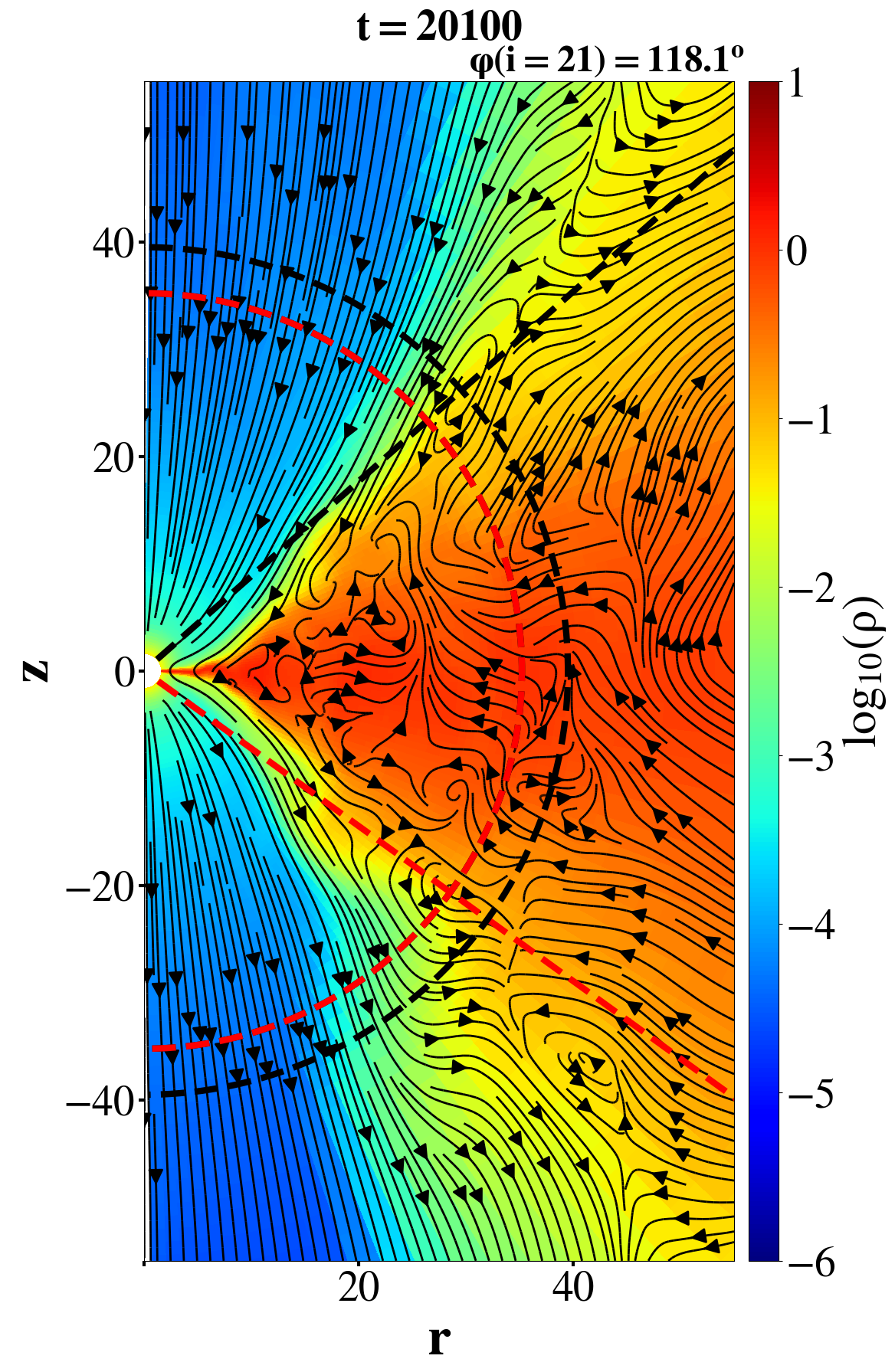}

\caption{{Three successive cases of magnetic islands formation at T=17500, 18500 and 19500$~r_{\mathrm g}/c$, just after the reconnection events shown in Fig.~\ref{islarec}, and their subsequent evolution. The data is based on SANE simulation. Colour denotes density while solid lines magnetic field. The dashed black and red lines are spherical coordinate lines passing through the centers of two magnetic islands located at the crossings of the circular and straight dashed lines of the same color. The value of $\varphi$ is indicated atop the upper right corner of each panel. }}
\label{fig:isla}
\end{figure*}

\begin{figure}[t]
     
     \includegraphics[width=.48\textwidth]{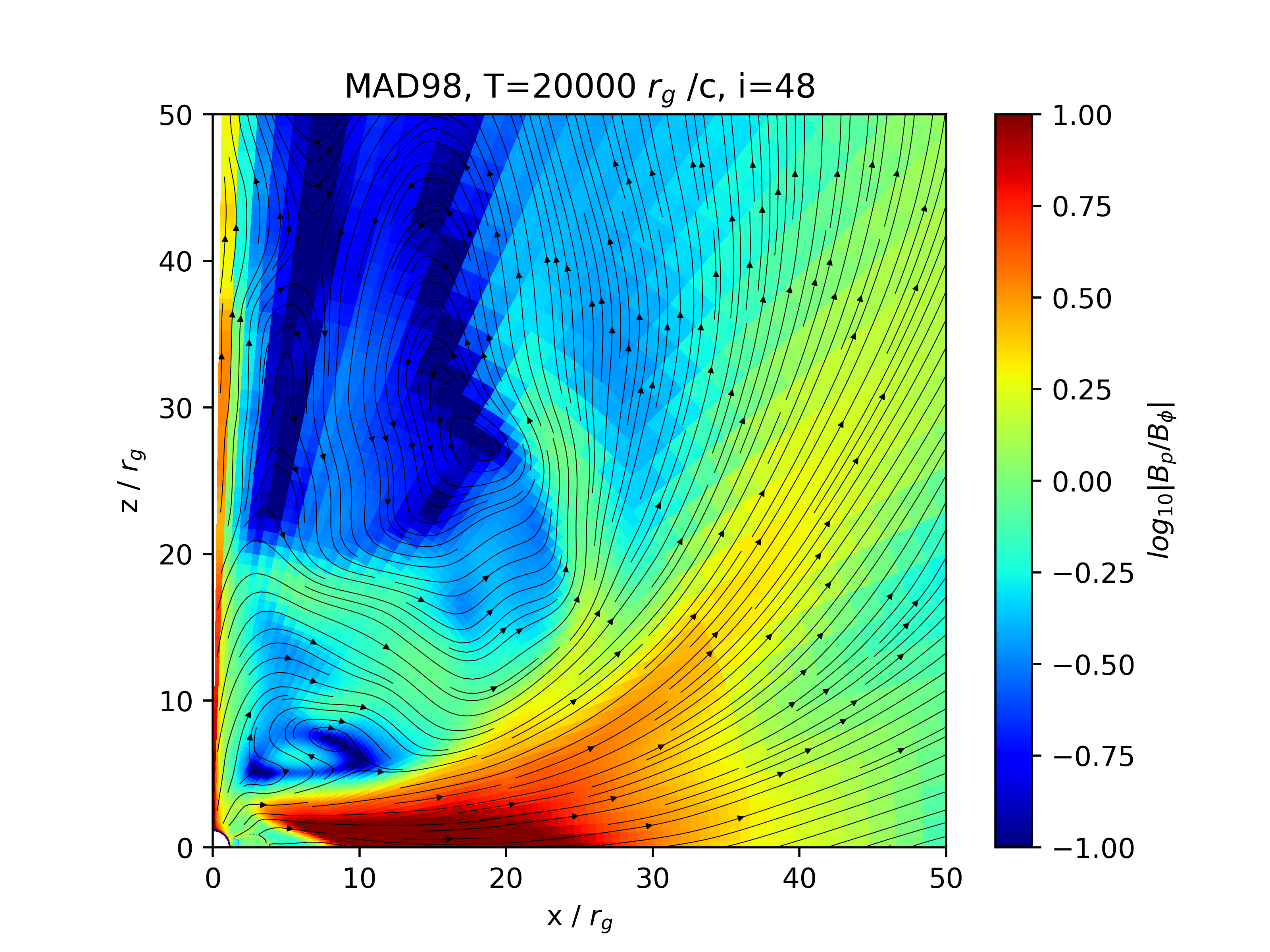}
     \includegraphics[width=.48\textwidth]{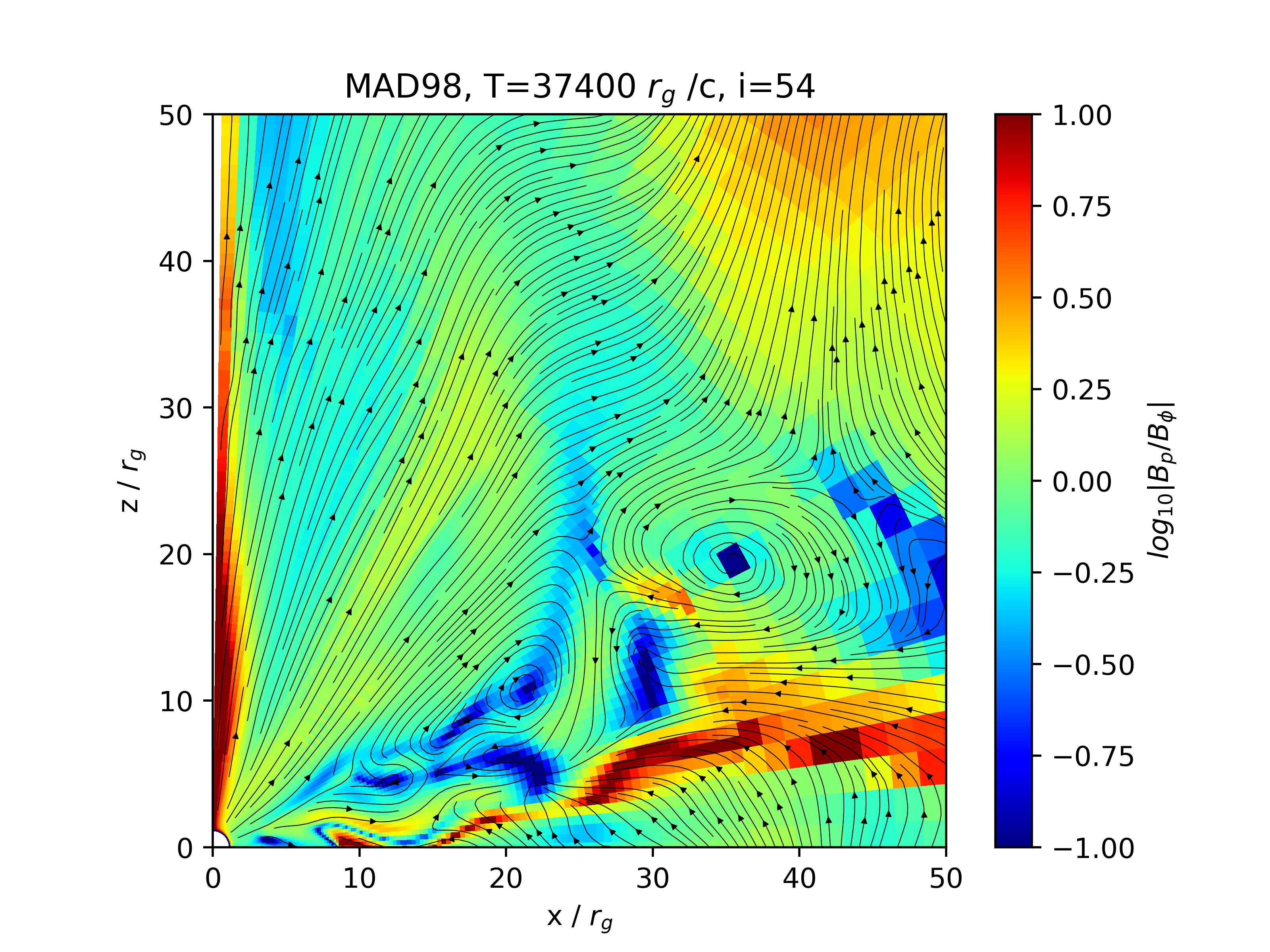}
     \caption{Properties of magnetic field at two different evolution times. The time (“T”) and the number of grid in the $\varphi$-direction (“i”) are shown at the top of each plot. The magnetic field is denoted by black solid lines with arrows indicating the direction of the field. The color indicates the logarithmic value of the ratio of poloidal and toroidal magnetic field strength. In the top panel, two flux ropes are located at $(x,z)=(8,7)$ and (20,28); in the bottom panel two flux ropes are located at (21,12) and (35,20). The magnetic field in the region where the two flux ropes are located is poloidally and toroidally dominated in the top and bottom panels, respectively. At the very center of the two flux ropes in the bottom panel, the poloidal field is weaker than the toroidal field because of the occurrence of reconnection (see text for details). The figure is plotted based on the simulation data of MAD.}
     \label{fig:fig5}
\end{figure}

\subsection{Periodicity of the formation and ejection of flux ropes}

We find from the simulation data that, in both SANE and MAD cases, every time when magnetic islands
are formed, their corresponding azimuthal angles $\varphi$ are usually different; but their corresponding radii are almost
always within the range of $20\la r\la 30 r_g$. More interestingly, the formation of flux rope seems to occur quasi-periodically, once the simulations have reached the steady state. Such a phenomenon does not depend on model parameters or numerical resolution. As an example, we show in Figure \ref{fig:isla}  three successive cases of magnetic island formation and their subsequent evolution. The formation time of the three islands is $T=17500, 18500$ and $19500${$~r_{\mathrm g}/c$}, respectively. More cases before $T=17500$ and after $T=19500${$~r_{\mathrm g}/c$} exist but are not shown in the figure due to the spatial limitation. This implies a period of $\sim 1000 r_g/c.$  Such a period matches the orbiting timescale of both
SANE and MAD at $20-30 r_g$ well, which are slightly larger than the Keplerian timescale since the angular velocities of both SANE and MAD are sub-Keplerian \citep{YuanNar14}.  Since the formation of flux rope at relatively large radii is accompanied by plasmoid ejection, such a periodicity implies that, in addition to flares, the ejection of plasmoids also occurs periodically. At smaller radii, the reconnection (thus flare) is also found to occur periodically, although not accompanied by ejections.

The presence of periodicity combined with its orbital timescale strongly implies that, compared to turbulence, the differential rotation of the accretion flow seems to play a more important role in twisting the magnetic field lines and resulting in the occurrence of magnetic reconnection. This scenario is enhanced by the fact that there seem to be no difference in terms of the periodicity of the
formation of flux ropes in SANE and MAD cases, although MAD is much less turbulent compared to SANE. In contrast, the occurrence of CMEs in the Sun mainly depends on solar activity. Although various periodicities exist for CMEs and flares \citep[e.g.,][]{Lou2003}, the periods are not related to the rotation period of the Sun. This is because, unlike the differentially rotating accretion flow, the geometry of Sun is spherical and the motion of the foot points of the field loops is driven by solar dynamo through local convection rather than by the rotation of the Sun.

In addition to the periodicity, another interesting finding is that in some cases, when a plasmoid moves outward, it may catch up and merge with a previously produced plasmoid having a slower speed. The merger of two magnetized islands must be accompanied by magnetic reconnection. This will results in  strong radiation so strong flares
are expected to occur. \citet{Yuan2012} have proposed this scenario as the mechanism of Gamma-Ray bursts.

\section{Applications to observations} \label{sec:observation}

\subsection{Periodic flares and the hot spot in Sgr A* and other black holes}

The formation of the flux rope occurs because of magnetic reconnection, which must be accompanied by electron acceleration. The synchrotron radiation of the accelerated electrons will be responsible for the observed flares in various wavebands, ranging from IR to X-ray and $\gamma$-ray flares in black hole sources mentioned above. 

Near-infrared flares in Sgr A* were recently resolved by high resolution GRAVITY observations, with a compact ``hot spot’’ detected orbiting around the black hole with orbiting radius of 6-10 $r_g$ \citep{Gravity2018} . This hot spot is produced by  magnetic reconnection found in our simulations at small radii of the accretion flow. The flares are found to be polarized, with the polarization signature being consistent with orbital motion of a hot spot in a strong poloidal magnetic field \citep{Gravity2018}. In our simulation, we find that at the place of reconnection (i.e., the place of ``hot spot’’), which is also roughly where the flux rope is located, the magnetic field could be dominated by either poloidal or toroidal components, depending on time. This is shown by the two panels of Figure~\ref{fig:fig5}. The top panel of this figure shows a case that the field is dominated by the poloidal component, which is consistent with the observation. But at another time shown by the bottom panel of this figure, the field is dominated by toroidal component. This prediction could be verified by future observations. The detailed phenomenological modeling of the IR and X-ray light curves of Sgr A* indicates that to explain the simultaneity and symmetry of the X-ray and near-infrared light curves, the magnetic field strength is required to significantly decrease during the flare \citep{Dodds2010}. This is consistent with the weakening of the magnetic field during reconnection which we have described above and explicitly shown in Figure~\ref{fig:banddensity}. 

From the over 2-hour long infrared light curve of Sgr A*, several periods of variability are identified by observations, with period ranging from 17 to 40 minute \citep{Genzel2003,Trippe2007,Genzel2010,Eckart2006}. The observed periods are consistent with orbital period at 3-5$r_g$ for the $3.6\times10^6$ solar mass black hole in Sgr A*, assuming the accretion flow in Sgr A* is described by an MAD. Given that the typical lifetime of a “hot spot” is shorter than one orbit timescale as we have explained before, the detected periodicity is best explained by the periodic occurrence of reconnection events found in our simulation at small radii.  The observed period is not a fixed value but has a range,  consistent with our finding that reconnection occurs in a range of radii.

\subsection{Ejected blobs in Sgr A*, M81, and M87: origin, velocity, and the possible periodicity}

In our simulations, the flux ropes (i.e., plasmoid) formed beyond 10-15 $r_g$ will be ejected outward from the surface of the accretion flow. This explains the origin of the radio-emitting blobs observed in Sgr A*, M81, M87, and other black hole sources. The observationally measured ejection velocity of 0.4, 0.5, and 0.2 of the speed of light for the ejected plasmoids in Sgr A*, M81 and M87 is also quantitatively similar to what we have obtained in the case of MAD, as shown in Figure~\ref{fig:blobvel}, which is about 0.3 of the speed of light. The ``MAD’’ nature of the accretion flow in M81 and M87 is consistent with most recent studies \citep{Shi2021,EHT2021,Yuan22}. 

Examining the periodicity of the ejection of plasmoids is difficult since it requires very high spatial resolution and sensitivity of the telescope and the suitable observational targets. The high-resolution Global Millimeter-VLBI-Array observation to M87 jet reveals some discrete blobs at distance of about 100 $r_g$ from the central black hole \citep{Algaba2021}. By measuring the spacing between two adjacent blobs from the image and adopting a projected moving speed of 0.2 of the speed of light \citep{Hada2014}, we can estimate that the time interval of two adjacent ejections is about one year. This is in good consistency with our predicted period of 1000 $r_g/c \sim$ 1 yr for the $6\times10^9$ solar mass supermassive black hole in M87 \citep{Gebhardt2011}. 

Most recently, two quasi-periodic oscillations (QPOs) are detected in the TeV blazar PKS 1510-089 by the Fermi Large Area Telescope \citep{Roy2021}. One is a 3.6-day QPO that lasted for five cycles with a moderate significance of $\sim 3.5 \sigma$, another is a 92-day QPO that lasted for seven cycles, with a significance of $\sim 7.0\sigma$. As discussed in \citet{Roy2021}, models like binary black holes and a procession jet are unlikely; two attractive models are a rotating hot spot in the inner region of the accretion flow and ``equispaced magnetic islands'' inside the jet. The remaining problem for the first model is that since the accretion flow in a blazar should have an almost face-on orientation, it may be hard for the rotating hot spot to produce enough variability. 

Obviously, these two scenarios are very similar to what we have proposed in the present work. The only difference is that, for the first ``hot spot'' model, we think the variability is not caused by a single hot spot orbiting around the black hole, but several hot spots periodically appeared in the inner region of the accretion flow since the typical lifetime of a hot spot is shorter than one orbital time thus difficult to explain the observed several cycles.  Moreover, our ``modified'' hot spot model has the advantage of overcoming the ``not-enough variability'' difficulty.   

\subsection{Possible application to protostellar accretion systems}

Although our simulations are done for hot accretion flows around black holes, we speculate that the basic physics relevant to flares and ejections should also work for accretion disks in protostars. This is why very similar phenomena have been observed in these systems. Specifically, in this case, the period should correspond to the orbital period at the inner edge of the accretion flow where it is truncated by the magnetosphere of the protostar, which is a few days. Periodic ejections are observed from the spacings of the jet knots \citep{Lee2020} in those systems, and the period can be estimated from the spatial intervals and the jet velocities. Unfortunately, the current instrumental limit of spatial resolution $\sim 0.1^{"}$ sets a limit on the period detection no shorter than about 1 yr \citep{Lee2020}. On the other hand, the Chandra light curve does show obvious periodicity with the period of $2\times10^5$ second or a few days, according to \citet{Wolk2005} and \citet{Flaccomio2012} (Ref. to Fig. 3 in \citet{Wolk2005} and Fig. 4 in \citet{Flaccomio2012}), consistent with our prediction. 

\section{Comparison with previous works}
\label{sec:comp}

The most relevant previous work is \citet{Yuan09}. In that work, by analogy with the coronal mass ejection in the Sun, an analytical model was proposed to explain the formation of episodic jets and their association with flares. The basic scenario proposed in that work, i.e., flux ropes are formed due to magnetic reconnection and they are then ejected out by the magnetic forces, are fully confirmed by the present numerical simulations. Due to the limitation of analytical calculations, although the whole physical processes involved are highly dynamic, many assumptions and simplifications have to be adopted in \citep{Yuan09}. For example, the formation of flux rope is purely an assumption there. The quantitative calculation to the ejected velocity of the flux rope is also based on assumptions on the distributions of magnetic field and density of the coronal gas. Compared to \citet{Yuan09}, our current work is much more realistic and comprehensive.  Moreover, important new findings are obtained, e.g., the periodic formation of the flux rope and its ejection, and the quantitative prediction of the velocity of the ejected blobs. 

Recently several MHD numerical simulation works focusing on the analysis of reconnection in accretion flows have been published. It is thus useful to compare our simulations with these works. In the two-dimensional general relativity ideal magnetohydrodynamic numerical simulations by \citet{Nathanail20}, the formation and evolution of current sheets are investigated. Special attention is paid to the effect of different initial magnetic field configurations. Similar to our work, they find that plasmoids are formed both within and at the surface of the accretion flow due to magnetic reconnection, which is driven by the turbulent motion of the accretion flow. Moreover, they find that plasmoids are formed during the reconnection, which is again consistent with our work. 

In the works by another group, the two and three dimensional MHD simulations including weak Ohmic resistivity are performed, focusing on the modeling of magnetic reconnection and the formation of palsmoids \citep{Ripperda20,Ripperda2021}. Since an explicit resistivity is included which can approximately mimic the kinetic effects, their simulations are more suitable than ours to describe magnetic reconnection. Both SANE and MAD are studied as in our work. They find that within about $20r_g$ of the accretion flow current sheets and plasmoids are ubiquitous features that form regardless of the initial magnetic field, the magnetization of accretion flows, and the spin of the black hole. 

Different from our work and \citet{Nathanail20}, another mechanism for the formation of plasmoid is identified in \citet{Ripperda20}, namely the tearing instability of the current sheet. When the reconnection layer becomes thin enough, the reconnection layer can break up and produce chains of plasmoids. These small plasmoids can merge and grow to macroscopic scales of the order of a few Schwarzschild radius, and be advected along the jet’s sheath or into the disk. We do not find such an instability in our simulations.  For this instability to occur, the Lundquist number $S=v_{\rm A} L/\eta$ needs to be larger than $10^4$. Here $v_{\rm A}$ being the Alfven speed, L being the typical length of the current sheet, and $\eta$ being the resitivity \citep{McKinney2012}. In our simulations, $S\lll10^4$   because of the high numerical resistivity caused by the relatively low resolution of our three-dimensional simulations. In reality both turbulence of the accretion flow and tearing instability in current sheet should be able to drive reconnection and the formation of plasmoids, as seems to be confirmed by a recent work \citep{Rosenberg2021}. In this work, resistive MHD equations are solved, reconnection and plasmoids are found to be formed both via merging current sheets due to turbulent motions of the accretion flow and through reconnection due to current sheet instabilities. 

\section{Summary}\label{summary}

Episodic ejections  of blobs (episodic jets) have been observed in both black hole and protostellar accretion systems and they are often associated with radiation flares. Notable sources include Sgr A*, M81, and M87, among others. \citet{Yuan09} has proposed an analytical MHD model for episodic jet and its association with flares. In the present work, we develop that work by performing three dimensional GRMHD  numerical simulation of hot accretion flows. Both SANE and MAD are considered. We have analyzed the simulation data and obtained the following results. 

\begin{itemize}
    \item[(1)] We find once the simulations have reached the steady state, flux ropes are keep forming both within and at the coronal region of the accretion flow, from radius as small as 4$r_g$ up to $\sim 30r_g$.  One example of a flux rope in the case of MAD is shown in Figure \ref{fig:fluxrope}. It extends in the $\varphi$ direction for about 120$^{\circ}$.
    
    \item[(2)] We find that the flux rope is formed due to the reconnection of magnetic field lines, driven by the differential rotation and turbulent motion of the accretion flow. This scenario is consistent with that proposed in \citet{Yuan09} and recent MHD numerical simulations. The detailed comparison between the present work and those previous ones are presented in Section \ref{sec:comp}.
    
    \item[(3)] We find that flux ropes formed at small and large radii have different fate. At small radii, inside of $\sim$ 10-15$r_g$, few flux ropes can be ejected out, likely because of the strong gravitational force of the black hole, similar to the absence of wind at small radii. Beyond that radius, flux ropes can be ejected out. The ejected flux ropes are found to break within one orbital timescale due to the kink instability. 
    
    \item[(4)] The broken flux ropes (i.e., plasmoids)  spiral outward, with radial velocities shown in Figure \ref{fig:blobvel}. A clear acceleration in both cases is found and the  velocity can reach $0.3c$ and $0.08c$ for MAD and SANE, respectively. We have analyzed the forces and found that the acceleration is mainly due to the gradient of the magnetic pressure, as shown by Figure \ref{fig:force}, again consistent with \citet{Yuan09}. 
    
    \item[(5)] The whole processes mentioned above, i.e., the formation of the flux ropes and their subsequent ejection, occur periodically, with the period being the orbiting time at the radius where the flux ropes are formed. This implies that differential rotation of the accretion flow plays a more important  role than turbulence in twisting the magnetic field lines and result in the formation of reconnection. 
    
    \item[(6)] Magnetic reconnection should accelerate some electrons, their synchrotron emission explains observed flares and the origin of the ``hot spot'' in Sgr A* detected by GRAVITY. 
    The periodicity detected from the IR light curves of Sgr A* and $\gamma-$ray light curves of blazar PKS 1510-089 are explained by the periodic formation of flux ropes. 
    
    \item[(7)] The observed ejection of blobs is explained by the ejection of flux ropes formed at relatively large radii. In Sgr A$^*$, M81, and M87, the measured velocity of the ejected blobs agrees with the velocity predicted in our simulations of MAD. The predicted periodicity of the  ejection seems to be supported by some observations of M87 and PKS 1510-089.  
    
    \item[(8)] Finally, we speculate that our results may also be able to interpreting flares and episodic ejection observed in protostellar systems. The measured period in the X-ray light curve is roughly consistent with the orbital period at the inner edge of the accretion flow where it is truncated by the magnetosphere of the protostar. 
    
\end{itemize}

\section*{Acknowledgments}
We thank the useful discussions with Drs. P.F. Chen, J. Lin, and Y.M. Wang on the solar flares and coronal mass ejections, and C. White for his help of using ATHENA++ code. The {anonymous} referee is acknowledged for 
constructive suggestions and comments. This work has made use of the High Performance Computing Resource in the Core Facility for Advanced Research Computing at Shanghai Astronomical Observatory. We also thank ASIAA, Taipei for the use of their visualization servers. This work was supported by CAS President’s International Fellowship for Visiting Scientists grant 2020VMC0002 (MC), Polish NCN grant 2019/33/B/ST9/01564 (MC), Natural Science Foundation of China grants 12133008, 12192220, and 12192223 (FY, HY), and Ministry of Science and Technology of Taiwan grants 108-2112-M-001-009 and 109-2112-M-001-028 (HS)

%




\appendix

\section{Three-dimensional GRMHD numerical simulations of a hot accretion flow}
We have performed numerical simulations in three-dimensions by solving the equations of ideal MHD describing the evolution of the accretion flow around a black hole in the Kerr metrics using the GRMHD code Athena++ \citep{White16,Stone2020}. Readers are referred to \citet{Yang21}  for details of the simulations, here we only present a brief overview. All our simulations are performed in Kerr-Schild (horizon penetrating) coordinates  ($t$, $r$, $\theta$, $\varphi$). 
The comoving rest-mass density is denoted by $\rho$, and  the component of the coordinate-frame 4-velocity by $u^{\mu}$. The equation of the state of the gas is $u=p_{\rm{ gas}}/(\Gamma-1)$, where $p_{\rm gas}$  is the gas pressure of the comoving mass, and $u$ is the internal energy of the gas, and $\Gamma$ is the adiabatic index, which we set to $\Gamma$ =4/3. The units we adopt are Heaviside-Lorentz with both the light speed and gravity constant set to unity, and the sign convection of metric (-, +, +, +). The metric is stationary and the self-gravity of the accretion flow is ignored in our simulation.

We have simulated four models: SANE00, SANE98, MAD00, and MAD98. They denote the SANE accretion flow around a black hole of spin $a = 0$ and 0.98, MAD accretion flow around a black hole of spins $a = 0$ and 0.98, respectively. The initial condition of all models is a rotating torus \citep{Fishbone1976} around a black hole, with the inner edge of the torus  at r = 40.5 $r_g$ and the radius of pressure maximum at r = 80 $r_g$. We have also added a poloidal magnetic field threading the torus in the way described by \citet{Penna2013}. For MAD cases, we set one poloidal loop threading the whole torus. For SANE cases, we initially set up a seed field consisting of multiple poloidal loops of magnetic field with changing polarity. We use the gas-to-magnetic pressure ratio $\beta \equiv p_{\rm gas}/p_{\rm mag}$ to normalize the magnetic field. For MAD, we set $\beta_{\rm min}=0.1$. For SANE, we set  $\beta_{\rm min}=0.05833$ $(a=0)$ and $\beta_{\rm min}=0.1$ $(a=0.98)$.

The grid we used is a static mesh refinement grid, which allows the use of higher resolution where it is needed in the simulation. We use logarithmically spaced radial grid  with inner edge 1.1\,${r_{\rm g}}$ and outer edge 1200\,${r_{\rm g}}$. The grid cells in the polar and azimuthal directions are uniform.  The final grid resolution in our simulation is $288\times128\times64$ cells in $r$, $\theta$, and $\varphi$ direction for SANE00 and MAD00, while for SANE98 and MAD98 it is $358\times128\times64$. We use the outflow boundary conditions at the inner and outer boundaries. For the $\theta$ and $\varphi$ directions, we use the polar axis boundary conditions and periodic boundary conditions respectively.

We have run simulations up to $t_f\,=\,40{,}000$, $80{,}000$,  $40{,}000$ and $40{,}000${~$r_{\mathrm g}/c$} for SANE00, SANE98, MAD00, and MAD98 respectively. They correspond to 8.9, 17.8, 8.9 and 8.9 orbital periods of the accretion flow at the pressure maximum. The ``inflow equilibrium’’ has been reached at roughly 26$r_g$, 40$r_g$, 80$r_g$, and 80$r_g$ for these four models respectively. Figure~\ref{fig:figs1} shows the snapshot distribution in the $x - z$ plane of the gas-to-magnetic pressure ratio $\beta$ (color) and the magnetic field lines for SANE and MAD.

The main physical implication of inflow equilibrium radius is that, within this radius the inflow has reached the steady state thus the radial profiles of density and velocity will not change with time so are reliable. For our problem, our main concern is not the density of accretion flow but the configuration of magnetic field, which is determined by the turbulence.  Beyond the inflow equilibrium radius, although density and velocity are not reliable, the level of turbulence is still reliable up to a much larger ``turbulence radius'', the limiting radius of turbulence steady state. This radius can be estimated as follows. Turbulence in accretion flow is because of MRI. The fastest growth rate of MRI at radius r is $\sim \Omega (r)$. More precisely, it takes 3–4 orbits for MRI to develop \citep{Hawley95}. At the end of our simulations to the four models, taking a timescale of 3 orbits, we can obtain that the “turbulence radii” are $\sim 130, 200, 120,$ and 120 $r_g$, respectively. Many analyses in the present paper are performed at  $t\sim 20000r_g/c$. At that time the ``turbulence radii'' are $\sim 100, 100, 80,$ and $80 r_g$, respectively.

\begin{figure*}
\includegraphics[width=.48\linewidth]{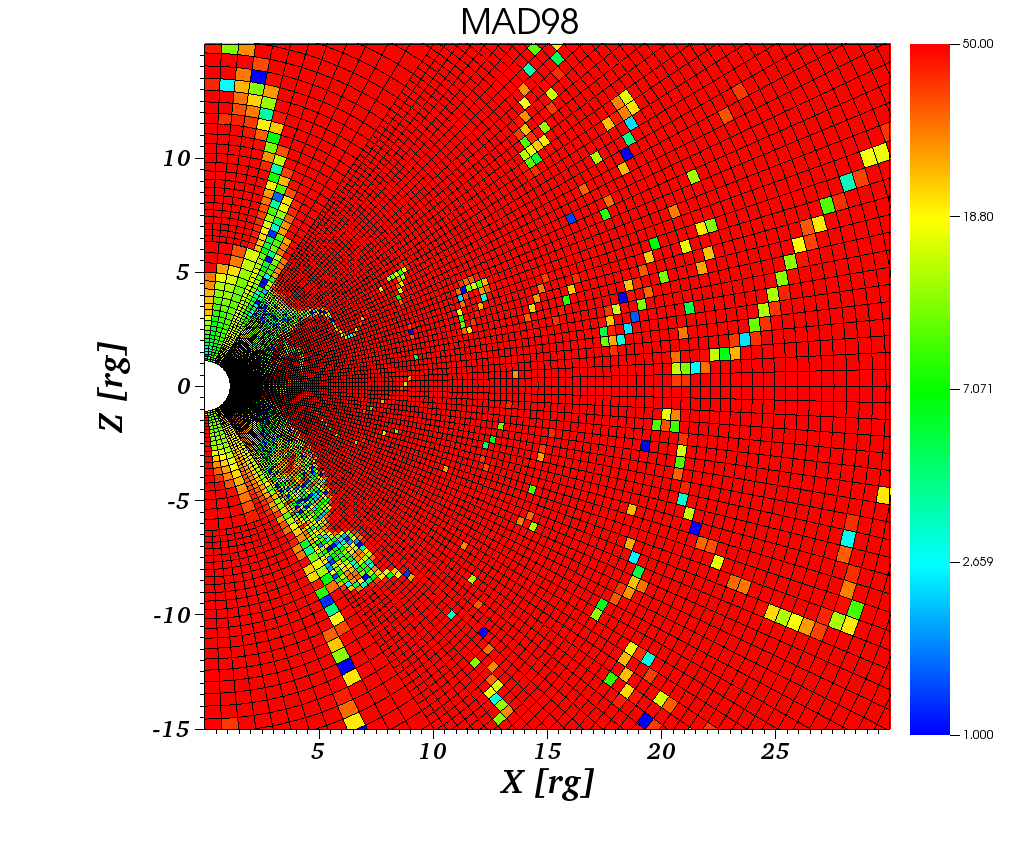}\quad\includegraphics[width=.48\linewidth]{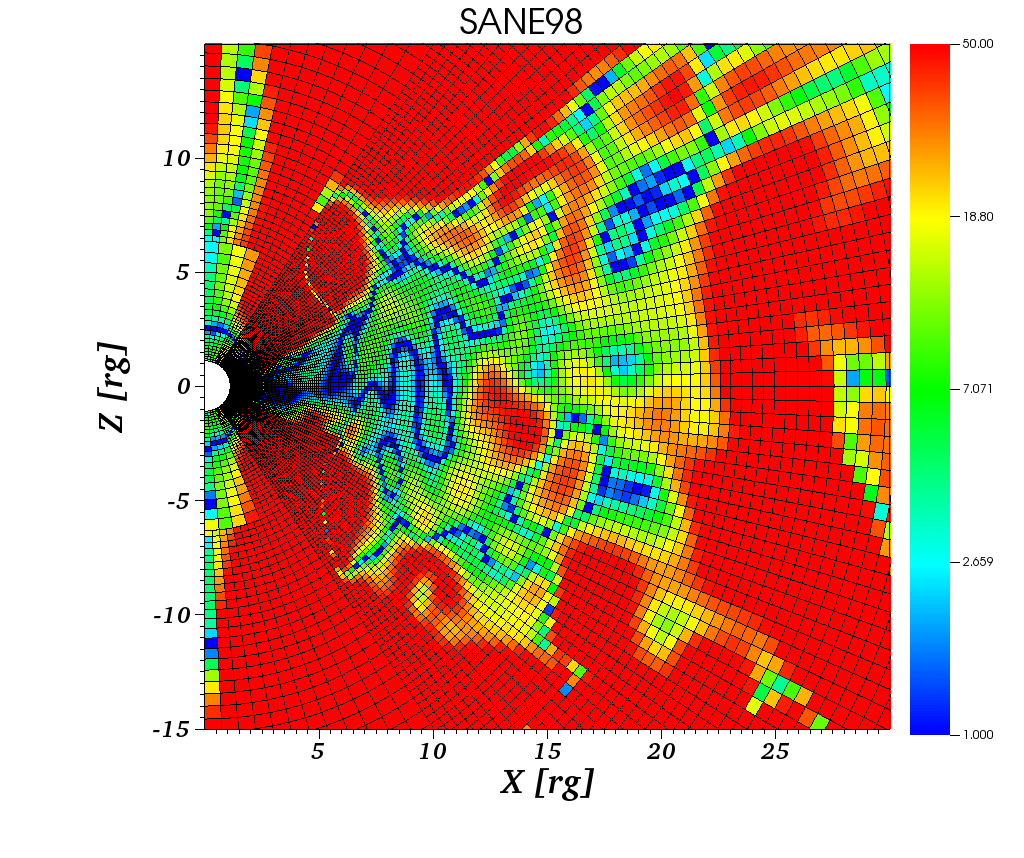}
\includegraphics[width=.48\linewidth]{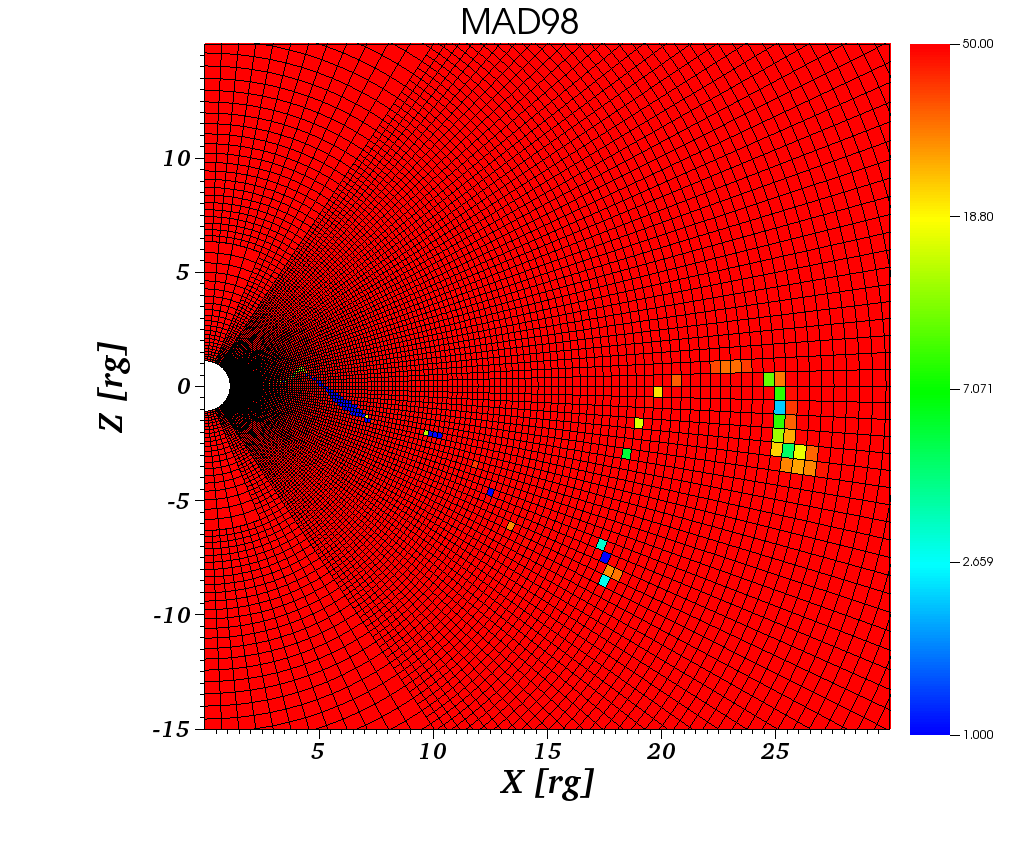}\quad\includegraphics[width=.48\linewidth]{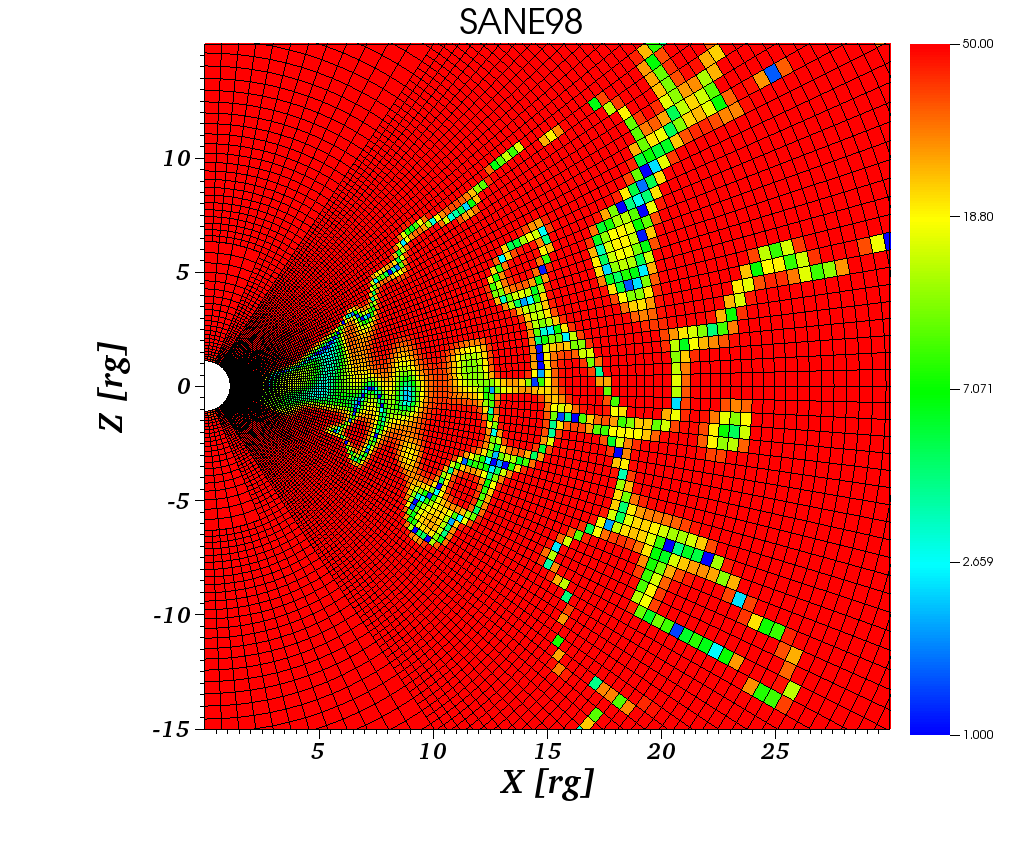}

    \caption{The values of $Q_{\theta}$ (upper) and $Q_{\varphi}$ (lower) in the inner region of MAD98 (left) and SANE98 (right), calculated based on our simulation data.}
    \label{fig:fig2s}
\end{figure*}

To check the resolution of our simulations, following \citet{Hawley2011}, we have calculated $Q_\theta$  and $Q_\varphi$ based on our simulation data, 

\begin{equation}
    {Q}_{\theta}=\frac{{\lambda}_{MRI}} {dx^{\theta}}=\frac{2\pi |v_{\theta,\rm A}|}{{\Omega}dx^{\theta}},
	\label{eq:qth}
\end{equation}
\begin{equation}
    {Q}_{\varphi}=\frac{{\lambda}_{MRI}} {dx^{\varphi}}=\frac{2\pi |v_{\varphi,\rm A}|}{{\Omega}dx^{\varphi}}.
	\label{eq:qph}
\end{equation}
Here $v_{\varphi,A}$ is the $\theta$- and $\varphi$-directed Alfven speed, $\Omega$ is the angular velocity of fluid. The physical meaning of these two parameters is the number of grid in one fastest growing wavelength of MRI in the $\theta$ and $\varphi$ directions, respectively. According to \citet{Hawley2011}, if $Q_{\theta}>10$ and $Q_\varphi>20$, the resolution of the simulation will be high enough to give quantitatively converged results for the accretion flow. The  calculated values of $Q_{\theta}$ and $Q_{\varphi}$ in the inner region of MAD98 and SANE98 are given in Figure~\ref{fig:fig2s}. We can see from the figure that  this criterion is well satisfied for MAD98 and marginally satisfied for SANE98. The values of $Q_{\theta}$ and $Q_{\varphi}$ are larger in  MAD than in SANE; this is because in the case of MAD the Alfven speed is larger and the angular velocity is smaller thus the fastest growing wavelength of MRI is larger.

The above approach is used to estimate convergence of nonlinear MHD turbulence in accretion flows resulted by the magnetorotational instability. In addition to this approach, we have also adopted another more straightforward approach to verify the convergence of our results. That is, we have simulated the SANE98 and MAD98 models with higher resolution of $704\times256\times128$. Each simulation costs us about 400K CPU hours. We have used these simulation data to repeat the analysis performed in the present paper as described below, including the occurrence of magnetic reconnection, formation of flux ropes, and the ejection speed of the plasmoids. We find that all the results remain quantitatively unchanged compared to the case of low-resolution simulations, indicating the convergence of our results.


\bibliography{episodicjet}{}

\begin{thebibliography}{}
\expandafter\ifx\csname natexlab\endcsname\relax\def\natexlab#1{#1}\fi
\providecommand{\url}[1]{\href{#1}{#1}}
\providecommand{\dodoi}[1]{doi:~\href{http://doi.org/#1}{\nolinkurl{#1}}}
\providecommand{\doeprint}[1]{\href{http://ascl.net/#1}{\nolinkurl{http://ascl.net/#1}}}
\providecommand{\doarXiv}[1]{\href{https://arxiv.org/abs/#1}{\nolinkurl{https://arxiv.org/abs/#1}}}

\bibitem[{{Baganoff} {et~al.}(2001){Baganoff}, {Bautz}, {Brandt}, {Chartas},
  {Feigelson}, {Garmire}, {Maeda}, {Morris}, {Ricker}, {Townsley}, \&
  {Walter}}]{Baganoff01}
{Baganoff}, F.~K., {Bautz}, M.~W., {Brandt}, W.~N., {et~al.} 2001, \nat, 413,
  45, \dodoi{10.1038/35092510}

\bibitem[{{Balbus} \& {Hawley}(1991)}]{Balbus1991}
{Balbus}, S.~A., \& {Hawley}, J.~F. 1991, \apj, 376, 214,
  \dodoi{10.1086/170270}

\bibitem[{{Broderick} \& {Loeb}(2006)}]{Broderick2006}
{Broderick}, A.~E., \& {Loeb}, A. 2006, \mnras, 367, 905,
  \dodoi{10.1111/j.1365-2966.2006.10152.x}

\bibitem[{{Chen} {et~al.}(2020){Chen}, {Xu}, \& {Ding}}]{Chenpf2020}
{Chen}, P.-F., {Xu}, A.-A., \& {Ding}, M.-D. 2020, Research in Astronomy and
  Astrophysics, 20, 166, \dodoi{10.1088/1674-4527/20/10/166}

\bibitem[{{Dexter} {et~al.}(2020){Dexter}, {Tchekhovskoy},
  {Jim{\'e}nez-Rosales}, {Ressler}, {Baub{\"o}ck}, {Dallilar}, {de Zeeuw},
  {Eisenhauer}, {von Fellenberg}, {Gao}, {Genzel}, {Gillessen}, {Habibi},
  {Ott}, {Stadler}, {Straub}, \& {Widmann}}]{Dexter20}
{Dexter}, J., {Tchekhovskoy}, A., {Jim{\'e}nez-Rosales}, A., {et~al.} 2020,
  \mnras, 497, 4999, \dodoi{10.1093/mnras/staa2288}

\bibitem[{{Dodds-Eden} {et~al.}(2010){Dodds-Eden}, {Sharma}, {Quataert},
  {Genzel}, {Gillessen}, {Eisenhauer}, \& {Porquet}}]{Dodds2010}
{Dodds-Eden}, K., {Sharma}, P., {Quataert}, E., {et~al.} 2010, \apj, 725, 450,
  \dodoi{10.1088/0004-637X/725/1/450}

\bibitem[{{Dodds-Eden} {et~al.}(2009){Dodds-Eden}, {Porquet}, {Trap},
  {Quataert}, {Haubois}, {Gillessen}, {Grosso}, {Pantin}, {Falcke}, {Rouan},
  {Genzel}, {Hasinger}, {Goldwurm}, {Yusef-Zadeh}, {Clenet}, {Trippe},
  {Lagage}, {Bartko}, {Eisenhauer}, {Ott}, {Paumard}, {Perrin}, {Yuan},
  {Fritz}, \& {Mascetti}}]{Dodds2009}
{Dodds-Eden}, K., {Porquet}, D., {Trap}, G., {et~al.} 2009, \apj, 698, 676,
  \dodoi{10.1088/0004-637X/698/1/676}

\bibitem[{{Eckart} {et~al.}(2006{\natexlab{a}}){Eckart}, {Sch{\"o}del},
  {Meyer}, {Trippe}, {Ott}, \& {Genzel}}]{Eckart2006}
{Eckart}, A., {Sch{\"o}del}, R., {Meyer}, L., {et~al.} 2006{\natexlab{a}},
  \aap, 455, 1, \dodoi{10.1051/0004-6361:20064948}

\bibitem[{{Eckart} {et~al.}(2006{\natexlab{b}}){Eckart}, {Baganoff},
  {Sch{\"o}del}, {Morris}, {Genzel}, {Bower}, {Marrone}, {Moran}, {Viehmann},
  {Bautz}, {Brandt}, {Garmire}, {Ott}, {Trippe}, {Ricker}, {Straubmeier},
  {Roberts}, {Yusef-Zadeh}, {Zhao}, \& {Rao}}]{Eckart06}
{Eckart}, A., {Baganoff}, F.~K., {Sch{\"o}del}, R., {et~al.}
  2006{\natexlab{b}}, \aap, 450, 535, \dodoi{10.1051/0004-6361:20054418}

\bibitem[{{EHT MWL Science Working Group} {et~al.}(2021){EHT MWL Science
  Working Group}, {Algaba}, {Anczarski}, {Asada}, {Balokovi{\'c}}, {Chandra},
  {Cui}, {Falcone}, {Giroletti}, {Goddi}, {Hada}, {Haggard}, {Jorstad}, {Kaur},
  {Kawashima}, {Keating}, {Kim}, {Kino}, {Komossa}, {Kravchenko}, {Krichbaum},
  {Lee}, {Lu}, {Lucchini}, {Markoff}, {Neilsen}, {Nowak}, {Park}, {Principe},
  {Ramakrishnan}, {Reynolds}, {Sasada}, {Savchenko}, {Williamson}, {Event
  Horizon Telescope Collaboration}, {Akiyama}, {Alberdi}, {Alef}, {Anantua},
  {Azulay}, {Baczko}, {Ball}, {Barrett}, {Bintley}, {Benson}, {Blackburn},
  {Blundell}, {Boland}, {Bouman}, {Bower}, {Boyce}, {Bremer}, {Brinkerink},
  {Brissenden}, {Britzen}, {Broderick}, {Broguiere}, {Bronzwaer}, {Byun},
  {Carlstrom}, {Chael}, {Chan}, {Chatterjee}, {Chatterjee}, {Chen}, {Chen},
  {Chesler}, {Cho}, {Christian}, {Conway}, {Cordes}, {Crawford}, {Crew},
  {Cruz-Osorio}, {Davelaar}, {de Laurentis}, {Deane}, {Dempsey}, {Desvignes},
  {Dexter}, {Doeleman}, {Eatough}, {Falcke}, {Farah}, {Fish}, {Fomalont},
  {Ford}, {Fraga-Encinas}, {Friberg}, {Fromm}, {Fuentes}, {Galison}, {Gammie},
  {Garc{\'\i}a}, {Gentaz}, {Georgiev}, {Gold}, {G{\'o}mez}, {G{\'o}mez-Ruiz},
  {Gu}, {Gurwell}, {Hecht}, {Hesper}, {Ho}, {Ho}, {Honma}, {Huang}, {Huang},
  {Hughes}, {Ikeda}, {Inoue}, {Issaoun}, {James}, {Jannuzi}, {Janssen},
  {Jeter}, {Jiang}, {Jim{\'e}nez-Rosales}, {Johnson}, {Jung}, {Karami},
  {Karuppusamy}, {Kettenis}, {Kim}, {Kim}, {Kim}, {Koay}, {Kofuji}, {Koch},
  {Koyama}, {Kramer}, {Kramer}, {Kuo}, {Lauer}, {Levis}, {Li}, {Li},
  {Lindqvist}, {Lico}, {Lindahl}, {Liu}, {Liu}, {Liuzzo}, {Lo}, {Lobanov},
  {Loinard}, {Lonsdale}, {MacDonald}, {Mao}, {Marchili}, {Marrone}, {Marscher},
  {Mart{\'\i}-Vidal}, {Matsushita}, {Matthews}, {Medeiros}, {Menten}, {Mizuno},
  {Mizuno}, {Moran}, {Moriyama}, {Moscibrodzka}, {M{\"u}ller}, {Musoke},
  {Mej{\'\i}as}, {Nagai}, {Nagar}, {Nakamura}, {Narayan}, {Narayanan},
  {Natarajan}, {Nathanail}, {Neri}, {Ni}, {Noutsos}, {Okino}, {Olivares},
  {Ortiz-Le{\'o}n}, {Oyama}, {{\"O}zel}, {Palumbo}, {Patel}, {Pen}, {Pesce},
  {Pi{\'e}tu}, {Plambeck}, {Popstefanija}, {Porth}, {P{\"o}tzl}, {Prather},
  {Preciado-L{\'o}pez}, {Psaltis}, {Pu}, {Rao}, {Rawlings}, {Raymond},
  {Rezzolla}, {Ricarte}, {Ripperda}, {Roelofs}, {Rogers}, {Ros}, {Rose},
  {Roshanineshat}, {Rottmann}, {Roy}, {Ruszczyk}, {Rygl}, {S{\'a}nchez},
  {S{\'a}nchez-Arguelles}, {Savolainen}, {Schloerb}, {Schuster}, {Shao},
  {Shen}, {Small}, {Sohn}, {Soohoo}, {Sun}, {Tazaki}, {Tetarenko}, {Tiede},
  {Tilanus}, {Titus}, {Toma}, {Torne}, {Trent}, {Traianou}, {Trippe}, {van
  Bemmel}, {van Langevelde}, {van Rossum}, {Wagner}, {Ward-Thompson}, {Wardle},
  {Weintroub}, {Wex}, {Wharton}, {Wielgus}, {Wong}, {Wu}, {Yoon}, {Young},
  {Young}, {Younsi}, {Yuan}, {Yuan}, {Zensus}, {Zhao}, {Zhao}, {Fermi Large
  Area Telescope Collaboration}, {Principe}, {Giroletti}, {D'Ammando},
  {Orienti}, {H.~E.~S.~S. Collaboration}, {Abdalla}, {Adam}, {Aharonian},
  {Benkhali}, {Ang{\"u}ner}, {Arcaro}, {Armand}, {Armstrong}, {Ashkar},
  {Backes}, {Baghmanyan}, {Barbosa Martins}, {Barnacka}, {Barnard},
  {Becherini}, {Berge}, {Bernl{\"o}hr}, {Bi}, {B{\"o}ttcher}, {Boisson},
  {Bolmont}, {de Lavergne}, {Breuhaus}, {Brun}, {Brun}, {Bryan}, {B{\"u}chele},
  {Bulik}, {Bylund}, {Caroff}, {Carosi}, {Casanova}, {Chand}, {Chen}, {Cotter},
  {Cury{\l}o}, {Damascene Mbarubucyeye}, {Davids}, {Davies}, {Deil}, {Devin},
  {Dewilt}, {Dirson}, {Djannati-Ata{\"\i}}, {Dmytriiev}, {Donath},
  {Doroshenko}, {Duffy}, {Dyks}, {Egberts}, {Eichhorn}, {Einecke}, {Emery},
  {Ernenwein}, {Feijen}, {Fegan}, {Fiasson}, {de Clairfontaine}, {Fontaine},
  {Funk}, {F{\"u}{\ss}ling}, {Gabici}, {Gallant}, {Giavitto}, {Giunti},
  {Glawion}, {Glicenstein}, {Gottschall}, {Grondin}, {Hahn}, {Haupt},
  {Hermann}, {Hinton}, {Hofmann}, {Hoischen}, {Holch}, {Holler}, {H{\"o}rbe},
  {Horns}, {Huber}, {Jamrozy}, {Jankowsky}, {Jankowsky}, {Jardin-Blicq},
  {Joshi}, {Jung-Richardt}, {Kasai}, {Kastendieck}, {Katarzy{\'n}ski}, {Katz},
  {Khangulyan}, {Kh{\'e}lifi}, {Klepser}, {Klu{\'z}niak}, {Komin}, {Konno},
  {Kosack}, {Kostunin}, {Kreter}, {Lamanna}, {Lemi{\`e}re}, {Lemoine-Goumard},
  {Lenain}, {Levy}, {Lohse}, {Lypova}, {Mackey}, {Majumdar}, {Malyshev},
  {Malyshev}, {Marandon}, {Marchegiani}, {Marcowith}, {Mares},
  {Mart{\'\i}-Devesa}, {Marx}, {Maurin}, {Meintjes}, {Meyer}, {Moderski},
  {Mohamed}, {Mohrmann}, {Montanari}, {Moore}, {Morris}, {Moulin}, {Muller},
  {Murach}, {Nakashima}, {Nayerhoda}, {de Naurois}, {Ndiyavala},
  {Niederwanger}, {Niemiec}, {Oakes}, {O'Brien}, {Odaka}, {Ohm},
  {Olivera-Nieto}, {de Ona Wilhelmi}, {Ostrowski}, {Panter}, {Panny},
  {Parsons}, {Peron}, {Peyaud}, {Piel}, {Pita}, {Poireau}, {Noel}, {Prokhorov},
  {Prokoph}, {P{\"u}hlhofer}, {Punch}, {Quirrenbach}, {Rauth}, {Reichherzer},
  {Reimer}, {Reimer}, {Remy}, {Renaud}, {Rieger}, {Rinchiuso}, {Romoli},
  {Rowell}, {Rudak}, {Ruiz-Velasco}, {Sahakian}, {Sailer}, {Sanchez},
  {Santangelo}, {Sasaki}, {Scalici}, {Schutte}, {Schwanke}, {Schwemmer},
  {Seglar-Arroyo}, {Senniappan}, {Seyffert}, {Shafi}, {Shiningayamwe},
  {Simoni}, {Sinha}, {Sol}, {Specovius}, {Spencer}, {Spir-Jacob}, {Stawarz},
  {Sun}, {Steenkamp}, {Stegmann}, {Steinmassl}, {Steppa}, {Takahashi},
  {Tavernier}, {Taylor}, {Terrier}, {Tiziani}, {Tluczykont}, {Tomankova},
  {Trichard}, {Tsirou}, {Tuffs}, {Uchiyama}, {van der Walt}, {van Eldik}, {van
  Rensburg}, {van Soelen}, {Vasileiadis}, {Veh}, {Venter}, {Vincent}, {Vink},
  {V{\"o}lk}, {Vuillaume}, {Wadiasingh}, {Wagner}, {Watson}, {Werner}, {White},
  {Wierzcholska}, {Wong}, {Yusafzai}, {Zacharias}, {Zanin}, {Zargaryan},
  {Zdziarski}, {Zech}, {Zhu}, {Zorn}, {Zouari}, {{\.Z}ywucka}, {MAGIC
  Collaboration}, {Acciari}, {Ansoldi}, {Antonelli}, {Engels}, {Artero},
  {Asano}, {Baack}, {Babi{\'c}}, {Baquero}, {de Almeida}, {Barrio}, {Becerra
  Gonz{\'a}lez}, {Bednarek}, {Bellizzi}, {Bernardini}, {Bernardos}, {Berti},
  {Besenrieder}, {Bhattacharyya}, {Bigongiari}, {Biland}, {Blanch}, {Bonnoli},
  {Bo{\v{s}}njak}, {Busetto}, {Carosi}, {Ceribella}, {Cerruti}, {Chai},
  {Chilingarian}, {Cikota}, {Colak}, {Colombo}, {Contreras}, {Cortina},
  {Covino}, {D'Amico}, {D'Elia}, {da Vela}, {Dazzi}, {de Angelis}, {de Lotto},
  {Delfino}, {Delgado}, {Delgado Mendez}, {Depaoli}, {di Pierro}, {di Venere},
  {Do Souto Espi{\~n}eira}, {Dominis Prester}, {Donini}, {Dorner}, {Doro},
  {Elsaesser}, {Ramazani}, {Fattorini}, {Ferrara}, {Fonseca}, {Font}, {Fruck},
  {Fukami}, {Garc{\'\i}a L{\'o}pez}, {Garczarczyk}, {Gasparyan}, {Gaug},
  {Giglietto}, {Giordano}, {Gliwny}, {Godinovi{\'c}}, {Green}, {Green},
  {Hadasch}, {Hahn}, {Heckmann}, {Herrera}, {Hoang}, {Hrupec}, {H{\"u}tten},
  {Inada}, {Inoue}, {Ishio}, {Iwamura}, {Jim{\'e}nez}, {Jormanainen}, {Jouvin},
  {Kajiwara}, {Karjalainen}, {Kerszberg}, {Kobayashi}, {Kubo}, {Kushida},
  {Lamastra}, {Lelas}, {Leone}, {Lindfors}, {Lombardi}, {Longo},
  {L{\'o}pez-Coto}, {L{\'o}pez-Moya}, {L{\'o}pez-Oramas}, {Loporchio}, {Machado
  de Oliveira Fraga}, {Maggio}, {Majumdar}, {Makariev}, {Mallamaci}, {Maneva},
  {Manganaro}, {Mannheim}, {Maraschi}, {Mariotti}, {Mart{\'\i}nez}, {Mazin},
  {Menchiari}, {Mender}, {Mi{\'c}anovi{\'c}}, {Miceli}, {Miener}, {Minev},
  {Miranda}, {Mirzoyan}, {Molina}, {Moralejo}, {Morcuende}, {Moreno},
  {Moretti}, {Neustroev}, {Nigro}, {Nilsson}, {Nishijima}, {Noda}, {Nozaki},
  {Ohtani}, {Oka}, {Otero-Santos}, {Paiano}, {Palatiello}, {Paneque},
  {Paoletti}, {Paredes}, {Pavleti{\'c}}, {Pe{\~n}il}, {Perennes}, {Persic},
  {Moroni}, {Prandini}, {Priyadarshi}, {Puljak}, {Rhode}, {Rib{\'o}}, {Rico},
  {Righi}, {Rugliancich}, {Saha}, {Sahakyan}, {Saito}, {Sakurai}, {Satalecka},
  {Saturni}, {Schleicher}, {Schmidt}, {Schweizer}, {Sitarek},
  {{\v{S}}nidari{\'c}}, {Sobczynska}, {Spolon}, {Stamerra}, {Strom}, {Strzys},
  {Suda}, {Suri{\'c}}, {Takahashi}, {Tavecchio}, {Temnikov}, {Terzi{\'c}},
  {Teshima}, {Tosti}, {Truzzi}, {Tutone}, {Ubach}, {van Scherpenberg}, {Vanzo},
  {Vazquez Acosta}, {Ventura}, {Verguilov}, {Vigorito}, {Vitale}, {Vovk},
  {Will}, {Wunderlich}, {Zari{\'c}}, {VERITAS Collaboration}, {Adams},
  {Benbow}, {Brill}, {Capasso}, {Christiansen}, {Chromey}, {Daniel}, {Errando},
  {Farrell}, {Feng}, {Finley}, {Fortson}, {Furniss}, {Gent}, {Giuri}, {Hassan},
  {Hervet}, {Holder}, {Hughes}, {Humensky}, {Jin}, {Kaaret}, {Kertzman},
  {Kieda}, {Kumar}, {Lang}, {Lundy}, {Maier}, {Moriarty}, {Mukherjee}, {Nieto},
  {Nievas-Rosillo}, {O'Brien}, {Ong}, {Otte}, {Patel}, {Pfrang}, {Pohl},
  {Prado}, {Pueschel}, {Quinn}, {Ragan}, {Reynolds}, {Ribeiro}, {Richards},
  {Roache}, {Rulten}, {Ryan}, {Santander}, {Sembroski}, {Shang}, {Weinstein},
  {Williams}, {Williamson}, {Eavn Collaboration}, {Hirota}, {Cui}, {Niinuma},
  {Ro}, {Sakai}, {Sawada-Satoh}, {Wajima}, {Wang}, {Liu}, \&
  {Yonekura}}]{Algaba2021}
{EHT MWL Science Working Group}, {Algaba}, J.~C., {Anczarski}, J., {et~al.}
  2021, \apjl, 911, L11, \dodoi{10.3847/2041-8213/abef71}

\bibitem[{{Event Horizon Telescope Collaboration} {et~al.}(2021){Event Horizon
  Telescope Collaboration}, {Akiyama}, {Algaba}, {Alberdi}, {Alef}, {Anantua},
  {Asada}, {Azulay}, {Baczko}, {Ball}, {Balokovi{\'c}}, {Barrett}, {Benson},
  {Bintley}, {Blackburn}, {Blundell}, {Boland}, {Bouman}, {Bower}, {Boyce},
  {Bremer}, {Brinkerink}, {Brissenden}, {Britzen}, {Broderick}, {Broguiere},
  {Bronzwaer}, {Byun}, {Carlstrom}, {Chael}, {Chan}, {Chatterjee},
  {Chatterjee}, {Chen}, {Chen}, {Chesler}, {Cho}, {Christian}, {Conway},
  {Cordes}, {Crawford}, {Crew}, {Cruz-Osorio}, {Cui}, {Davelaar}, {De
  Laurentis}, {Deane}, {Dempsey}, {Desvignes}, {Dexter}, {Doeleman}, {Eatough},
  {Falcke}, {Farah}, {Fish}, {Fomalont}, {Ford}, {Fraga-Encinas}, {Friberg},
  {Fromm}, {Fuentes}, {Galison}, {Gammie}, {Garc{\'\i}a}, {Gelles}, {Gentaz},
  {Georgiev}, {Goddi}, {Gold}, {G{\'o}mez}, {G{\'o}mez-Ruiz}, {Gu}, {Gurwell},
  {Hada}, {Haggard}, {Hecht}, {Hesper}, {Himwich}, {Ho}, {Ho}, {Honma},
  {Huang}, {Huang}, {Hughes}, {Ikeda}, {Inoue}, {Issaoun}, {James}, {Jannuzi},
  {Janssen}, {Jeter}, {Jiang}, {Jimenez-Rosales}, {Johnson}, {Jorstad}, {Jung},
  {Karami}, {Karuppusamy}, {Kawashima}, {Keating}, {Kettenis}, {Kim}, {Kim},
  {Kim}, {Kim}, {Kino}, {Koay}, {Kofuji}, {Koch}, {Koyama}, {Kramer}, {Kramer},
  {Krichbaum}, {Kuo}, {Lauer}, {Lee}, {Levis}, {Li}, {Li}, {Lindqvist}, {Lico},
  {Lindahl}, {Liu}, {Liu}, {Liuzzo}, {Lo}, {Lobanov}, {Loinard}, {Lonsdale},
  {Lu}, {MacDonald}, {Mao}, {Marchili}, {Markoff}, {Marrone}, {Marscher},
  {Mart{\'\i}-Vidal}, {Matsushita}, {Matthews}, {Medeiros}, {Menten}, {Mizuno},
  {Mizuno}, {Moran}, {Moriyama}, {Moscibrodzka}, {M{\"u}ller}, {Musoke}, {Mus
  Mej{\'\i}as}, {Michalik}, {Nadolski}, {Nagai}, {Nagar}, {Nakamura},
  {Narayan}, {Narayanan}, {Natarajan}, {Nathanail}, {Neilsen}, {Neri}, {Ni},
  {Noutsos}, {Nowak}, {Okino}, {Olivares}, {Ortiz-Le{\'o}n}, {Oyama},
  {{\"O}zel}, {Palumbo}, {Park}, {Patel}, {Pen}, {Pesce}, {Pi{\'e}tu},
  {Plambeck}, {PopStefanija}, {Porth}, {P{\"o}tzl}, {Prather},
  {Preciado-L{\'o}pez}, {Psaltis}, {Pu}, {Ramakrishnan}, {Rao}, {Rawlings},
  {Raymond}, {Rezzolla}, {Ricarte}, {Ripperda}, {Roelofs}, {Rogers}, {Ros},
  {Rose}, {Roshanineshat}, {Rottmann}, {Roy}, {Ruszczyk}, {Rygl},
  {S{\'a}nchez}, {S{\'a}nchez-Arguelles}, {Sasada}, {Savolainen}, {Schloerb},
  {Schuster}, {Shao}, {Shen}, {Small}, {Sohn}, {SooHoo}, {Sun}, {Tazaki},
  {Tetarenko}, {Tiede}, {Tilanus}, {Titus}, {Toma}, {Torne}, {Trent},
  {Traianou}, {Trippe}, {van Bemmel}, {van Langevelde}, {van Rossum}, {Wagner},
  {Ward-Thompson}, {Wardle}, {Weintroub}, {Wex}, {Wharton}, {Wielgus}, {Wong},
  {Wu}, {Yoon}, {Young}, {Young}, {Younsi}, {Yuan}, {Yuan}, {Zensus}, {Zhao},
  \& {Zhao}}]{EHT2021}
{Event Horizon Telescope Collaboration}, {Akiyama}, K., {Algaba}, J.~C.,
  {et~al.} 2021, \apjl, 910, L13, \dodoi{10.3847/2041-8213/abe4de}

\bibitem[{{Feigelson} \& {Montmerle}(1999)}]{Feigelson1999}
{Feigelson}, E.~D., \& {Montmerle}, T. 1999, \araa, 37, 363,
  \dodoi{10.1146/annurev.astro.37.1.363}

\bibitem[{{Fender} \& {Belloni}(2004)}]{Fender04}
{Fender}, R., \& {Belloni}, T. 2004, \araa, 42, 317,
  \dodoi{10.1146/annurev.astro.42.053102.134031}

\bibitem[{{Fishbone} \& {Moncrief}(1976)}]{Fishbone1976}
{Fishbone}, L.~G., \& {Moncrief}, V. 1976, \apj, 207, 962,
  \dodoi{10.1086/154565}

\bibitem[{{Flaccomio} {et~al.}(2012){Flaccomio}, {Micela}, \&
  {Sciortino}}]{Flaccomio2012}
{Flaccomio}, E., {Micela}, G., \& {Sciortino}, S. 2012, \aap, 548, A85,
  \dodoi{10.1051/0004-6361/201219362}

\bibitem[{{Gebhardt} {et~al.}(2011){Gebhardt}, {Adams}, {Richstone}, {Lauer},
  {Faber}, {G{\"u}ltekin}, {Murphy}, \& {Tremaine}}]{Gebhardt2011}
{Gebhardt}, K., {Adams}, J., {Richstone}, D., {et~al.} 2011, \apj, 729, 119,
  \dodoi{10.1088/0004-637X/729/2/119}

\bibitem[{{Genzel} {et~al.}(2010){Genzel}, {Eisenhauer}, \&
  {Gillessen}}]{Genzel2010}
{Genzel}, R., {Eisenhauer}, F., \& {Gillessen}, S. 2010, Reviews of Modern
  Physics, 82, 3121, \dodoi{10.1103/RevModPhys.82.3121}

\bibitem[{{Genzel} {et~al.}(2003){Genzel}, {Sch{\"o}del}, {Ott}, {Eckart},
  {Alexander}, {Lacombe}, {Rouan}, \& {Aschenbach}}]{Genzel2003}
{Genzel}, R., {Sch{\"o}del}, R., {Ott}, T., {et~al.} 2003, \nat, 425, 934,
  \dodoi{10.1038/nature02065}

\bibitem[{{Gou} {et~al.}(2019){Gou}, {Liu}, {Kliem}, {Wang}, \&
  {Veronig}}]{Gou2019}
{Gou}, T., {Liu}, R., {Kliem}, B., {Wang}, Y., \& {Veronig}, A.~M. 2019,
  Science Advances, 5, 7004, \dodoi{10.1126/sciadv.aau7004}

\bibitem[{{Gravity Collaboration} {et~al.}(2018){Gravity Collaboration},
  {Abuter}, {Amorim}, {Baub{\"o}ck}, {Berger}, {Bonnet}, {Brandner},
  {Cl{\'e}net}, {Coud{\'e} Du Foresto}, {de Zeeuw}, {Deen}, {Dexter}, {Duvert},
  {Eckart}, {Eisenhauer}, {F{\"o}rster Schreiber}, {Garcia}, {Gao}, {Gendron},
  {Genzel}, {Gillessen}, {Guajardo}, {Habibi}, {Haubois}, {Henning}, {Hippler},
  {Horrobin}, {Huber}, {Jim{\'e}nez-Rosales}, {Jocou}, {Kervella}, {Lacour},
  {Lapeyr{\`e}re}, {Lazareff}, {Le Bouquin}, {L{\'e}na}, {Lippa}, {Ott},
  {Panduro}, {Paumard}, {Perraut}, {Perrin}, {Pfuhl}, {Plewa}, {Rabien},
  {Rodr{\'\i}guez-Coira}, {Rousset}, {Sternberg}, {Straub}, {Straubmeier},
  {Sturm}, {Tacconi}, {Vincent}, {von Fellenberg}, {Waisberg}, {Widmann},
  {Wieprecht}, {Wiezorrek}, {Woillez}, \& {Yazici}}]{Gravity2018}
{Gravity Collaboration}, {Abuter}, R., {Amorim}, A., {et~al.} 2018, \aap, 618,
  L10, \dodoi{10.1051/0004-6361/201834294}

\bibitem[{{Hada} {et~al.}(2014){Hada}, {Giroletti}, {Kino}, {Giovannini},
  {D'Ammando}, {Cheung}, {Beilicke}, {Nagai}, {Doi}, {Akiyama}, {Honma},
  {Niinuma}, {Casadio}, {Orienti}, {Krawczynski}, {G{\'o}mez}, {Sawada-Satoh},
  {Koyama}, {Cesarini}, {Nakahara}, \& {Gurwell}}]{Hada2014}
{Hada}, K., {Giroletti}, M., {Kino}, M., {et~al.} 2014, \apj, 788, 165,
  \dodoi{10.1088/0004-637X/788/2/165}

\bibitem[{{Hawley} {et~al.}(1995){Hawley}, {Gammie}, \& {Balbus}}]{Hawley95}
{Hawley}, J.~F., {Gammie}, C.~F., \& {Balbus}, S.~A. 1995, \apj, 440, 742,
  \dodoi{10.1086/175311}

\bibitem[{{Hawley} {et~al.}(2011){Hawley}, {Guan}, \& {Krolik}}]{Hawley2011}
{Hawley}, J.~F., {Guan}, X., \& {Krolik}, J.~H. 2011, \apj, 738, 84,
  \dodoi{10.1088/0004-637X/738/1/84}

\bibitem[{{Ho}(2008)}]{Ho2008}
{Ho}, L.~C. 2008, \araa, 46, 475,
  \dodoi{10.1146/annurev.astro.45.051806.110546}

\bibitem[{{Igumenshchev} {et~al.}(2003){Igumenshchev}, {Narayan}, \&
  {Abramowicz}}]{Igum03}
{Igumenshchev}, I.~V., {Narayan}, R., \& {Abramowicz}, M.~A. 2003, \apj, 592,
  1042, \dodoi{10.1086/375769}

\bibitem[{{King} {et~al.}(2016){King}, {Miller}, {Bietenholz}, {G{\"u}ltekin},
  {Reynolds}, {Mioduszewski}, {Rupen}, \& {Bartel}}]{King2016}
{King}, A.~L., {Miller}, J.~M., {Bietenholz}, M., {et~al.} 2016, Nature
  Physics, 12, 772, \dodoi{10.1038/nphys3724}

\bibitem[{{Lee}(2020)}]{Lee2020}
{Lee}, C.-F. 2020, \aapr, 28, 1, \dodoi{10.1007/s00159-020-0123-7}

\bibitem[{{Lin} \& {Forbes}(2000)}]{2000JGR...105.2375L}
{Lin}, J., \& {Forbes}, T.~G. 2000, \jgr, 105, 2375,
  \dodoi{10.1029/1999JA900477}

\bibitem[{{Lou} {et~al.}(2003){Lou}, {Wang}, {Fan}, {Wang}, \&
  {Wang}}]{Lou2003}
{Lou}, Y.-Q., {Wang}, Y.-M., {Fan}, Z., {Wang}, S., \& {Wang}, J.~X. 2003,
  \mnras, 345, 809, \dodoi{10.1046/j.1365-8711.2003.06993.x}

\bibitem[{{Markoff} {et~al.}(2001){Markoff}, {Falcke}, {Yuan}, \&
  {Biermann}}]{Markoff2001}
{Markoff}, S., {Falcke}, H., {Yuan}, F., \& {Biermann}, P.~L. 2001, \aap, 379,
  L13, \dodoi{10.1051/0004-6361:20011346}

\bibitem[{{Marrone} {et~al.}(2008){Marrone}, {Baganoff}, {Morris}, {Moran},
  {Ghez}, {Hornstein}, {Dowell}, {Mu{\~n}oz}, {Bautz}, {Ricker}, {Brandt},
  {Garmire}, {Lu}, {Matthews}, {Zhao}, {Rao}, \& {Bower}}]{Marrone2008}
{Marrone}, D.~P., {Baganoff}, F.~K., {Morris}, M.~R., {et~al.} 2008, \apj, 682,
  373, \dodoi{10.1086/588806}

\bibitem[{{McKinney} {et~al.}(2012){McKinney}, {Tchekhovskoy}, \&
  {Blandford}}]{McKinney2012}
{McKinney}, J.~C., {Tchekhovskoy}, A., \& {Blandford}, R.~D. 2012, \mnras, 423,
  3083, \dodoi{10.1111/j.1365-2966.2012.21074.x}

\bibitem[{{Narayan} {et~al.}(2003){Narayan}, {Igumenshchev}, \&
  {Abramowicz}}]{Nar03}
{Narayan}, R., {Igumenshchev}, I.~V., \& {Abramowicz}, M.~A. 2003, \pasj, 55,
  L69, \dodoi{10.1093/pasj/55.6.L69}

\bibitem[{{Nathanail} {et~al.}(2020){Nathanail}, {Fromm}, {Porth}, {Olivares},
  {Younsi}, {Mizuno}, \& {Rezzolla}}]{Nathanail20}
{Nathanail}, A., {Fromm}, C.~M., {Porth}, O., {et~al.} 2020, \mnras, 495, 1549,
  \dodoi{10.1093/mnras/staa1165}

\bibitem[{{Nathanail} {et~al.}(2021){Nathanail}, {Mpisketzis}, {Porth},
  {Fromm}, \& {Rezzolla}}]{Nathanail2021}
{Nathanail}, A., {Mpisketzis}, V., {Porth}, O., {Fromm}, C.~M., \& {Rezzolla},
  L. 2021, arXiv e-prints, arXiv:2111.03689.
\newblock \doarXiv{2111.03689}

\bibitem[{{Park} {et~al.}(2019){Park}, {Lee}, {Kim}, {Hodgson}, {Trippe},
  {Kim}, {Algaba}, {Kino}, {Zhao}, {Lee}, \& {Gurwell}}]{Park2019}
{Park}, J., {Lee}, S.-S., {Kim}, J.-Y., {et~al.} 2019, \apj, 877, 106,
  \dodoi{10.3847/1538-4357/ab1b27}

\bibitem[{{Penna} {et~al.}(2013){Penna}, {Kulkarni}, \& {Narayan}}]{Penna2013}
{Penna}, R.~F., {Kulkarni}, A., \& {Narayan}, R. 2013, \aap, 559, A116,
  \dodoi{10.1051/0004-6361/201219666}

\bibitem[{{Rauch} {et~al.}(2016){Rauch}, {Ros}, {Krichbaum}, {Eckart},
  {Zensus}, {Shahzamanian}, \& {Mu{\v{z}}i{\'c}}}]{Rauch2016}
{Rauch}, C., {Ros}, E., {Krichbaum}, T.~P., {et~al.} 2016, \aap, 587, A37,
  \dodoi{10.1051/0004-6361/201527286}

\bibitem[{{Ripperda} {et~al.}(2020){Ripperda}, {Bacchini}, \&
  {Philippov}}]{Ripperda20}
{Ripperda}, B., {Bacchini}, F., \& {Philippov}, A.~A. 2020, \apj, 900, 100,
  \dodoi{10.3847/1538-4357/ababab}

\bibitem[{{Ripperda} {et~al.}(2021){Ripperda}, {Liska}, {Chatterjee}, {Musoke},
  {Philippov}, {Markoff}, {Tchekhovskoy}, \& {Younsi}}]{Ripperda2021}
{Ripperda}, B., {Liska}, M., {Chatterjee}, K., {et~al.} 2021, arXiv e-prints,
  arXiv:2109.15115.
\newblock \doarXiv{2109.15115}

\bibitem[{{Rosenberg} \& {Ebrahimi}(2021)}]{Rosenberg2021}
{Rosenberg}, J., \& {Ebrahimi}, F. 2021, \apjl, 920, L29,
  \dodoi{10.3847/2041-8213/ac2b2e}

\bibitem[{{Roy} {et~al.}(2021){Roy}, {Sarkar}, {Chatterjee}, {Gupta},
  {Chitnis}, \& {Wiita}}]{Roy2021}
{Roy}, A., {Sarkar}, A., {Chatterjee}, A., {et~al.} 2021, arXiv e-prints,
  arXiv:2112.08955.
\newblock \doarXiv{2112.08955}

\bibitem[{{Scepi} {et~al.}(2021){Scepi}, {Dexter}, \& {Begelman}}]{Dexter2021}
{Scepi}, N., {Dexter}, J., \& {Begelman}, M.~C. 2021, arXiv e-prints,
  arXiv:2107.08056.
\newblock \doarXiv{2107.08056}

\bibitem[{{Shi} {et~al.}(2021){Shi}, {Li}, {Yuan}, \& {Zhu}}]{Shi2021}
{Shi}, F., {Li}, Z., {Yuan}, F., \& {Zhu}, B. 2021, Nature Astronomy, 5, 928,
  \dodoi{10.1038/s41550-021-01394-0}

\bibitem[{{Sklodowski} {et~al.}(2021){Sklodowski}, {Tripathi}, \&
  {Carter}}]{2021JPlPh..87f9016S}
{Sklodowski}, K.~D., {Tripathi}, S., \& {Carter}, T. 2021, Journal of Plasma
  Physics, 87, 905870616, \dodoi{10.1017/S0022377821001239}

\bibitem[{{Stone} {et~al.}(2020){Stone}, {Tomida}, {White}, \&
  {Felker}}]{Stone2020}
{Stone}, J.~M., {Tomida}, K., {White}, C.~J., \& {Felker}, K.~G. 2020, \apjs,
  249, 4, \dodoi{10.3847/1538-4365/ab929b}

\bibitem[{{Tchekhovskoy} {et~al.}(2011){Tchekhovskoy}, {Narayan}, \&
  {McKinney}}]{Tchek11}
{Tchekhovskoy}, A., {Narayan}, R., \& {McKinney}, J.~C. 2011, \mnras, 418, L79,
  \dodoi{10.1111/j.1745-3933.2011.01147.x}

\bibitem[{{Trippe} {et~al.}(2007){Trippe}, {Paumard}, {Ott}, {Gillessen},
  {Eisenhauer}, {Martins}, \& {Genzel}}]{Trippe2007}
{Trippe}, S., {Paumard}, T., {Ott}, T., {et~al.} 2007, \mnras, 375, 764,
  \dodoi{10.1111/j.1365-2966.2006.11338.x}

\bibitem[{{Wang} {et~al.}(2016){Wang}, {Zhuang}, {Hu}, {Liu}, {Shen}, \&
  {Chi}}]{Wang2016}
{Wang}, Y., {Zhuang}, B., {Hu}, Q., {et~al.} 2016, Journal of Geophysical
  Research (Space Physics), 121, 9316, \dodoi{10.1002/2016JA023075}

\bibitem[{{White} {et~al.}(2016){White}, {Stone}, \& {Gammie}}]{White16}
{White}, C.~J., {Stone}, J.~M., \& {Gammie}, C.~F. 2016, \apjs, 225, 22,
  \dodoi{10.3847/0067-0049/225/2/22}

\bibitem[{{Wilms} {et~al.}(2007){Wilms}, {Pottschmidt}, {Pooley}, {Markoff},
  {Nowak}, {Kreykenbohm}, \& {Rothschild}}]{Wilms2007}
{Wilms}, J., {Pottschmidt}, K., {Pooley}, G.~G., {et~al.} 2007, \apjl, 663,
  L97, \dodoi{10.1086/520508}

\bibitem[{{Wolk} {et~al.}(2005){Wolk}, {Harnden}, {Flaccomio}, {Micela},
  {Favata}, {Shang}, \& {Feigelson}}]{Wolk2005}
{Wolk}, S.~J., {Harnden}, F.~R., J., {Flaccomio}, E., {et~al.} 2005, \apjs,
  160, 423, \dodoi{10.1086/432099}

\bibitem[{{Yang} {et~al.}(2021){Yang}, {Yuan}, {Yuan}, \& {White}}]{Yang21}
{Yang}, H., {Yuan}, F., {Yuan}, Y.-F., \& {White}, C.~J. 2021, \apj, 914, 131,
  \dodoi{10.3847/1538-4357/abfe63}

\bibitem[{{Yuan} {et~al.}(2009){Yuan}, {Lin}, {Wu}, \& {Ho}}]{Yuan09}
{Yuan}, F., {Lin}, J., {Wu}, K., \& {Ho}, L.~C. 2009, \mnras, 395, 2183,
  \dodoi{10.1111/j.1365-2966.2009.14673.x}

\bibitem[{{Yuan} \& {Narayan}(2014)}]{YuanNar14}
{Yuan}, F., \& {Narayan}, R. 2014, \araa, 52, 529,
  \dodoi{10.1146/annurev-astro-082812-141003}

\bibitem[{{Yuan} {et~al.}(2004){Yuan}, {Quataert}, \& {Narayan}}]{Yuan2004}
{Yuan}, F., {Quataert}, E., \& {Narayan}, R. 2004, \apj, 606, 894,
  \dodoi{10.1086/383117}

\bibitem[{{Yuan} {et~al.}(2022){Yuan}, {Wang}, \& {Yang}}]{Yuan22}
{Yuan}, F., {Wang}, H., \& {Yang}, H. 2022, \apj, 924, 124,
  \dodoi{10.3847/1538-4357/ac4714}

\bibitem[{{Yuan} \& {Zhang}(2012)}]{Yuan2012}
{Yuan}, F., \& {Zhang}, B. 2012, \apj, 757, 56,
  \dodoi{10.1088/0004-637X/757/1/56}

\bibitem[{{Yusef-Zadeh} {et~al.}(2006){Yusef-Zadeh}, {Roberts}, {Wardle},
  {Heinke}, \& {Bower}}]{Yusef-Zadeh2006}
{Yusef-Zadeh}, F., {Roberts}, D., {Wardle}, M., {Heinke}, C.~O., \& {Bower},
  G.~C. 2006, \apj, 650, 189, \dodoi{10.1086/506375}

\bibitem[{{Zhang} \& {Low}(2005)}]{Zhang2005}
{Zhang}, M., \& {Low}, B.~C. 2005, \araa, 43, 103,
  \dodoi{10.1146/annurev.astro.43.072103.150602}

\bibitem[{{Zhao} {et~al.}(2020){Zhao}, {Yuan}, \&
  {Kumar}}]{2020MNRAS.499.1561Z}
{Zhao}, T.-L., {Yuan}, Y.-F., \& {Kumar}, R. 2020, \mnras, 499, 1561,
  \dodoi{10.1093/mnras/staa2600}

\end{thebibliography}
\bibliographystyle{aasjournal}



\end{document}